# Sparse Subspace Clustering via Two-Step Reweighted L1-Minimization: Algorithm and Provable Neighbor Recovery Rates

Jwo-Yuh Wu[*], Liang-Chi Huang, Ming-Hsun Yang, and Chun-Hung Liu

*Abstract*—Sparse subspace clustering (SSC) relies on sparse regression for accurate neighbor identification. Inspired by recent progress in compressive sensing, this paper proposes a new sparse regression scheme for SSC via two-step reweighted $\ell_1$-minimization, which also generalizes a two-step $\ell_1$-minimization algorithm introduced by E. J. Candès et al in [*The Annals of Statistics*, vol. 42, no. 2, pp. 669–699, 2014] without incurring extra algorithmic complexity. To fully exploit the prior information offered by the computed sparse representation vector in the first step, our approach places a weight on each component of the regression vector, and solves a weighted LASSO in the second step. We propose a data weighting rule suitable for enhancing neighbor identification accuracy. Then, under the formulation of the dual problem of weighted LASSO, we study in depth the theoretical neighbor recovery rates of the proposed scheme. Specifically, an interesting connection between the locations of nonzeros of the optimal sparse solution to the weighted LASSO and the indexes of the active constraints of the dual problem is established. Afterwards, under the semi-random model, analytic probability lower/upper bounds for various neighbor recovery events are derived. Our analytic results confirm that, with the aid of data weighting and if the prior neighbor information is enough accurate, the proposed scheme with a higher probability can produce many correct neighbors and few incorrect neighbors as compared to the solution without data weighting. Computer simulations are provided to validate our analytic study and evidence the effectiveness of the proposed approach.

*Index Terms*—Subspace clustering, sparse representation, sparse subspace clustering, neighbor recovery, discovery rate, performance guarantees, LASSO, weighted LASSO, compressive sensing.

## I. INTRODUCTION

### A. Overview

Consider a data set $\mathcal{Y} = \{\mathbf{y}_1, \cdots, \mathbf{y}_N\} \subset \mathbb{R}^n$ obeying the union-of-subspace model [1] under noise corruption. That is, the ground truth of $\mathcal{Y}$ is a disjoint union

$$\mathcal{Y} = \mathcal{Y}_1 \cup \mathcal{Y}_2 \cup \cdots \cup \mathcal{Y}_L, \tag{1.1}$$

where cluster $\mathcal{Y}_k \subset \mathbb{R}^n$ consists of $|\mathcal{Y}_k| > 0$ noisy data points originating from the $k$th subspace $\mathcal{S}_k$ of dimension $d_k$, and $|\mathcal{Y}_1| + \cdots + |\mathcal{Y}_L| = N$. Given $\mathcal{Y}$ with unknown $L$ and $d_k$, $1 \leq k \leq L$, the task of subspace clustering is to uncover the partition (1.1) [1]. Sparse subspace clustering (SSC) [1-2], which is built on the state-of-the-art compressive sensing (CS) and sparse representation techniques [3-7], is an

---

† This work is sponsored by the Ministry of Science and Technology of Taiwan under grants MOST 107-2634-F-009-002, MOST 108-2634-F-009-002, and MOST 108-2221-E-009-025-MY3, and is supported by MOST Joint Research Center for AI Technology and All Vista Healthcare. This paper is presented in part at *IEEE International Workshop on Machine Learning for Signal Processing*, Aalborg, Denmark, Sept. 2018.

J. Y. Wu, L. C. Huang, and M. S. Yang are with the Department of Electrical and Computer Engineering, the Institute of Communications Engineering, and the College of Electrical Engineering, National Chiao Tung University, Taiwan. Emails: jywu@cc.nctu.edu.tw; uuranus001@gmail.com; archenemy310017@hotmail.com.

C. H. Liu is with the Department of Electrical and Computer Engineering, Mississippi State University, Starkville MS, USA. Email: chliu@ece.msstate.edu.

* Contact author.

Table I. Outline of the SSC algorithm.

| SSC algorithm using $\ell_1$-minimization |
|---|
| **Input:** Observed data set $\mathcal{Y} = \{\mathbf{y}_1, \cdots, \mathbf{y}_N\}$ |
| 1. Neighbor identification for each $\mathbf{y}_i$ via solving (1.2) or (1.3) to obtain $\mathbf{c}_i^*$. |
| 2. Normalize the column vector $\mathbf{c}_i^*$ to be unit-norm, and let $\bar{\mathbf{c}}_i^* = [c_{i,1}^* \cdots c_{i,i-1}^* \ 0 \ c_{i,i+1}^* \cdots c_{i,N}^*]^T \in \mathbb{R}^N$. |
| 3. Set $\mathbf{C} = [c_{i,j}] = [\bar{\mathbf{c}}_1^* \cdots \bar{\mathbf{c}}_N^*]$, and $\mathbf{G} = [g_{i,j}]$, where $g_{i,j} = |c_{i,j}| + |c_{j,i}|$. |
| 4. Form a similarity graph with $N$ nodes, with the weight on the edge between the $(i, j)$ node pair equal to $g_{i,j}$. |
| 5. Apply spectral clustering to the similarity graph. |
| **Output:** Partition $\mathcal{Y} = \hat{\mathcal{Y}}_1 \cup \cdots \cup \hat{\mathcal{Y}}_{\hat{L}}$ |

effective means for dealing with this task (see Table I for an outline of the SSC algorithm). A crucial part of SSC is to identify the neighbors of each $\mathbf{y}_i$ via finding its sparse representation with respect to all other $\mathbf{y}_j$'s, $j \neq i$. Specifically, let us stack the elements of $\mathcal{Y}$ without $\mathbf{y}_i$ to form the matrix $\mathbf{Y}_{-i} = [\mathbf{y}_1 \cdots \mathbf{y}_{i-1} \ \mathbf{y}_{i+1} \cdots \mathbf{y}_N] \in \mathbb{R}^{n \times (N-1)}$, and try to express $\mathbf{y}_i$ as a linear combination of columns of $\mathbf{Y}_{-i}$, say, $\mathbf{y}_i = \mathbf{Y}_{-i}\mathbf{c}_i$, where[1] $\mathbf{c}_i = [c_{i,1} \cdots c_{i,i-1} \ c_{i,i+1} \cdots c_{i,N}]^T \in \mathbb{R}^{N-1}$. SSC then seeks a sparse solution $\mathbf{c}_i^*$, with the hope that the columns of $\mathbf{Y}_{-i}$ supported on the non-zeros of $\mathbf{c}_i^*$ are correct neighbors of $\mathbf{y}_i$ (i.e., they belong to the same cluster as $\mathbf{y}_i$). Computing $\mathbf{c}_i^*$ is commonly done via $\ell_1$-minimization [2]. For example, one would like to solve

$$\mathbf{c}_i^* = \arg\min \ \|\mathbf{c}_i\|_1 \ \text{subject to} \ \|\mathbf{y}_i - \mathbf{Y}_{-i}\mathbf{c}_i\|_2 \leq \tau, \tag{1.2}$$

where $\tau > 0$ is proportional to the noise level, or to solve the popular alternative

$$\mathbf{c}_i^* = \arg\min \ \lambda\|\mathbf{c}_i\|_1 + \frac{1}{2}\|\mathbf{y}_i - \mathbf{Y}_{-i}\mathbf{c}_i\|_2^2, \tag{1.3}$$

which is the LASSO estimator [8-9]. Note that, in the LASSO regression (1.3), the regularization parameter $\lambda > 0$ accounts for the tradeoff between sparsity promotion (large $\lambda$) and noise reduction (small $\lambda$). With the optimal $\mathbf{c}_i^* = [c_{i,1}^* \cdots c_{i,i-1}^* \ c_{i,i+1}^* \cdots c_{i,N}^*]^T$, following the terminology in [10] we say an index pair $(i,j)$, $1 \leq i \neq j \leq N$, is a *discovery* if $c_{i,j}^* \neq 0$. A pair $(i,j)$ is a *true* (*false*, respectively) *discovery* if it is a discovery and, moreover, $\mathbf{y}_i$ and $\mathbf{y}_j$ originate from the same (distinct, respectively) cluster(s). In other words, by a true (false, respectively) discovery $(i,j)$ we mean the algorithm correctly identifies (misidentifies, respectively) $\mathbf{y}_j$ as a neighbor of $\mathbf{y}_i$. Many existing works on noisy SSC were dedicated to investigating the conditions guaranteeing the subspace detection property (SDP) [10-11], i.e., the computed sparse representation $\mathbf{c}_i^*$ produces only true discoveries, thus returning a subset of neighbors from the correct cluster; related studies can be found in, e.g., [10-14].

---

1. Towards consistent discussions, the entries of $\mathbf{c}_i$ are indexed in the same way as for columns of $\mathbf{Y}_{-i}$, namely, the index $i$ is removed.

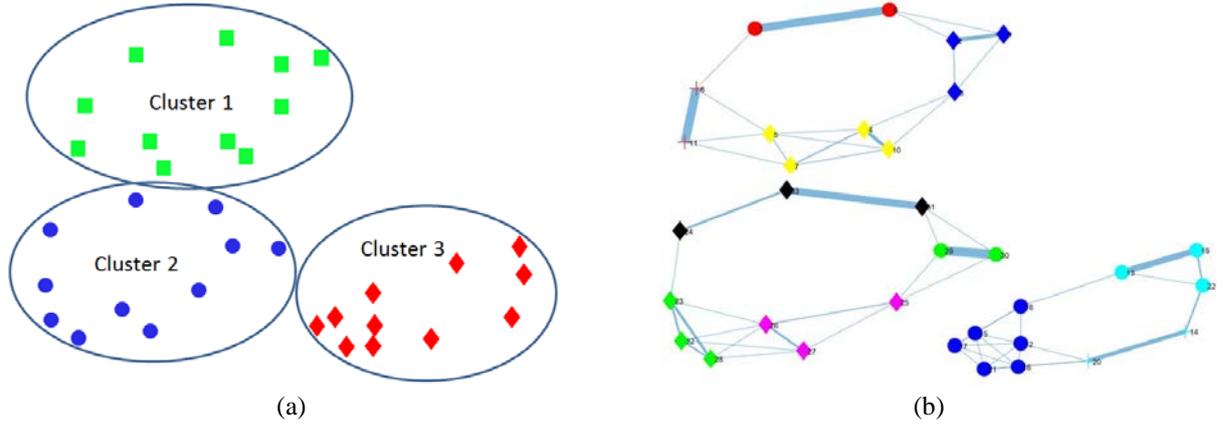

| (a) | (b) |

Fig. 1. Noisy SSC classification result of a 3-cluster data set; the LASSO estimator (1.3) with a large $\lambda$ is used for computing the sparse representation of each data point. (a) The ground truth partition. (b) The clustering outcome, in which the data set is incorrectly classified into 10 groups, even though SDP is satisfied; thick (thin, respectively) edges correspond to large (small, respectively) weights $g_{i,j}$.

*B. Research Motivation*

According to [10, p.675], fulfillment of SDP alone does not necessarily ensure good data classification accuracy. This can be simply illustrated via the example shown in Fig. 1, in which the data set obeys a ground truth partition into three clusters (see Fig. 1-(a)). To implement SSC, we use the LASSO sparse regression (1.3) with a large $\lambda$ for neighbor identification. Fig. 1-(b) depicts the clustering outcome, wherein data points marked by the same color are reported as residing in the same cluster. The result clearly demonstrates that SDP is satisfied since for each data point the corresponding edges are directed to those all belonging to the same cluster. However, this data set is incorrectly classified into a total number of 10 clusters (whereas the ground truth is 3). This example indicates that the data classification accuracy could be quite poor even if SDP is satisfied. This is mainly because the LASSO estimator (1.3) with a large $\lambda$ overly promotes sparsity, resulting in "too few" true discoveries. One may argue the classification failure can be well compensated by decreasing the value of $\lambda$, so as to obtain a "not overly sparse" $\mathbf{c}_i^*$. Even though this can increase true discoveries, more undesirable false discoveries will meanwhile be produced, thereby rendering data clustering still prone to errors. Fig. 2 illustrates a case of misclassification caused by the presence of many false discoveries (due to the use of a small $\lambda$). It is noted that accurate data classification is possible when there are sufficiently many true discoveries and few false discoveries, thanks to the use of spectral clustering that follows the sparse regression step [10, p.675]. Fig. 3 goes on to illustrate one such example, using a properly tuned moderate $\lambda$. We remark that, in the context of SSC with LASSO, how to find precise rules regarding the design of an appropriate $\lambda$ is still a fairly difficult problem.

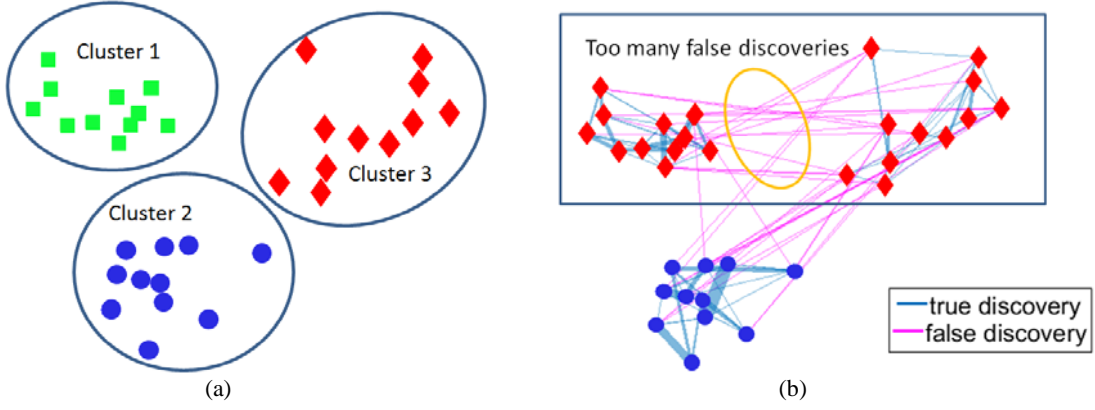

(a)                (b)

Fig. 2. Noisy SSC classification result of a 3-cluster data set, using the LASSO regression (1.3) with a small $\lambda$. (a) The ground truth partition. (b) The clustering outcome, in which the data set is incorrectly classified into just 2 groups due to too many false discoveries; thick (thin, respectively) edges correspond to large (small, respectively) weights $g_{i,j}$.

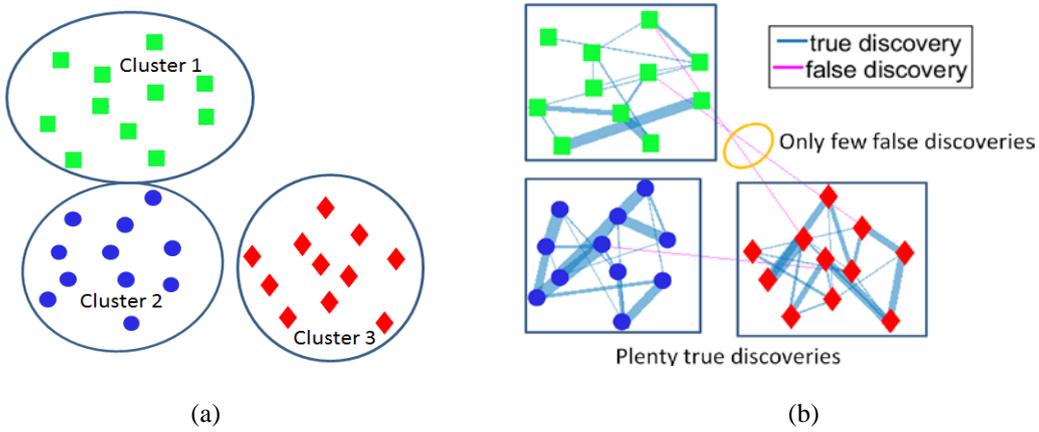

(a)                (b)

Fig. 3. Noisy SSC classification result of a 3-cluster data set, using the LASSO regression (1.3) with a tuned moderate $\lambda$. (a) The ground truth partition. (b) Correct data clustering is achieved as there are sufficiently many true discoveries and few false discoveries; thick (thin, respectively) edges correspond to large (small, respectively) weights $g_{i,j}$.

Motivated by the above facts, one shall devise sparse regression schemes able to systematically regulate the sparsity-promoting action, in a way that *many true discoveries are constructed, while the presence of false recoveries, if any, is somehow under control (thus, not that harmful)*. Among the existing proposals capable of tackling this challenge, a data-driven approach based on two-step $\ell_1$-minimization was proposed in [10], whereby (i) a quadratically-constrained $\ell_1$-minimization algorithm (1.2) is first conducted for each data point $\mathbf{y}_i$ to find a "coarse" sparse representation, say, $\tilde{\mathbf{c}}_i = [\tilde{c}_{i,1} \cdots \tilde{c}_{i,i-1} \ \tilde{c}_{i,i+1} \cdots \tilde{c}_{i,N}]^T$, based on which an estimate of the subspace dimension, and in turn a good regularization factor $\lambda$, are obtained[2]; (ii) a LASSO regression (1.3) with the computed $\lambda$ is then performed to find an updated sparse representation that can provably produce many true discoveries.

---

2. According to [10], if a data point $\mathbf{y}_i$ originates from a $d$-dimensional subspace, a good choice of $\lambda$ is around $\sqrt{(1/d)}$.

Table II. Proposed two-step reweighted $\ell_1$-minimization based sparse regression for noisy SSC.

| Sparse regression via two-step reweighted $\ell_1$-minimization |
|---|
| **Input:** Observed data set $\mathcal{Y} = \{\mathbf{y}_1, \cdots, \mathbf{y}_N\}$ |
| **for** $i = 1 \cdots N$ **do** <br> 1. Solve <br> $\quad\quad \tilde{\mathbf{c}}_i = \arg\min \|\mathbf{c}_i\|_1$ subject to $\|\mathbf{y}_i - \mathbf{Y}_{-i}\mathbf{c}_i\|_2 \leq \tau, 1 \leq i \leq N$. <br> 2. Set $\lambda = f(\|\tilde{\mathbf{c}}_i\|_1)$, for some pre-determined function $f(\cdot)$. <br> 3. Set the weighting coefficients $w_j^{(i)}$ according to (1.4). <br> 4. With $\lambda = f(\|\tilde{\mathbf{c}}_i\|_1)$, solve the weighted LASSO (1.5) for $\mathbf{c}_i^*$, $1 \leq i \leq N$. <br> **end** |
| **Output:** sparse representation vectors $\{\mathbf{c}_1^*, \cdots, \mathbf{c}_N^*\}$ |

## C. Paper Contribution

*I. Algorithm Development:* Inspired by the two-step method in [10], this paper proposes a new SSC scheme based on reweighted $\ell_1$-minimization [15]. The key idea behind is that the availability of $\tilde{\mathbf{c}}_i$ in the first step not just reveals the knowledge of the subspace dimension. More importantly, the amplitude of $\tilde{c}_{i,j}$ further offers certain side information about the neighbors of $\mathbf{y}_i$; indeed, data points corresponding to a large amplitude $|\tilde{c}_{i,j}|$ are likely to originate from the same cluster. Such side information, when properly exploited, can potentially improve the clustering accuracy. Towards this end, we propose to place a weight $w_j^{(i)} > 0$, chosen to be

$$w_j^{(i)} = \frac{\varepsilon}{|\tilde{c}_{i,j}| + \varepsilon}, \quad (1.4)$$

where $\varepsilon > 0$ is a small constant, on the $j$th entry of $\mathbf{c}_i$ to reflect the level of $|\tilde{c}_{i,j}|$ (i.e., confidence in $\mathbf{y}_j$ being a neighbor). With data weighting, in the second step we instead solve a weighted LASSO

$$\mathbf{c}_i^* = \arg\min\ \lambda\|\mathbf{W}_i\mathbf{c}_i\|_1 + \frac{1}{2}\|\mathbf{y}_i - \mathbf{Y}_{-i}\mathbf{c}_i\|_2^2, \ 1 \leq i \leq N, \quad (1.5)$$

where $\mathbf{W}_i = diag\{w_1^{(i)}, \cdots, w_{i-1}^{(i)}, w_{i+1}^{(i)}, \cdots w_N^{(i)}\} \in \mathbb{R}^{(N-1)\times(N-1)}$ is a diagonal weighting matrix; see Table II for an outline of the proposed algorithm. In the conference version of this paper [16], we have discussed in detail why the weighting rule (1.4) is suitable for the SSC purpose. The proposed approach is reminiscent of the reweighted $\ell_1$-minimization in the CS literature [15, 17-20], which has been a popular technique for improving sparse signal reconstruction performance.

*II. Theoretical Neighbor Recovery Rate Analysis:* We study in depth the analytic neighbor recovery rates of the proposed scheme. Our main purpose is to make the following fact precise: When employing the proposed weighting rule (1.4), the weighted LASSO in the second step is more likely to produce many true discoveries and few false discoveries, as compared to the method in [10]. This explains why the proposed approach can yield a better data classification performance, as illustrated by the simulation results. Our analysis is built on the dual problem of the weighted LASSO. Specifically, under the duality formulation, an explicit connection between the discoveries, which are determined by the support of the optimal $\mathbf{c}_i^*$ to the primal problem (1.5), and the indexes of the active constraints of the dual problem is established; in particular, true discoveries can be identified as those "correctly activated" constraints. Under the semi-random model [13] and by exploiting the polyhedron geometry underpinning the dual problem, we derive analytic probability lower/upper bounds for three important events, namely, the recovered outcome contains: (Event 1) less than $k_t(>0)$ true discoveries and at most $k_f(\geq 0)$ false discoveries, (Event 2) at most $k_f$ false discoveries, and (Event 3) at least $k_t$ true discoveries and at most $k_f$ false discoveries. Our analytic results confirm that, with the aid of data weighting (1.4) and if the prior neighbor information is accurate enough, the proposed approach can promote many true discoveries and few false discoveries with a higher probability, overall leading to better data clustering accuracy, as compared to the method in [10] without data weighting. We then provide experimental results to validate our theoretical study and evidence the effectiveness of the proposed method.

*D. Connection to Existing Works*

Since the publication of the seminal work [2], SSC has been intensively studied in the literature. However, theoretical neighbor recovery rate analysis and mathematical performance guarantees were relatively less explored. In the noiseless case, [2, 11] derived sufficient conditions for SDP using metrics and tools from convex geometry (e.g., subspace affinity, in-radius, dual direction, polytopes, etc.). For noisy SSC, [10] then conducted deep recovery rate analyses for the two-step $\ell_1$-minimization algorithm without data weighting in the second step. By means of various probability lower bounds, it was proven in [10] that, under mild assumptions on subspace orientation, sample density, and noise levels, the two-step algorithm in [10] is very likely to produce many true discoveries. The work [12] also considered the noisy case, and derived both deterministic and probabilistic performance guarantees for standard SSC, whereby a single LASSO without data weighting is conducted for neighbor identification. Our current study can be regarded as an extension of [10] to the case when data weighting is employed in the second step, in order to better exploit the prior neighbor information acquired in the first step. As far as we know, our work is the first in the SSC literature that investigates provable neighbor recovery rates under the two-step reweighted $\ell_1$-minimization framework. A very recent work [13] also considered noisy SSC employing LASSO sparse regression. The study therein mainly focused on performance guarantees with dimensionality-reduced data; still, no data weighting is considered since the problem formulation is under standard SSC without prior neighbor information. It is noted that greedy-based sparse regression using, e.g., (orthogonal) matching pursuit, is a popular low-complexity alternative to $\ell_1$-minimization [5-6]. Performance guarantees for SSC using greedy-based neighbor identification have been addressed in, e.g., [14] and [21]. Also, the work [22] proposed a simple thresholding-based neighbor identification algorithm for SSC; statistical performance analysis was provided to justify the robustness of the algorithm against noise.

Finally, we would like to remark that weighted/reweighted $\ell_1$-minimization has been as an effective means for improving conventional $\ell_1$-minimization based sparse signal recovery in the CS literature. Notably, a similar two-step reweighted $\ell_1$-minimization algorithm (a basis pursuit with a quadratic constraint in the first step, followed by a LASSO in the second step) for sparse signal reconstruction and its performance guarantee have been proposed and analyzed in [20]. Related applications of reweighted $\ell_1$-minimization in sparse signal estimation/detection have also been found in many areas, such as channel estimation in MIMO wireless communications [23-24], data gathering and signal reconstruction in wireless sensor networks [25-26], computational/medical imagining [27-28], remote sensing [29-31], smart grids [32], and switched control system design [33], etc.

The rest of this paper is organized as follows. Section II presents the problem formulation under the dual problem of weighted LASSO as well as the underlying polyhedron geometry, which clearly motivates why data weighting can improve the recovery rate performance. Section III presents the main theoretical results of this paper. Section IV provides numerical simulations to validate our analysis and illustrate the performance of the proposed approach. Section V presents the proofs of the main mathematical theorems. Finally, Section VI is the conclusion. To ease reading, many detailed technical derivations are relegated to appendix.

## II. DUAL OF WEIGHTRED LASSO AND ITS GEOMETRY

This section presents the basics for our recovery rate analysis. Section II-A first overviews the dual problem of the weighted LASSO, and establishes an associated important property that is central to our problem formulation. Section II-B then shows the polyhedron geometry underpinning the dual problem. This facilitates the understanding of why data weighting can improve the accuracy of neighbor identification, as will be discussed in Section II-C. Finally, Section II-D presents a mathematical characterization of the elements of the polyhedron sets that will be used in the subsequent analyses.

*A. Dual Problem of Weighted LASSO*

Since the proposed algorithm identifies the neighbors on a sample-by-sample basis, there is no loss of generality to consider the last data sample $\mathbf{y}_N$; that is, we seek a sparse linear combination of $\mathbf{y}_1, \cdots, \mathbf{y}_{N-1}$ to best approximate $\mathbf{y}_N$ using weighted LASSO (1.5). We first note that the optimal sparse solution $\mathbf{c}_N^*$ to the primal problem (1.5) is hard to analytically characterize, not even mention to identify the non-zero entries, in particular, which among them yield true/false discoveries. To get rid of this difficulty, we instead consider the dual problem of weighted LASSO, namely (e.g., [34, eq. (2)]),

$$\mathbf{z}^*(\mathbf{y}_N) = \underset{\mathbf{z} \in \mathcal{P}}{\operatorname{argmin}} \ \|\mathbf{z} - \mathbf{y}_N\|_2^2, \text{ where } \mathcal{P} \triangleq \{\mathbf{z} \in \mathbb{R}^n \mid \|\mathbf{W}_N^{-1} \mathbf{Y}_{-N}^T \mathbf{z}\|_\infty \leq \lambda\}. \tag{2.1}$$

Notably, $\mathbf{z}^*(\mathbf{y}_N)$ in (2.1) is the optimal projection of the data point $\mathbf{y}_N$ onto the feasible polyhedron $\mathcal{P}$. Observe that the constraint in the dual problem (2.1) is essentially a set of $N-1$ scalar inequality constraints, namely, $(1/w_j^{(N)})|\mathbf{y}_j^T \mathbf{z}| \leq \lambda$, $1 \leq j \leq N-1$. For $\mathbf{z} \in \mathcal{P}$, we will simply call the $j$th constraint active[3] if $(1/w_j^{(N)})|\mathbf{y}_j^T \mathbf{z}| = \lambda$, and inactive if $(1/w_j^{(N)})|\mathbf{y}_j^T \mathbf{z}| < \lambda$. The next lemma pins down an interesting connection between the optimal sparse solution $\mathbf{c}_N^*$ to the primal problem (1.5) and the active constraints of the dual problem (2.1). This lays the foundation behind our recovery rate analysis.

*Lemma 2.1:* Let $\mathbf{c}_N^* = [c_{N,1}^* \ c_{N,2}^* \cdots c_{N,N-1}^*]^T \in \mathbb{R}^{N-1}$ and $\mathbf{z}^*(\mathbf{y}_N) \in \mathbb{R}^n$ be the optimal solutions to, respectively, the primal problem (1.5) and the dual problem (2.1). The following results hold.

(a) If $c_{N,j}^* \neq 0$, then $(1/w_j^{(N)})|\mathbf{y}_j^T \mathbf{z}^*(\mathbf{y}_N)| = \lambda$.

(b) Assume that $(1/w_j^{(N)})|\mathbf{y}_j^T \mathbf{z}^*(\mathbf{y}_N)| = \lambda$. If $c_{N,j}^* = 0$, then $\mathbf{y}_N \in \mathcal{H} \subset \mathbb{R}^n$, where $\mathcal{H}$ is a translation of a subspace with dimension less than $n$.

*[Proof]:* See Appendix A. □

Part (a) of Lemma 2.1 shows that $c_{N,j}^* \neq 0$, i.e., the index pair $(N, j)$ is a discovery, only if the $j$th scalar constraint of the dual problem (2.1) is active. Despite the converse is not true (cf. Part (b)), a null entry $c_{N,j}^* = 0$ occurs only if the data point $\mathbf{y}_N$ falls within a "very small" subset of the ambient space $\mathbb{R}^n$. The significance of Lemma 2.1 is two-fold. Firstly, recall that in the setting of noisy SSC the data points are noise-corrupted and typically assumed to obey certain probability distributions (the considered assumptions on signal and noise distributions are quite standard and will be made precise in Section III). Hence, conditioned on the $j$th constraint being active, $\Pr(\{c_{N,j}^* = 0 \mid (1/w_j^{(N)})|\mathbf{y}_j^T \mathbf{z}^*(\mathbf{y}_N)| = \lambda\})$ never exceeds $\Pr(\{\mathbf{y}_N \in \mathcal{H} \mid (1/w_j^{(N)})|\mathbf{y}_j^T \mathbf{z}^*(\mathbf{y}_N)| = \lambda\})$, which is identically zero since the subset $\mathcal{H}$, as a translation of a low-dimensional subspace of $\mathbb{R}^n$, is (Lebesgue) measure zero [35]. This immediately implies $\Pr(\{c_{N,j}^* \neq 0 \mid (1/w_j^{(N)})|\mathbf{y}_j^T \mathbf{z}^*(\mathbf{y}_N)| = \lambda\}) = 1$. That is to say, the $j$th constraint being active guarantees $(N, j)$ is a discovery with probability one (kinds of a "weak" converse result of part (a) of Lemma 2.1). With the above in mind, we will hereafter identify a discovery $(N, j)$ with the $j$th constraint of (2.1) being active. As a result, the formulation of identifying true discoveries recovered by $\mathbf{c}_N^*$ amounts to checking which constraints of the dual problem (2.1) are "correctly activated". Secondly, the dual problem (2.1) enjoys a nice geometric structure: it is essentially about finding the optimal projection of $\mathbf{y}_N$ onto a polyhedron. Such a geometric perspective can offer a rigorous and systematic way of characterizing the active/inactive constraints, in turn facilitating our neighbor recovery rate analysis as will be shown later.

---

3. Formally speaking, the $j$th constraint is active if $(1/w_j^{(N)})\mathbf{y}_j^T \mathbf{z} = \lambda$ or $(1/w_j^{(N)})\mathbf{y}_j^T \mathbf{z} = -\lambda$; in our case, one and only one of the two equations can hold. Our current treatment, which condenses two equations into a single one involving the absolute value of $\mathbf{y}_j^T \mathbf{z}$, can ease discussions without causing any technical errors.

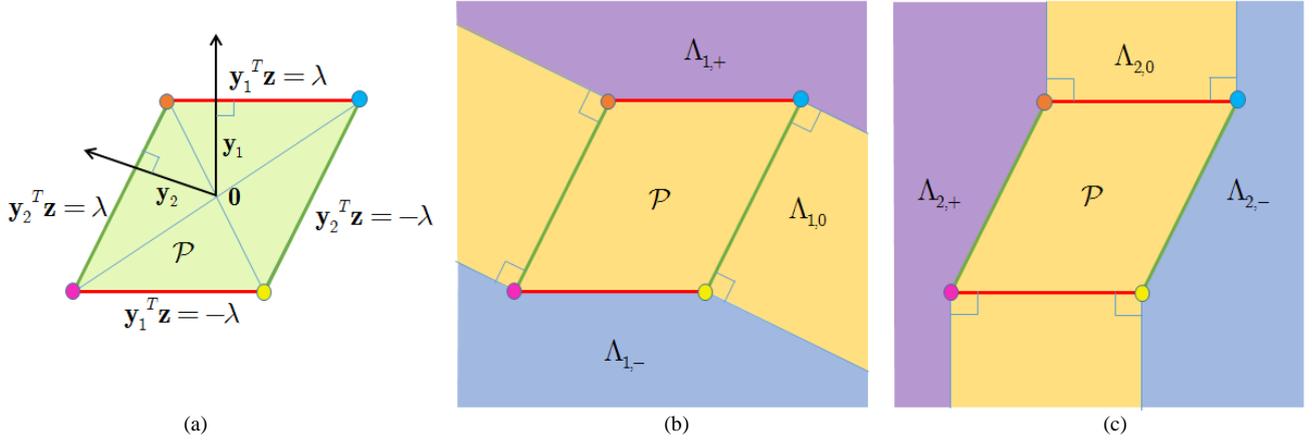

Fig. 4. Consider a three-point data set $\mathcal{Y} = \{\mathbf{y}_1, \mathbf{y}_2, \mathbf{y}_3\}$ in the ambient space $\mathbb{R}^2$. (a) The feasible polyhedron $\mathcal{P}$ of the dual problem (2.1). (b) Partition of the ambient space into $\mathbb{R}^2 = \Lambda_{1,+} \cup \Lambda_{1,0} \cup \Lambda_{1,-}$. (c) Partition of the ambient space into $\mathbb{R}^2 = \Lambda_{2,+} \cup \Lambda_{2,0} \cup \Lambda_{2,-}$.

### B. Geometry of Projection onto Polyhedron

We go on to introduce necessary details to formalize the geometry underpinning the dual problem (2.1). To better manifest the idea and also to simplify exposure, we will first consider the case without weighting, that is, $\mathbf{W}_N = \mathbf{I}$; the result can be directly extended to the general case, offering insights into why the proposed weighting rule (1.4) is better able to promote true discoveries.

To proceed, for $\mathbf{y} \in \mathbb{R}^n$ we denote by $\mathbf{z}^*(\mathbf{y})$ the optimal solution to (2.1) with $\mathbf{y}_N$ replaced by $\mathbf{y}$. Associated with each constraint $|\mathbf{y}_j^T \mathbf{z}^*(\mathbf{y})| \leq \lambda$, we can partition $\mathbb{R}^n$ into a disjoint union of the following three regions

$$\Lambda_{j,+} = \{\mathbf{y} \in \mathbb{R}^n \mid \mathbf{y}_j^T \mathbf{z}^*(\mathbf{y}) = \lambda\}, \quad \Lambda_{j,-} = \{\mathbf{y} \in \mathbb{R}^n \mid \mathbf{y}_j^T \mathbf{z}^*(\mathbf{y}) = -\lambda\}, \quad \Lambda_{j,0} = \{\mathbf{y} \in \mathbb{R}^n \mid |\mathbf{y}_j^T \mathbf{z}^*(\mathbf{y})| < \lambda\}. \quad (2.2)$$

Hence, the $j$th constraint is activated if and only if $\mathbf{y} \in \Lambda_{j,+} \cup \Lambda_{j,-}$, and stays inactive whenever $\mathbf{y} \in \Lambda_{j,0}$. Consider a simple example[4] with ambient space $\mathbb{R}^2$ and a data set $\mathcal{Y} = \{\mathbf{y}_1, \mathbf{y}_2, \mathbf{y}_3\}$ (i.e., $N = 3$), and we wish to express $\mathbf{y}_3$ as linear combination of $\mathbf{y}_1$ and $\mathbf{y}_2$. Fig. 4 depicts the partition of the ambient space into $\mathbb{R}^2 = \Lambda_{j,+} \cup \Lambda_{j,-} \cup \Lambda_{j,0}$, for $j = 1, 2$. Now, by jointly considering all the totally $N - 1$ constraints, the space $\mathbb{R}^n$ is partitioned into a disjoint union of $3^{N-1}$ polyhedrons, namely,

$$\mathbb{R}^n = \bigcup_{q=1}^{3^{N-1}} \Phi_q, \text{ where } \Phi_q \triangleq \bigcap_{j=1}^{N-1} \Lambda_{j,p_{q,j}}, \ p_{q,j} \in \{+, -, 0\}. \tag{2.3}$$

---

4. This example is only for ease of illustration. As considered in many previous works [2, 10-11] the ambient dimension $n$ is typically larger than the data size $N$.

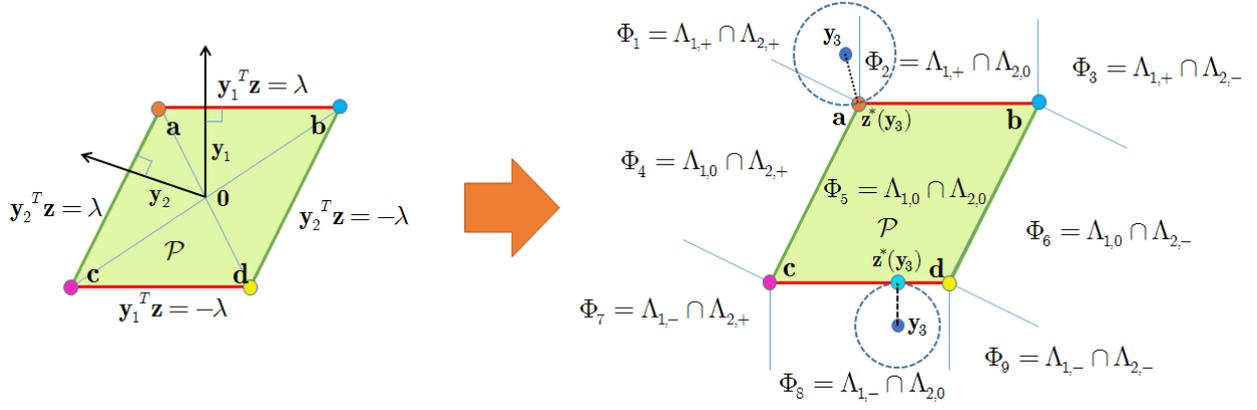

Fig. 5. A schematic depiction of the partition (2.3) for the three-point data set $\mathcal{Y} = \{\mathbf{y}_1, \mathbf{y}_2, \mathbf{y}_3\}$ in $\mathbb{R}^2$. The ambient domain is decomposed into a disjoint union of 9 polyhedrons, and the feasible polyhedron set $\mathcal{P}$ is the closure of $\Phi_5$. The optimal projection $\mathbf{z}^*(\mathbf{y}_3)$ is the vertex $\mathbf{a}$ when $\mathbf{y}_3 \in \Phi_1$, and lies on the line segment connecting vertexes $\mathbf{c}$ and $\mathbf{d}$ when $\mathbf{y}_3 \in \Phi_8$.

Consider again the above illustrative example. The decomposition (2.3) is then a union of 9 regions $\{\Phi_1, \cdots, \Phi_9\}$ as depicted in Fig. 5, in which the feasible polyhedron $\mathcal{P}$ onto which $\mathbf{y}_3$ is to be projected is the closure of $\Phi_5$. If $\mathbf{y}_3 \in \Phi_1$, then the optimal projection $\mathbf{z}^*(\mathbf{y}_3)$ is the vertex $\mathbf{a}$; in this case, we have $\mathbf{y}_1^T \mathbf{z}^*(\mathbf{y}_3) = \lambda$ and $\mathbf{y}_2^T \mathbf{z}^*(\mathbf{y}_3) = \lambda$, i.e., the first and second constraints are active, and thus $c_{3,1}^* \neq 0$ and $c_{3,2}^* \neq 0$. Similarly, if $\mathbf{y}_3 \in \Phi_8$, then the optimal $\mathbf{z}^*(\mathbf{y}_3)$ lies on the line segment connecting vertexes $\mathbf{c}$ and $\mathbf{d}$; as such, it follows $\mathbf{y}_1^T \mathbf{z}^*(\mathbf{y}_3) = -\lambda$ and $|\mathbf{y}_2^T \mathbf{z}^*(\mathbf{y}_3)| < \lambda$, leading to $c_{3,1}^* \neq 0$ and $c_{3,2}^* = 0$. Associated with each $1 \leq q \leq 3^{N-1}$, we go on to define the following three index sets

$$\mathcal{I}_{q,+} \triangleq \{j \mid \mathbf{y} \in \Phi_q, \mathbf{y}_j^T \mathbf{z}^*(\mathbf{y}) = \lambda\}, \tag{2.4}$$

$$\mathcal{I}_{q,-} \triangleq \{j \mid \mathbf{y} \in \Phi_q, \mathbf{y}_j^T \mathbf{z}^*(\mathbf{y}) = -\lambda\}, \tag{2.5}$$

$$\mathcal{I}_{q,0} \triangleq \{j \mid \mathbf{y} \in \Phi_q, |\mathbf{y}_j^T \mathbf{z}^*(\mathbf{y})| < \lambda\}. \tag{2.6}$$

Notably, $\mathcal{I}_{q,+} \cup \mathcal{I}_{q,-}$ consists of the indexes of the active constraints associated with $\Phi_q$, whereas $\mathcal{I}_{q,0}$ is for the indexes of the inactive constraints. For the example depicted in Fig. 5, associated with $q = 1$ we have $\mathcal{I}_{1,+} = \{1, 2\}$ and $\mathcal{I}_{1,-} = \mathcal{I}_{1,0} = \phi$, while for $q = 8$ we have $\mathcal{I}_{8,+} = \phi$, $\mathcal{I}_{8,-} = \{1\}$, and $\mathcal{I}_{8,0} = \{2\}$.

### C. Impact of Data Weighting

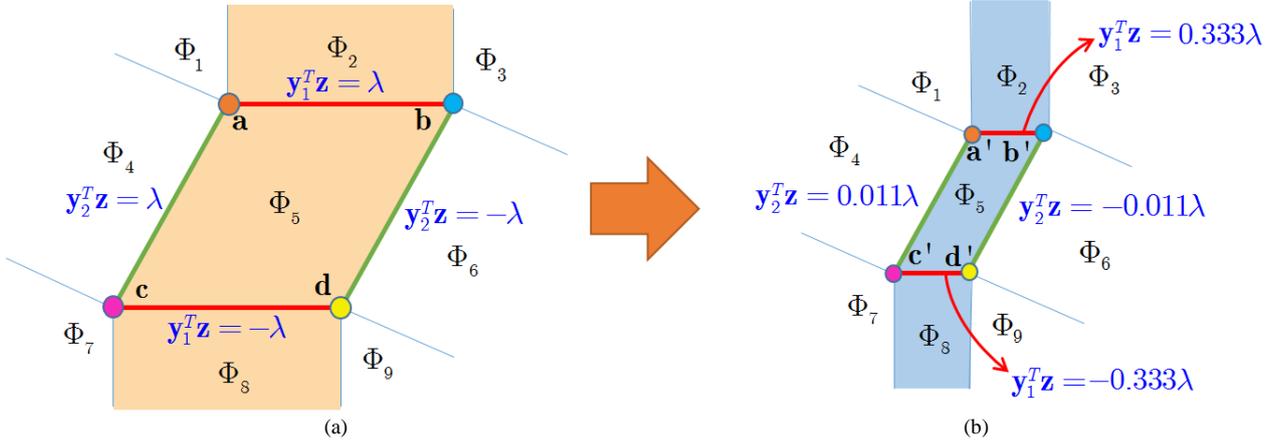

Fig. 6. Illustration of impact of data weighting on the boundaries of polyhedrons. With data weighting, the two sets of hyperplanes move closer to the origin, causing the region $\Phi_2 \cup \Phi_5 \cup \Phi_8$ to shrink.

When data weighting is employed, the $j$th scalar constraint associated with the dual problem (2.1) becomes $|\mathbf{y}_j^T\mathbf{z}| \leq w_j^{(N)}\lambda$. To examine the impact of weighting, consider again the example in Fig. 4. Suppose that the computed sparse representation $\tilde{\mathbf{c}}_3 = [\tilde{c}_{3,1}\ \tilde{c}_{3,2}]^T$ in the first step yields $|\tilde{c}_{3,1}| = 0.02$ and $|\tilde{c}_{3,2}| = 0.91$. In this case, $\mathbf{y}_2$ is more likely a neighbor of $\mathbf{y}_3$, and the weighted LASSO in the second step is expected to promote the event $c_{3,2}^* \neq 0$, i.e., the index pair $(3,2)$ is a true discovery. With $\varepsilon = 0.01$, the proposed weighting coefficients (1.4) are $w_1^{(3)} = 0.333$ and $w_2^{(3)} = 0.011$, and the two scalar constraints in (2.1) accordingly become $|\mathbf{y}_1^T\mathbf{z}| \leq 0.333\lambda$ and $|\mathbf{y}_2^T\mathbf{z}| \leq 0.011\lambda$. Compared to the uniform weighting case with scalar constraints $|\mathbf{y}_1^T\mathbf{z}| \leq \lambda$ and $|\mathbf{y}_2^T\mathbf{z}| \leq \lambda$ (see Fig. 6-(a)), the two pairs of hyperplanes corresponding to $\mathbf{y}_1^T\mathbf{z} = \pm 0.333\lambda$ and $\mathbf{y}_2^T\mathbf{z} = \pm 0.011\lambda$ (see the red and green line segment pairs, respectively, in the Fig. 6-(b)) "move towards the origin", as depicted in Fig. 6-(b). The hyperplane pair defined by $\mathbf{y}_2^T\mathbf{z} = \pm 0.011\lambda$ (the green one) is located much closer to the origin than the pair for $\mathbf{y}_1^T\mathbf{z} = \pm 0.333\lambda$ (the red one). Such a shift of the hyperplanes alters the boundaries of $\Phi_q$'s, and, in particular, the region $\Phi_2 \cup \Phi_5 \cup \Phi_8$ shrinks (see Fig. 6-(b) for an illustration). Hence, if we assign certain probability distributions on the data and noise, $\Pr(\{\mathbf{y}_3 \in \Phi_2 \cup \Phi_5 \cup \Phi_8\})$ is accordingly decreased. Note that the inactive constraint index sets (defined in (2.6)) associated with the three regions $\Phi_2$, $\Phi_5$, $\Phi_8$ are, respectively, $\mathcal{I}_{2,0} = \{2\}$, $\mathcal{I}_{5,0} = \{1,2\}$, and $\mathcal{I}_{8,0} = \{2\}$, all containing the index $\{2\}$. Since $\Pr(\{\mathbf{y}_3 \in \Phi_2 \cup \Phi_5 \cup \Phi_8\})$ is decreased, the index $\{2\}$ less frequently stays inactive, or put differently, is more likely to be activated, and consequently the event $c_{3,2}^* \neq 0$ is promoted.

### D. Mathematical Characterization of $\Phi_q$

We end this section by providing a technical lemma, which establishes an explicit expression for the elements of $\Phi_q$ in terms of the data points $\{\mathbf{y}_1, \cdots, \mathbf{y}_{N-1}\}$; the lemma will be used for proving our theoretical results in the next section.

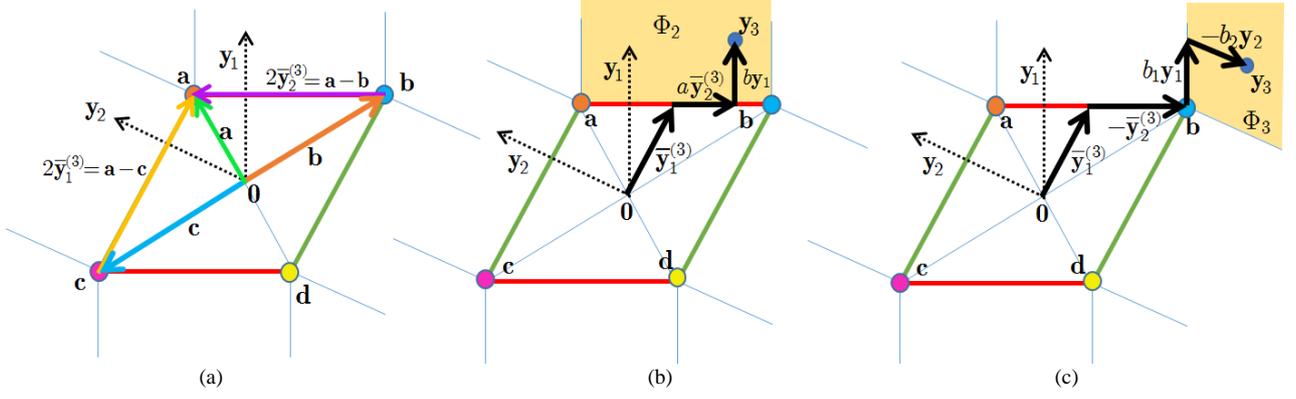

Fig. 7 (a). Schematic depiction of $2\bar{\mathbf{y}}_1^{(3)}$ ($2\bar{\mathbf{y}}_2^{(3)}$, respectively) as the difference of vertexes $\mathbf{a}$ and $\mathbf{c}$ ($\mathbf{a}$ and $\mathbf{b}$, respectively). (b) Illustration of $\mathbf{y}_3 \in \Phi_2$ as a linear combination of $\bar{\mathbf{y}}_1^{(3)}, \bar{\mathbf{y}}_2^{(3)}, \mathbf{y}_1$. (c) Illustration of $\mathbf{y}_3 \in \Phi_3$ as a linear combination of $\bar{\mathbf{y}}_1^{(3)}, \bar{\mathbf{y}}_2^{(3)}, \mathbf{y}_1, \mathbf{y}_2$.

Since the ambient dimension $n$ is larger than the data size $N$, let $\{\mathbf{n}_1, \cdots, \mathbf{n}_{n-(N-1)}\}$ be an orthonormal basis for the null space of $\mathbf{Y}_{-N}^T \in \mathbb{R}^{(N-1) \times n}$, and define

$$\mathbf{A}_N \triangleq [\mathbf{Y}_{-N} \ \mathbf{n}_1 \cdots \mathbf{n}_{n-N+1}] \in \mathbb{R}^{n \times n}, \quad (2.7)$$

which is nonsingular[5]. Denote by $\bar{\mathbf{y}}_j^{(N)}$ the $j$th column of the matrix $\lambda (\mathbf{A}_N^T)^{-1}$, so that

$$\mathbf{y}_i^T \bar{\mathbf{y}}_j^{(N)} = \begin{cases} \lambda, & i = j, \\ 0, & i \neq j. \end{cases} \quad (2.8)$$

Note that $\bar{\mathbf{y}}_j^{(N)}$, when properly scaled, acts as the "boundary vector" of the feasible polyhedron $\mathcal{P}$ in the sense that it can be written as the difference of two adjacent vertexes of $\mathcal{P}$. To see this, consider again the simple example in the previous subsection; in this case we have $\mathbf{A}_3 = [\mathbf{y}_1 \ \mathbf{y}_2] \in \mathbb{R}^{2 \times 2}$. Since the vertexes $\mathbf{a}$ and $\mathbf{c}$ in Fig. 5 obey, respectively, $\mathbf{A}_3^T \mathbf{a} = [\lambda \ \lambda]^T$ and $\mathbf{A}_3^T \mathbf{c} = [-\lambda \ \lambda]^T$ (this can be directly seen from the left diagram in Fig. 5), it follows $\mathbf{a} - \mathbf{c} = (\mathbf{A}_3^T)^{-1} [\lambda \ \lambda]^T - (\mathbf{A}_3^T)^{-1} [-\lambda \ \lambda]^T = 2\bar{\mathbf{y}}_1^{(3)}$ (see Fig. 7-(a) for a depiction); similarly, for vertex $\mathbf{b}$ we have $\mathbf{A}_3^T \mathbf{b} = [\lambda \ -\lambda]^T$, and it is easy to check $\mathbf{a} - \mathbf{b} = 2\bar{\mathbf{y}}_2^{(3)}$ (see also Fig. 7-(a)). The following lemma characterizes the elements of $\Phi_q$.

**Lemma 2.2:** Let $\bar{\mathbf{y}}_j^{(N)}$ be defined as in (2.8). Also, let $\mathcal{I}_{q,+}$, $\mathcal{I}_{q,-}$, and $\mathcal{I}_{q,0}$ be defined in, respectively, (2.4), (2.5), and (2.6). Then we have

$$\Phi_q = \left\{ \sum_{j \in \mathcal{I}_{q,0}} a_j \bar{\mathbf{y}}_j^{(N)} + \sum_{j \in \mathcal{I}_{q,+}} \left( \bar{\mathbf{y}}_j^{(N)} + b_j \mathbf{y}_j \right) - \sum_{j \in \mathcal{I}_{q,-}} \left( \bar{\mathbf{y}}_j^{(N)} + b_j \mathbf{y}_j \right) + \sum_{k=1}^{n-N-1} h_k \mathbf{n}_k, \ |a_j| < 1, \ b_j \geq 0, \ h_k \in \mathbb{R} \right\}. \quad (2.9)$$

*[Proof]:* See Appendix B. □

For a given data set $\{\mathbf{y}_j\}_{1 \leq j \leq N-1}$, we can regard $\{\mathbf{y}_j\}_{1 \leq j \leq N-1}$, $\{\bar{\mathbf{y}}_j^{(N)}\}_{1 \leq j \leq N-1}$, and $\{\mathbf{n}_k\}_{1 \leq k \leq n-N+1}$ as the built-in "anchor vectors" for the ambient space $\mathbb{R}^n$. Lemma 2.2 pins down an explicit expression characterizing how elements of $\Phi_q$ can be constructed based on this anchor dictionary. Representation of elements of $\Phi_q$ in terms of the anchor dictionary, but not the basis $\{\mathbf{y}_j\}_{1\leq j\leq N-1} \cup \{\mathbf{n}_k\}_{1\leq k\leq n-N+1}$ (i.e., columns of $\mathbf{A}_N$), enjoys the following distinctive advantage: it allows us to incorporate the statistical distributions of the data (signal points and noise) for deriving provable neighbor recovery rates, meanwhile keeping the analysis tractable, as will be shown later. Finally, we again use the simple example in the previous subsection to illustrate the result of Lemma 2.2. For $q=2$, we have $\mathcal{I}_{2,+} = \{1\}$ and $\mathcal{I}_{2,0} = \{2\}$; if $\mathbf{y}_3 \in \Phi_2$, Lemma 2.2 asserts $\mathbf{y}_3 = a\overline{\mathbf{y}}_2^{(3)} + (\overline{\mathbf{y}}_1^{(3)} + b\mathbf{y}_1)$, where $|a| < 1$ and $b > 0$ (see Fig. 7-(b) for a depiction). For $q=3$, we have $\mathcal{I}_{3,+} = \{1\}$ and $\mathcal{I}_{3,-} = \{2\}$; if $\mathbf{y}_3 \in \Phi_3$, it follows $\mathbf{y}_3 = (\overline{\mathbf{y}}_1^{(3)} + b_1\mathbf{y}_1) - (\overline{\mathbf{y}}_2^{(3)} + b_2\mathbf{y}_2)$, where $b_1 > 0$ and $b_2 > 0$ (see Fig. 7-(c)). An interesting geometric feature demonstrated from Fig. 7-(b) and 7-(c) is that, we can first traverse along the boundary of $\mathcal{P}$, and then move forward along the directions of certain data points $\mathbf{y}_j$'s ($j \neq N$) to reach $\mathbf{y}_N$. The following corollary provides a mathematical expression for a translation of $\Phi_q$ that will also be used in our subsequent analysis.

---

5. $\mathbf{A}_N$ is nonsingular as long as the columns of $\mathbf{Y}_{-N}$ are linearly independent, which is typically true in the setting of noisy SSC.

*Corollary 2.3:* Under the same setup as in Lemma 2.2, the translation $\Phi_q + \mathbf{w} \triangleq \{\mathbf{y} + \mathbf{w} \mid \mathbf{y} \in \Phi_q\}$, where $\mathbf{w} \in \mathbb{R}^n$, can be expressed as

$$\Phi_q + \mathbf{w} = \left\{ \sum_{j \in \mathcal{I}_{q,0}} a_j \overline{\mathbf{y}}_j^{(N)} + \sum_{j \in \mathcal{I}_{q,+}} b_j \mathbf{y}_j - \sum_{j \in \mathcal{I}_{q,-}} b_j \mathbf{y}_j + \sum_{k=1}^{n-N+1} h_k \mathbf{n}_k, \ -1+\alpha_j < a_j < 1+\alpha_j, \ b_j \geq \beta_j, \ h_k \in \mathbb{R} \right\} \quad (2.10)$$

for some real $\alpha_j$ and $\beta_j$.

*[Proof]:* See Appendix B. □

## III. THEORETICAL RESULTS

Consider the two-step reweighted $\ell_1$-minimization algorithm outlined in Table II. Let us rewrite the weighted Lasso functional in (1.5) as

$$J_i = \lambda \|\mathbf{W}_i \mathbf{c}_i\|_1 + \frac{1}{2} \|\mathbf{y}_i - \mathbf{Y}_{-i} \mathbf{c}_i\|_2 = \lambda \|\mathbf{d}_i\|_1 + \frac{1}{2} \|\mathbf{y}_i - \mathbf{Y}_{-i} \mathbf{W}_i^{-1} \mathbf{d}_i\|_2, \quad (3.1)$$

where the second equality holds through a change of variable $\mathbf{d}_i \triangleq \mathbf{W}_i \mathbf{c}_i = [d_{i,1} \cdots d_{i,N}]^T$. Minimization of $J_i$ in (3.1) amounts to seeking a sparse linear combination of the columns of $\mathbf{Y}_{-i} \mathbf{W}_i^{-1} = [(1/w_1^{(i)})\mathbf{y}_1 \cdots (1/w_N^{(i)})\mathbf{y}_N]$ to best approximate $\mathbf{y}_i$. With $w_j^{(i)}$ given in (1.4), we have $1/w_j^{(i)} > 1$, and hence all weighted data points $(1/w_j^{(i)})\mathbf{y}_j$'s are "stretched". As such, more column vectors of $\mathbf{Y}_{-i} \mathbf{W}_i^{-1}$ are allowed to participate in the linear combination $\mathbf{Y}_{-i} \mathbf{W}_i^{-1} \mathbf{d}_i$ with small non-zeros in $\mathbf{d}_i$, in an attempt to better approximate $\mathbf{y}_i$ for noise robustness, while the cost function $J_i$ is still rendered small. Hence, compared to the conventional LASSO without data weighting, the weighted LASSO sparse regressor tends to output a denser $\mathbf{c}_i = \mathbf{W}_i^{-1} \mathbf{d}_i$, thus more discoveries. It is our hope that most newly produced nonzeros yield true discoveries. Leveraging the polyhedron geometry of the dual problem of weighted

LASSO, our purpose in this section is to make the following fact precise: When employing the proposed weighting rule (1.4), the weighted LASSO can more likely produce many true discoveries and few false discoveries, as compared to the case without data weighting. Section III-A presents the main mathematical theorems. Related discussions are then given in Section III-B.

*A. Recovery Rate Analysis*

To formalize matters, we assume each data vector in the set $\mathcal{Y} = \{\mathbf{y}_1, \cdots, \mathbf{y}_N\}$ obeys the standard additive noise model, that is,

$$\mathbf{y}_i = \mathbf{x}_i + \mathbf{e}_i, \quad 1 \leq i \leq N, \tag{3.2}$$

where $\mathbf{x}_i \in \mathbb{R}^n$ is a unit-norm[6] signal vector, and $\mathbf{e}_i \in \mathbb{R}^n$ is the noise term. Our analysis is developed under the semi-random model[7] [12]; that is, the ground truth subspaces $\mathcal{S}_1, \cdots, \mathcal{S}_L$ are considered to be fixed but unknown, whereas the signal points and noise are assumed to be random. Such a model is widely used in the theoretical study of SSC, thanks to its interpretability and amiability to analysis. Similar to [10], the following assumptions are made in the sequel.

*Assumption 1:* For each $1 \leq i \leq N$, the signal vector $\mathbf{x}_i \in \mathcal{S}_k$ is uniformly sampled from the unit sphere in subspace $\mathcal{S}_k$. □

*Assumption 2:* For each $1 \leq i \leq N$, the noise $\mathbf{e}_i \in \mathbb{R}^n$ are i.i.d. Gaussian random vectors with zero mean and covariance matrix $(\sigma^2/n)\mathbf{I}$, i.e., $\mathbf{e}_i \sim \mathcal{N}(0, \sigma^2/n\mathbf{I})$, and are independent of the signal vectors $\mathbf{x}_i$'s. □

*Assumption 3:* The subspace affinity[8] $\mathit{aff}(i,j)$ satisfies $\max_{i:i\neq j} \mathit{aff}(i,j) \leq \frac{K}{\log N} - \sigma\sqrt{\frac{d_j}{n \log N}}$ for $1 \leq j \leq L$, where $K$ is a constant and $d_j$ is the dimension of the *j*th subspace. □

*Assumption 4:* For each $1 \leq k \leq L$, the sample density $\rho_k \triangleq |\mathcal{Y}_k|/d_k$ of the *k*th subspace satisfies $\rho_k \geq \rho^*$ for $1 \leq k \leq L$, where $\rho^* > 0$ is a constant. □

Notably, Assumptions 3 and 4 are made (see [10]) so as to guarantee (i) different subspaces are well separated from each other, and (ii) there are sufficiently many data points in each cluster/subspace.

For a fixed data set $\{\mathbf{y}_1, \mathbf{y}_2, \cdots, \mathbf{y}_{N-1}\}$, our discussions in Section II-C based on the polyhedron geometry of the weighted LASSO have revealed how data weighting can promote true discoveries. With the aid of Lemma 2.2, we can now go one step further to formalize this observation in a probabilistic setting. Specifically, the following proposition holds.

---

6. The unit-norm assumption is quite standard and has been assumed in many previous works, e.g., [10-11].
7. Notably, two alternative model assumptions are the fully deterministic model and fully random model [12]. Extension of our study to these two data models is to be further investigated.
8. The affinity between subspaces $\mathcal{S}_i$ and $\mathcal{S}_j$ is defined to be $\mathit{aff}(i,j) \triangleq \|\mathbf{U}_i^T \mathbf{U}_j\|_F / \sqrt{\min(d_i, d_j)}$, where columns of $\mathbf{U}_i$ ($\mathbf{U}_j$, respectively) consist of an orthonormal basis for $\mathcal{S}_i$ ($\mathcal{S}_j$, respectively) [11].

***Proposition 3.1:*** Let $\mathcal{I}_{q,0}$ defined in (2.6) be the inactive constraint index set associated with the polyhedron $\Phi_q$. Then we have

$$\Pr(\{\mathbf{y}_N \in \Phi_q \mid \mathbf{y}_1, \mathbf{y}_2, \cdots, \mathbf{y}_{N-1}\}) \leq C_q \prod_{j \in \mathcal{I}_{q,0}} erf\left(g_j w_j^{(N)}\right), \tag{3.3}$$

where $erf(\cdot)$ is the error function, $C_q$ and $g_j$ are some positive universal constants.

*[Proof]:* See Section V-A. □

The conditional probability upper bound (3.3) depends on the weighting coefficients $w_j^{(N)}$'s, where $j$'s belong to the inactive constraint index set $\mathcal{I}_{q,0}$. To see the interpretation of the bound (3.3), suppose that, for some $1 \leq j \leq N-1$, the corresponding $w_j^{(N)}$ is very small. Hence, the data point $\mathbf{y}_j$ is highly likely a neighbor, and the index pair $(N, j)$ should be promoted as a true discovery. It is then expected that the $j$th constraint shall often be activated, that is to say, $\mathbf{y}_N$ shall seldom lie in those $\Phi_q$ with the $j$th constraint being inactive. The upper bound (3.3) therefore offers a quantitative support of the above point: with data weighting, the probability that $\mathbf{y}_N \in \Phi_q$ with $j \in \mathcal{I}_{q,0}$ is reduced (penalized) by the factor $erf(g_j w_j^{(N)})$, which decreases if the value of $w_j^{(N)}$ gets smaller, i.e., $\mathbf{y}_j$ is more certain to be a neighbor.

The main theorem of this paper, which is developed based on Prop. 3.1, is presented below. To state the theorem, we define the following two index sets $\mathcal{T} \triangleq \{j \mid (N, j) \text{ is a true discovery}\}$ and $\mathcal{F} \triangleq \{j \mid (N, j) \text{ is a false discovery}\}$. Also, denote by $\{|\mathcal{T}| < k_t\}$ ($\{|\mathcal{T}| \geq k_t\}$, respectively) the event that there are less than (at least, respectively) $k_t > 0$ true discoveries, and by $\{|\mathcal{F}| \leq k_f\}$ the event with at most $k_f \geq 0$ false discoveries. We have the following theorem.

***Theorem 3.2:*** Assume that $\{\mathbf{y}_{t_1}, \cdots, \mathbf{y}_{t_{m_N}}\}$ are in the same cluster as $\mathbf{y}_N$, and assume without loss of generality that the corresponding weighting coefficients are sorted in an increasing order $w_{t_1}^{(N)} \leq w_{t_2}^{(N)} \leq \cdots \leq w_{t_{m_N}}^{(N)}$; similarly, for those data points $\{\mathbf{y}_{f_1}, \cdots, \mathbf{y}_{f_{(N-1)-m_N}}\}$ from different clusters the corresponding weighting coefficients are also sorted according to $w_{f_1}^{(N)} \leq w_{f_2}^{(N)} \leq \cdots \leq w_{f_{(N-1)-m_N}}^{(N)}$. Suppose that the noise level satisfies $\sigma \leq \sigma^*$, where $\sigma^*$ is a sufficiently small numerical constant, and take $\tau = 2\sigma$ and $f(t) \geq 0.707\sigma t^{-1}$ in the two-step reweighted $\ell_1$-minimization algorithm (see Table II). Under the above setting and Assumptions 1~4, the following results hold.

*(a) (Many discoveries)*

$$\Pr(\{|\mathcal{T}| < k_t\} \cap \{|\mathcal{F}| \leq k_f\}) \leq C \prod_{j \in \mathcal{I}_{k_t, k_f}} erf\left(\gamma_j w_j^{(N)}\right) + 10^{-(N-1)}, \tag{3.4}$$

where $C > 0$ is some universal constant and

$$\mathcal{I}_{k_t, k_f} \triangleq \{1, \cdots, N-1\} \setminus [\{t_1, \cdots, t_{k_t-1}\} \cup \{f_1, \cdots, f_{k_f}\}]. \tag{3.5}$$

*(b) (Few false discoveries)*

$$\Pr(\{|\mathcal{F}| \leq k_f\}) \geq 1 - e^{-\alpha_0(n-d_N)\left(1-(w_{t_1}^{(N)})^2/2\right)^2} - 2Ne^{-\alpha_1 n} - e^{-\sqrt{m_N d_N}} - N^{-2} - N^{-3} - 8N^{-4w_{f_{k_f}+1}^{(N)}+1}, \quad (3.6)$$

where $m_N$, $d_N$ are the number of data points and dimension of the subspace that attached to $\mathbf{y}_N$, and $\gamma_j$, $\alpha_0$, and $\alpha_1$ are constants.

*(c) (Many true discoveries and few false discoveries)*

$$\Pr(\{|\mathcal{T}| \geq k_t\} \cap \{|\mathcal{F}| \leq k_f\})$$

$$\geq 1 - C \prod_{j \in \mathcal{I}_{k_t,k_f}} erf\left(\gamma_j w_j^{(N)}\right) - e^{-\alpha_0(n-d_N)\left(1-(w_{t_1}^{(N)})^2/2\right)^2} - 2Ne^{-\alpha_1 n} - e^{-\sqrt{m_N d_N}} - N^{-2} - N^{-3} - 8N^{-4w_{f_{k_f}+1}^{(N)}+1} - 10^{-(N-1)}$$

(3.7)

*[Proof]:* See Section V-B. □

## B. Discussions

Theorem 3.2 establishes explicit neighbor recovery rates which can uncover the impact of data weighting on neighbor identification via the proposed weighted LASSO scheme (1.5). Specifically, we can observe the following:

1. Recall that data weighting (using (1.4)) tends to increase the number of discoveries. The upper bound (3.4) supports this fact. Indeed, the event $\{|\mathcal{T}| < k_t\} \cap \{|\mathcal{F}| \leq k_f\}$ (if it occurs) yields at most $k_t + k_f - 1$ discoveries. For small $k_t$ and $k_f$, i.e., few discoveries, $\Pr(\{|\mathcal{T}| < k_t\} \cap \{|\mathcal{F}| \leq k_f\})$ will be small. This is because, according to (3.5), the cardinality $|\mathcal{I}_{k_t,k_f}| = (N-1) - (k_t + k_f - 1) = N - k_t - k_f$ is large, and the first term on the right-hand-side (RHS) of (3.4) thus involves a product of many "small" error function terms. It is noteworthy that, with $0 < w_j^{(N)} < 1$, the upper bound (3.4) decreases as compared to the case without data weighting (i.e., $w_j^{(N)} = 1$ for all $1 \leq j \leq N-1$). Hence, with the aid of data weighting, events with few discoveries rarely happen. In particular, compared to the method in [10], the proposed algorithm can more often produce many discoveries.

2. Regarding the technical aspect of the bound (3.4), we first note that data points assigned with a small weight are more likely to be selected as neighbors. Since the two sets of weighting coefficients are sorted according to $w_{t_1}^{(N)} \leq w_{t_2}^{(N)} \leq \cdots \leq w_{t_{m_N}}^{(N)}$ and $w_{f_1}^{(N)} \leq w_{f_2}^{(N)} \leq \cdots \leq w_{f_{(N-1)-m_N}}^{(N)}$, the recovered neighbors in the event $\{|\mathcal{T}| < k_t\} \cap \{|\mathcal{F}| \leq k_f\}$ most likely come from $\{\mathbf{y}_{t_1}, \cdots, \mathbf{y}_{t_{k_t-1}}\} \cup \{\mathbf{y}_{f_1}, \cdots, \mathbf{y}_{f_{k_f}}\}$. This amounts to saying that the data point $\mathbf{y}_N$ most likely lies in some polyhedron $\Phi_q$ with active constraint index set $\mathcal{I}_{q,+} \cup \mathcal{I}_{q,-} \subset \{t_1, \cdots, t_{k_t-1}\} \cup \{f_1, \cdots, f_{k_f}\} = \{1, \cdots, N-1\} \setminus \mathcal{I}_{k_t,k_f}$ (see (3.5)), hence with an inactive constraint index set $\mathcal{I}_{q,0} = \{1, \cdots, N-1\} \setminus (\mathcal{I}_{q,+} \cup \mathcal{I}_{q,-}) \supset \mathcal{I}_{k_t,k_f}$. It is therefore straightforward to see

$$\prod_{j \in \mathcal{I}_{q,0}} erf\left(\gamma_j w_j^{(N)}\right) \leq \prod_{j \in \mathcal{I}_{k_t,k_f}} erf\left(\gamma_j w_j^{(N)}\right) \text{ for any } \mathcal{I}_{q,0} \supset \mathcal{I}_{k_t,k_f}. \tag{3.8}$$

Conceptually speaking, the upper bound (3.4) is obtained by first averaging the conditional probability upper bound (3.3) with respect to the assumed data distributions, and then computing the summation of the resultant averaged probabilities over those indexes $q$'s contributing to the occurrence of the event $\{|\mathcal{T}| < k_t\} \cap \{|\mathcal{F}| \leq k_f\}$, meanwhile also taking into account inequality (3.8) and sorting of the weighting coefficients (details referred to Section V-B). Notably, since $\mathcal{I}_{k_t,k_f}$ is the smallest possible inactive constraint index set (among all admissible $\mathcal{I}_{q,0}$), the upper bound (3.4) is kinds of a worst-case bound. Nevertheless, since the cardinality $|\mathcal{I}_{k_t,k_f}| = N - k_t - k_f$ is typically large (especially when $k_t$ and $k_f$ are small), the upper bound (3.4), as a product of $N - k_t - k_f$ small error function terms plus the very small term $10^{-(N-1)}$, can remain small and is a fairly good bound.

3. The lower bound (3.6) asserts that the proposed approach with a high probability yields just few false discoveries. Indeed, assuming that the prior information acquired in the first step (through knowledge of the amplitude of $\tilde{c}_{i,j}$) is enough accurate so that $w_{t_1}^{(N)} \approx 0$ and $w_{f_k f+1}^{(N)} \approx 1$, the probability lower bound (3.6) can be kept fairly large. In addition, as the quality of prior neighbor information improves so that $w_{t_1}^{(N)}$ decreases to zero and $w_{f_k f+1}^{(N)}$ increases to one, the lower bound (3.6) is enlarged. Hence, the more reliable the priori information is, the higher the probability the proposed algorithm can produce few false discoveries.

4. For the special case without data weighting, i.e., $w_j^{(N)} = 1$ for all $1 \leq j \leq N-1$, and no false discoveries, namely, $k_f = 0$, the bound (3.6) simplifies to

$$\Pr(\{|\mathcal{F}| = 0\}) \geq 1 - e^{-\alpha_0(n-d_N)/4} - 2Ne^{-\alpha_1 n} - e^{-\sqrt{m_N d_N}} - N^{-2} - 9N^{-3}. \tag{3.9}$$

We note that in [10], a lower bound for $\Pr(\{|\mathcal{F}| = 0\})$ analogue to (3.9) has been derived to show that the two-step algorithm proposed in [10] yields zero false discoveries with a high probability. We would like to mention that our lower bound (3.6) can serve as an extension of the result in [10] to the general case when data weighting is employed in the second step and the number of false discoveries is allowed to be $k_f \geq 0$.

5. Parts (a) and (b) of Theorem 3.2 show that the proposed method is highly likely to produce many discoveries, only few among which are false discoveries. This then confirms that, with a high probability (exceeding the lower bound (3.7)), there are many true discoveries and few false discoveries. In particular, with the aid of data weighting and good quality of the prior neighbor information, the proposed approach achieves better neighbor recovery rate performance as compared to the method in [10].

# IV. EXPERIMENTAL RESULTS

In this section, we provide numerical simulations using both synthetic and real human face data to validate our theoretical study, and illustrate the performance of the proposed approach. To implement the proposed scheme and towards fair comparison with the method in [10], certain parameter setting just follows that in [10]. Specifically, in the first step the $\ell_2$-norm error upper bound $\tau$ is set to be $\tau = 2\sigma$, whereas in the second step, the regularization factor $\lambda$ used in the weighted LASSO is chosen to be $\lambda = 0.707\sigma / \|\tilde{\mathbf{c}}_i\|_1$, where $\tilde{\mathbf{c}}_i$ is the computed optimal solution in the first step. Synthetic data are generated similar to [12]. The ground truth is a union of three 4-dimensional subspaces $\mathcal{S}_1$, $\mathcal{S}_2$ and $\mathcal{S}_3$ of $\mathbb{R}^{100}$. Signal points are randomly and uniformly drawn from each subspace, and are corrupted by Gaussian noise with zero mean and standard deviation $\sigma$. The sampling density of each subspace is set to be 5. The subspace affinity $aff(i,j)$ is used to gauge the separation between $\mathcal{S}_i$ and $\mathcal{S}_j$. In order to test the recovery rate performance, we consider the following two metrics: the *discovery rate* (DCR)

$$DCR \triangleq \left(\sum_{i=1}^{N} \|\mathbf{c}_i^*\|_0\right) / (N(N-1)), \tag{4.1}$$

where $\mathbf{c}_i^*$ is the optimal sparse solution in the second step, and the *true discovery rate* (TDR)

$$TDR \triangleq (\# \text{ of true discovery}) / \left(\sum_{i=1}^{N} \|\mathbf{c}_i^*\|_0\right). \tag{4.2}$$

As the global clustering performance measure, we consider as in [10] the *correct clustering rate* (CCR), defined as

$$CCR \triangleq (\# \text{ of correctly classified data points}) / N. \tag{4.3}$$

## A. Synthetic Data

The performance of the proposed method is tested by using the synthetic data. For simplicity of illustration we consider the scenario $aff(i,j) = \rho$, $1 \leq i,j \leq 3$, i.e., all subspaces are equally separated from each other. First of all, we shall determine a good parameter $\varepsilon$ that is to be used in the weighting coefficient (1.4). For this we consider various pairs of noise standard deviation $\sigma$ and subspace affinity $\rho$, and then search for each pair $(\sigma, \rho)$ the best $\varepsilon$ that yields the highest CCR. Fig. 8 plots the computed optimal $\varepsilon$ (in the log scale) as the graph over different pairs of $(\sigma, \rho)$. As the figure shows, the optimal $\varepsilon$ for small $(\sigma, \rho)$ assumes small values. This is because, if $(\sigma, \rho)$ is small (hence, small noise corruption and subspaces separated far away from each other), the prior neighbor information acquired in the first step is very reliable. As a result, a small $\varepsilon$ is preferred since it yields a small weighting coefficient (see (1.4)) to reflect more confidence in the prior information, overall leading to improved clustering accuracy. For large $(\sigma, \rho)$, the optimal $\varepsilon$ is seen to assume large values. This is reasonable since, if $(\sigma, \rho)$ is large (meaning that noise corruption is severe and subspaces are close to each other), the quality of prior neighbor information degrades. As such, a large $\varepsilon$ can result in a large weight so as to demote the impact of uncertainty in the prior information. We go on to compare the proposed scheme with the two-step algorithm in [10] without data weighting; in implementing our algorithm, the parameter $\varepsilon$ in the weighting rule (1.4) is chosen according to the results in Fig. 8. For subspace affinity $\rho = 0.02, 0.5,$ and $0.86$, Fig. 9 shows the CCR curves of the two methods with respect to different noise standard deviation $\sigma$. As the figure shows, the proposed method with data weighting in all cases outperforms the two-step algorithm in [10]. For large $\rho = 0.86$, i.e., subspaces are close to each other, performance improvement is significant when noise level $\sigma$ is from small to medium. This is because, when noise is not large, our method can benefit from the reliable prior neighbor information acquired in the first step. However, as $\sigma$ increases, the prior knowledge is less accurate and does not help much. For small and medium $\rho = 0.02$ and $0.5$, i.e., subspaces are well separated from each other, our method only slightly outperforms [10] when noise is small, but it achieves a large performance improvement when noise is from medium to large. This is expected since, with small noise and well-separated subspaces, data points from different subspaces are potentially more discernable, even without the aid of reliable prior information. Nevertheless, as the noise level increases, our method then benefits from the prior information and improves CCR. Fig. 10 further compares the DCR and TDR of the two methods. It can be seen from Fig. 10-(a) that the proposed method produces more discoveries, and overall yields higher TDR as illustrated in Fig. 10-(b). This confirms our analytical findings in Section III.

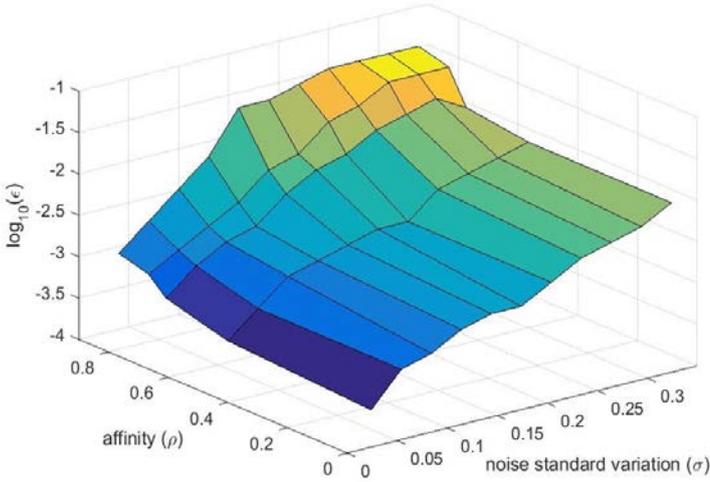
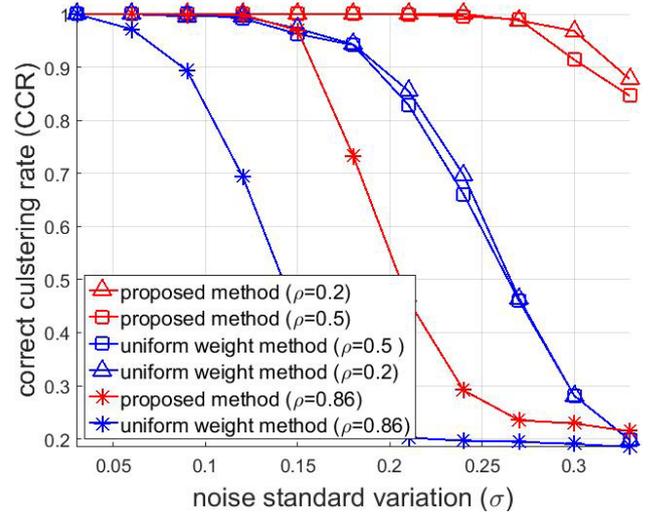

Fig. 8. Optimal $\varepsilon$ achieving highest CCR for different $(\sigma, \rho)$ pairs.

Fig. 9. CCR versus $\sigma$ curves of two methods for different subspace affinity $\rho$.

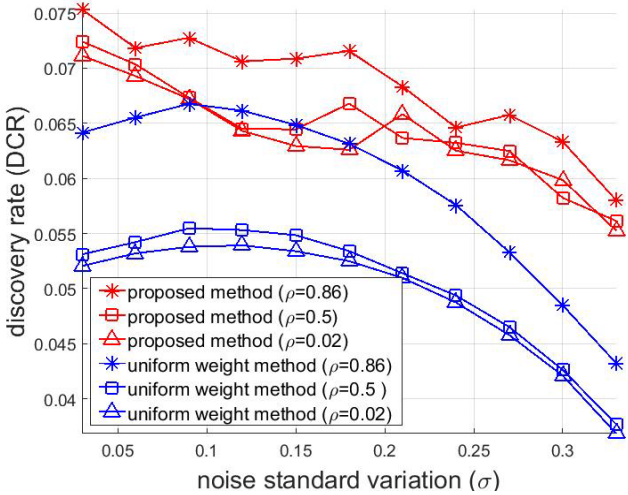
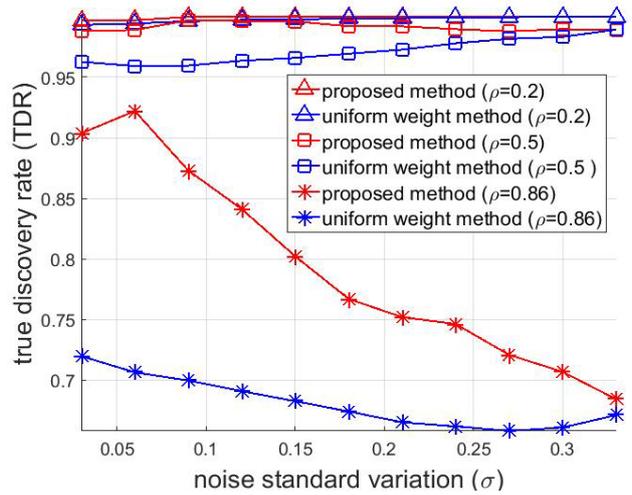

Fig. 10-(a). DCR versus noise standard deviation $\sigma$.

Fig. 10-(b). TDR versus noise standard deviation $\sigma$.

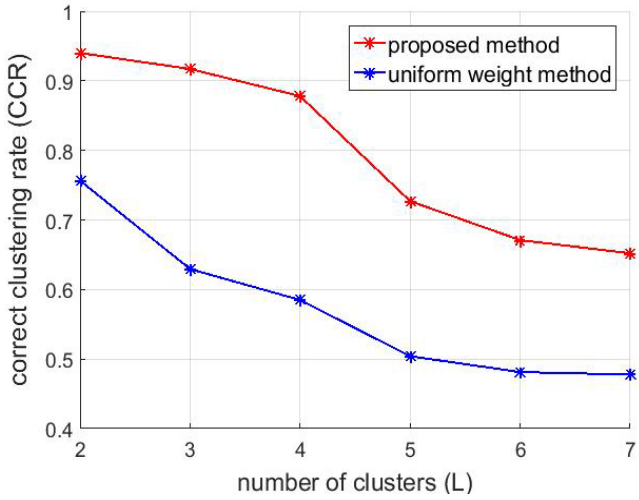 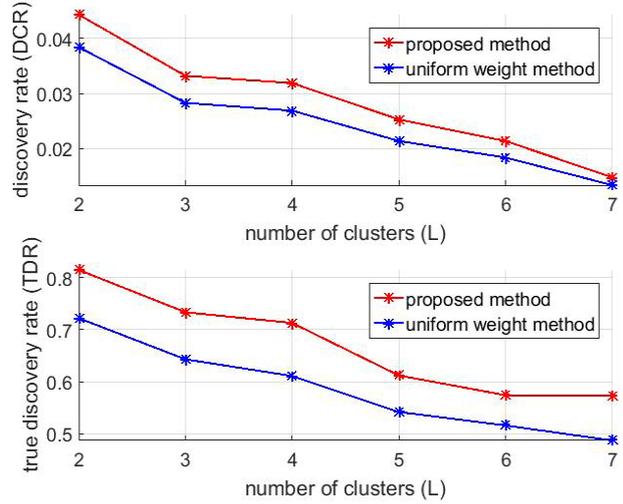

Fig. 11-(a). CCR versus number $L$ of clusters for Extended Yale human face dataset.

Fig. 11-(b). TDR and DCR versus number $L$ of clusters for Extended Yale B human face dataset.

## B. Real Human Face Data

Next, we examine the performance of the proposed algorithm using the Extended Yale B human face dataset [2]. Before applying SSC for data segmentation, the same dimensionality reduction technique as in [2] is used for reducing the computational complexity. Since the noise standard deviation $\sigma$ is not known, we consider it as a tunable parameter, and for each method conduct exhaustive search over the interval $[0 \ 0.5]$ to determine a $\sigma_0$ yielding highest CCR; as before, the $\ell_2$-norm error upper bound in the first step is set to be $\tau = 2\sigma_0$, and the regularization factor $\lambda$ used in the second step is $\lambda = 0.707\sigma_0 / \|\tilde{\mathbf{c}}_i\|_1$. Regarding the selection of the parameter $\varepsilon$, we also conduct exhaustive search for each $L$ (the number of clusters/subspaces) to determine a good solution; it is found that $\varepsilon \approx 0.01$ can achieve very good performance. Fig. 11-(a) plots the CCR as the number $L$ of people (subspaces) ranges from 2 to 7. The value of CCR at each $L$ is obtained by averaging over 200 independent trials, where in each trial $L$ clusters of human faces is randomly drawn from totally 38 clusters (each containing 64 data points). We observe from the figure that, thanks to data weighting, the proposed approach still outperforms the algorithm in [10]. Fig. 11-(b) further compares the DCR and TDR of the two methods. The proposed scheme is observed to achieve higher TDR and this again confirms our analytical study.

# V. PROOFS

This section presents the proofs of the two main theorems of our recovery rate analysis, namely, Proposition 3.1 and Theorem 3.2.

## A. *Proof of Proposition 3.1*

Assume for the moment that $w_j^{(N)} = 1$ for all $1 \leq j \leq N-1$; the results can be directly generalized to the case with data weighting (by simply scaling the data points). Note that $\Phi_q$ is determined once the data points $\{\mathbf{y}_1, \mathbf{y}_2, \cdots, \mathbf{y}_{N-1}\}$ are given. Denote by $\mathcal{B}_N$ the intersection of the unit sphere of $\mathbb{R}^n$ with the subspace containing the signal point $\mathbf{x}_N$. The conditional probability can be expressed as

$$
\begin{aligned}
&\Pr(\{\mathbf{y}_N \in \Phi_q \mid \mathbf{y}_1, \mathbf{y}_2, \cdots, \mathbf{y}_{N-1}\}) \\
&= \int_{\Phi_q} p(\mathbf{y}_N \mid \mathbf{y}_1, \mathbf{y}_2, \cdots, \mathbf{y}_{N-1}) d\mathbf{y}_N \\
&\stackrel{(a)}{=} \int_{\Phi_q} p(\mathbf{y}_N) d\mathbf{y}_N \\
&= \int_{\Phi_q} \left[ \int_{\mathcal{B}_N} p(\mathbf{y}_N \mid \mathbf{x}_N) p(\mathbf{x}_N) d\mathbf{x}_N \right] d\mathbf{y}_N \\
&= \int_{\mathcal{B}_N} \left[ \int_{\Phi_q} p(\mathbf{y}_N \mid \mathbf{x}_N) d\mathbf{y}_N \right] p(\mathbf{x}_N) d\mathbf{x}_N \\
&\stackrel{(b)}{=} \int_{\mathcal{B}_N} \left[ \int_{\Phi_q} \det\left(\frac{2\pi\sigma^2 \mathbf{I}}{n}\right)^{-1/2} \exp\left(\frac{-(\mathbf{y}_N - \mathbf{x}_N)^T [(\sigma^2/n)\mathbf{I}]^{-1} (\mathbf{y}_N - \mathbf{x}_N)}{2}\right) d\mathbf{y}_N \right] p(\mathbf{x}_N) d\mathbf{x}_N \\
&= \left(\sqrt{\frac{2\pi\sigma^2}{n}}\right)^{-n} \int_{\mathcal{B}_N} \left[ \int_{\Phi_q} \exp\left(\frac{-(\mathbf{y}_N - \mathbf{x}_N)^T [(\sigma^2/n)\mathbf{I}]^{-1} (\mathbf{y}_N - \mathbf{x}_N)}{2}\right) d\mathbf{y}_N \right] p(\mathbf{x}_N) d\mathbf{x}_N \\
&= \left(\sqrt{\frac{2\pi\sigma^2}{n}}\right)^{-n} \int_{\mathcal{B}_N} \left[ \int_{\Phi_q} \exp\left(\frac{-n(\mathbf{y}_N - \mathbf{x}_N)^T (\mathbf{y}_N - \mathbf{x}_N)}{2\sigma^2}\right) d\mathbf{y}_N \right] p(\mathbf{x}_N) d\mathbf{x}_N \\
&\stackrel{(c)}{=} \left(\sqrt{\frac{2\pi\sigma^2}{n}}\right)^{-n} \int_{\mathcal{B}_N} \left[ \int_{\Phi_q - \mathbf{x}_N} \exp\left(\frac{-n(\mathbf{y}_N)^T (\mathbf{y}_N)}{2\sigma^2}\right) d\mathbf{y}_N \right] p(\mathbf{x}_N) d\mathbf{x}_N \\
&= \left(\sqrt{\frac{2\pi\sigma^2}{n}}\right)^{-n} \int_{\mathcal{B}_N} \underbrace{\left[ \int_{\Phi_q - \mathbf{x}_N} \exp\left(\frac{-n \|\mathbf{y}_N\|_2^2}{2\sigma^2}\right) d\mathbf{y}_N \right]}_{\triangleq \rho(\mathbf{x}_N)} p(\mathbf{x}_N) d\mathbf{x}_N,
\end{aligned}
$$

(5.1)

where (a) holds since $\mathbf{y}_i$'s are independent and (b) holds by invoking Assumptions 1 and 2. Equality (c) in (5.1) follows from a direct change of variable, say, $\mathbf{w}_N = \mathbf{y}_N - \mathbf{x}_N$, while we still use $\mathbf{y}_N$ as the dummy variable to conserve notation; for a fixed $\mathbf{x}_N$ the domain of integration in the inner integral is the translated set

$$\Phi_q - \mathbf{x}_N \triangleq \{\mathbf{y} - \mathbf{x}_N \mid \mathbf{y} \in \Phi_q\}. \tag{5.2}$$

To derive the upper bound for $\Pr(\{\mathbf{y}_N \in \Phi_q \mid \mathbf{y}_1, \mathbf{y}_2, \cdots, \mathbf{y}_{N-1}\})$, we shall first find a more explicit expression for $\rho(\mathbf{x}_N)$, which involves an integration of the exponential function over $\Phi_q - \mathbf{x}_N$. Towards this end, the basic idea is to express $\mathbf{y}_N \in \Phi_q - \mathbf{x}_N$ in the form specified in Corollary 2.3, and leverage the change-of-variable theorem in integration theory [36], stated as below.

***Theorem 5.1 [36, Theorem 7.26]:*** Assume that

(i) $X \subset V \subset \mathbb{R}^n$, $V$ is open, $T : V \to \mathbb{R}^n$ is continuous;

(ii) $X$ is Lebesgue measurable, $T$ is one-to-one on $X$, and $T$ is differentiable at every point of $X$;

(iii) The set $T(V - X)$ is measure zero[9], i.e., $|T(V - X)| = 0$.

Then, setting $Y = T(X)$,

$$\int_Y f dm = \int_X (f \circ T) |\det(T')| dm \tag{5.3}$$

for every measurable $f : \mathbb{R}^n \to [0, \infty]$, where $T'$ is the Jacobian of $T$. $\square$

To derive an expression for $\rho(\mathbf{x}_N)$ under the setting of Theorem 5.1, suppose that $\mathcal{I}_{q,0} = \{s_1, \cdots, s_{N_1}\}$, $\mathcal{I}_{q,+} = \{t_1, \cdots t_{N_2}\}$, $\mathcal{I}_{q,-} = \{t_{N_2+1}, \cdots t_{N_3}\}$ (hence, there are $N_1$ inactive constraints and $N_3$ active constraints[10] so that $N_1 + N_3 = N - 1$). We then define the two subsets $\mathcal{S}_1, \mathcal{S}_2 \subset \mathbb{R}^n$ to be

$$\mathcal{S}_1 \triangleq \underbrace{(-1 + \alpha_{s_1}, 1 + \alpha_{s_1}) \times \cdots \times (-1 + \alpha_{s_{N_1}}, 1 + \alpha_{s_{N_1}})}_{N_1-\text{folds}} \times \\ \underbrace{(\beta_{t_1}, \infty) \times \cdots \times (\beta_{t_{N_2}}, \infty) \times (-\infty, -\beta_{t_{N_2+1}}) \times \cdots \times (-\infty, -\beta_{t_{N_3}})}_{N_3-\text{fold}} \times \underbrace{\mathbb{R} \times \cdots \times \mathbb{R}}_{(n-N+1)-\text{fold}} \tag{5.4}$$

$$\mathcal{S}_2 \triangleq \underbrace{(-1 + \alpha_{s_1}, 1 + \alpha_{s_1}) \times \cdots \times (-1 + \alpha_{s_{N_1}}, 1 + \alpha_{s_{N_1}})}_{N_1-\text{fold}} \times \\ \underbrace{[\beta_{t_1}, \infty) \times \cdots \times [\beta_{t_{N_2}}, \infty) \times (-\infty, -\beta_{t_{N_2+1}}] \times \cdots \times (-\infty, -\beta_{t_{N_3}}]}_{N_3-\text{fold}} \times \underbrace{\mathbb{R} \times \cdots \times \mathbb{R}}_{(n-N+1)-\text{fold}}, \tag{5.5}$$

for some real $\alpha_{s_1}, \cdots, \alpha_{s_{N_1}}$ and $\beta_{t_1}, \cdots, \beta_{t_{N_3}}$. Since $\mathcal{S}_1$ is open, it is Lebesgue measurable. Consider the linear transformation $T : \mathcal{S}_2 \to \mathbb{R}^n$ given by

$$T(\mathbf{x}) \triangleq \overline{\mathbf{A}}_q \mathbf{x}, \tag{5.6}$$

where

$$\overline{\mathbf{A}}_q = \left[ \overline{\mathbf{y}}_{s_1}^{(N)}, \cdots, \overline{\mathbf{y}}_{s_{N_1}}^{(N)}, \mathbf{y}_{t_1}, \cdots, \mathbf{y}_{t_{N_3}}, \mathbf{n}_1, \cdots, \mathbf{n}_{n-N+1} \right] \in \mathbb{R}^{n \times n} \tag{5.7}$$

is nonsingular[11], and the columns $\overline{\mathbf{y}}_{s_j}^{(N)}$'s and $\mathbf{n}_i$'s are defined in (2.8) and (2.7), respectively. Note that $T$ in (5.6) is one-to-one, differentiable, and Lipschitz continuous; it is also straightforward to see (using Corollary 2.3) the image of $T$ is $T(\mathcal{S}_2) = \Phi_q - \mathbf{x}_N$. Note that

$$\underbrace{\mathcal{S}_2 \setminus \mathcal{S}_1 = (-1+\alpha_{s_1}, 1+\alpha_{s_1}) \times \cdots \times (-1+\alpha_{s_{N_1}}, 1+\alpha_{s_{N_1}})}_{N_1-\text{fold}} \times \underbrace{\{\beta_{t_1}\} \times \cdots \times \{\beta_{t_{N_3}}\}}_{N_3-\text{fold}} \times \underbrace{\mathbb{R} \times \cdots \times \mathbb{R}}_{(n-N+1)-\text{fold}} \quad (5.8)$$

is contained in the subset $\mathbb{R}^{N_1} \times \underbrace{\{\beta_{t_1}\} \times \cdots \times \{\beta_{t_{N_3}}\}}_{N_3-\text{fold}} \times \mathbb{R}^{(n-N+1)}$, which is a translation of the subspace $\mathbb{R}^{N_1} \times \underbrace{\{0\} \times \cdots \times \{0\}}_{N_3-\text{fold}} \times \mathbb{R}^{(n-N+1)}$ with dimension $n - N_3 < n$. Since a subspace (of $\mathbb{R}^n$) with a dimension less than $n$ is measure zero [35, Chap. 2] and the measure of a set is invariant under translation [35, p-60], it follows

$$|\mathcal{S}_2 \setminus \mathcal{S}_1| = 0. \quad (5.9)$$

---

9. To conserve notation, we keep the notation $|\mathcal{S}|$ for the Lebesgue measure of a general Lebesgue measurable subset $\mathcal{S} \subset \mathbb{R}^n$.
10. The integers $N_1, N_2, N_3$ depend on the index $q$, which is omitted for simplicity and without causing confusion.
11. As we have mentioned, the matrix $\overline{\mathbf{A}}_q$ being nonsingular is typically true in the setting of noisy SSC.

Also, since $(\Phi_q - \mathbf{x}_N) \setminus T(\mathcal{S}_1) = T(\mathcal{S}_2) \setminus T(\mathcal{S}_1) = T(\mathcal{S}_2 \setminus \mathcal{S}_1)$, we have

$$|(\Phi_q - \mathbf{x}_N) \setminus T(\mathcal{S}_1)| = |T(\mathcal{S}_2) \setminus T(\mathcal{S}_1)| = |T(\mathcal{S}_2 \setminus \mathcal{S}_1)| = 0, \quad (5.10)$$

where the last equality holds since the image of a measure-zero set under Lipschitz transformation is again with measure zero [35, p-55]. With $T$ defined as in (5.6) and setting $V = X = \mathcal{S}_1$, assumptions (i), (ii), and (iii) in Theorem 5.1 are fulfilled. Therefore, based on (5.3) an explicit formula for $\rho(\mathbf{x}_N)$ is obtained as follows

$$\begin{aligned}
\rho(\mathbf{x}_N) &= \int_{\Phi_q - \mathbf{x}_N} \exp\left(\frac{-n\|\mathbf{y}_N\|_2^2}{2\sigma^2}\right) d\mathbf{y}_N = \int_{T(\mathcal{S}_2)} \exp\left(\frac{-n\|\mathbf{y}_N\|_2^2}{2\sigma^2}\right) d\mathbf{y}_N \\
&\stackrel{(a)}{=} \int_{T(\mathcal{S}_2) \setminus T(\mathcal{S}_2 \setminus \mathcal{S}_1)} \exp\left(\frac{-n\|\mathbf{y}_N\|_2^2}{2\sigma^2}\right) d\mathbf{y}_N \\
&= \int_{T(\mathcal{S}_1)} \exp\left(\frac{-n\|\mathbf{y}_N\|_2^2}{2\sigma^2}\right) d\mathbf{y}_N \\
&\stackrel{(b)}{=} \int_{\mathcal{S}_1} e^{\frac{-n}{2\sigma^2}\left\|\sum_{j=1}^{N_1} a_{s_j} \overline{\mathbf{y}}_{s_j}^{(N)} + \sum_{j=1}^{N_2} b_{t_j} \mathbf{y}_{t_j} - \sum_{j=N_2+1}^{N_3} b_{t_j} \mathbf{y}_{t_j} + \sum_{k=1}^{n-N+1} h_k \mathbf{n}_k\right\|_2^2} \left|\det(\overline{\mathbf{A}}_q)\right| da_{s_1} \cdots da_{s_{N_1}} db_{t_1} \cdots db_{t_{N_3}} dh_1 \cdots dh_{n-(N-1)} \\
&\stackrel{(c)}{=} \left|\det(\overline{\mathbf{A}}_q)\right| \int_{\mathcal{S}_1} e^{\frac{-n}{2\sigma^2}\left(\left\|\sum_{j=1}^{N_1} a_{s_j} \overline{\mathbf{y}}_{s_j}^{(N)}\right\|_2^2 + \left\|\sum_{j=1}^{N_2} b_{t_j} \mathbf{y}_{t_j} - \sum_{j=N_2+1}^{N_3} b_{t_j} \mathbf{y}_{t_j}\right\|_2^2 + \left\|\sum_{k=1}^{n-N+1} h_k \mathbf{n}_k\right\|_2^2\right)} da_{s_1} \cdots da_{s_{N_1}} db_{t_1} \cdots db_{t_{N_3}} dh_1 \cdots dh_{n-(N-1)} \\
&\stackrel{(d)}{=} \left|\det(\overline{\mathbf{A}}_q)\right| \delta_1 \delta_2 \delta_3,
\end{aligned}$$
$$(5.11)$$

where (a) holds since $T(\mathcal{S}_2 \setminus \mathcal{S}_1)$ is measure zero, (b) follows from (5.3), (c) is due to (2.8), and (d) is obtained with some manipulations leading to

$$\delta_1 \triangleq \int_{-\infty}^{\infty} \cdots \int_{-\infty}^{\infty} \exp\left(\frac{-n}{2\sigma^2}\left\|\sum_{k=1}^{n-N+1} h_k \mathbf{n}_k\right\|_2^2\right) dh_1 \cdots dh_{n-N+1}, \quad (5.12)$$

$$\delta_2 \triangleq \int_{-1+\alpha_{s_1}}^{1+\alpha_{s_1}} \cdots \int_{-1+\alpha_{s_{N_1}}}^{1+\alpha_{s_{N_1}}} \exp\left(\frac{-n}{2\sigma^2}\left\|\sum_{j=1}^{N_1} a_{s_j} \overline{\mathbf{y}}_{s_j}^{(N)}\right\|_2^2\right) da_{s_{N_1}} \cdots da_{s_1}, \quad (5.13)$$

$$\delta_3 \triangleq \int_{\beta_{t_1}}^{\infty} \cdots \int_{\beta_{t_{N_3}}}^{\infty} \exp\left(\frac{-n}{2\sigma^2}\left\|\sum_{j=1}^{N_2} b_{t_j} \mathbf{y}_{t_j} - \sum_{j=N_2+1}^{N_3} b_{t_j} \mathbf{y}_{t_j}\right\|_2^2\right) db_{t_{N_3}} \cdots db_{t_1}. \quad (5.14)$$

Based on (5.11), it remains to derive upper bounds for $\delta_i$, $1 \leq i \leq 3$, defined as above. An upper bound for $\delta_1$ can be immediately obtained as

$$\delta_1 = \int_{-\infty}^{\infty} \cdots \int_{-\infty}^{\infty} \exp\left(\frac{-n}{2\sigma^2}\left\|\sum_{k=1}^{n-N+1} h_k \mathbf{n}_k\right\|_2^2\right) dh_1 \cdots dh_{n-N+1}$$

$$\stackrel{(a)}{=} \left(\int_{-\infty}^{\infty} \exp\left(\frac{-nh^2}{2\sigma^2}\right) dh\right)^{n-N+1} \quad (5.15)$$

$$\stackrel{(b)}{=} \left(\sqrt{2\sigma^2/n} \int_{-\infty}^{\infty} \exp(-v^2) dv\right)^{n-N+1}$$

$$= \left(\sqrt{2\pi\sigma^2/n}\right)^{n-N+1},$$

where (a) holds since $\mathbf{n}_k$'s are orthonormal, and (b) follows through a direct change of variable. Upper bounds for $\delta_2$ and $\delta_3$ are obtained as below (detailed proof is induction based and is given in Appendix C)

$$\delta_2 \leq \prod_{j=1}^{N_1} \left(\sqrt{\frac{2\pi\sigma^2}{nc_{j,q}\left\|\overline{\mathbf{y}}_{s_j}^{(N)}\right\|_2^2}} erf\left(\sqrt{\frac{nc_{j,q}}{2\sigma^2}}\left\|\overline{\mathbf{y}}_{s_j}^{(N)}\right\|_2\right)\right), \quad (5.16)$$

and

$$\delta_3 \leq \prod_{j=1}^{N_3} \sqrt{\frac{2\pi\sigma^2}{n\left\|\mathbf{y}_{t_j}\right\|_2^2}} \quad (5.17)$$

in which $c_{j,q}$'s are some constants. Combining (5.11) and (5.15)~(5.17) yields

$$\rho(\mathbf{x}_N) = \left|\det\left(\overline{\mathbf{A}}_q\right)\right|\delta_1\delta_2\delta_3$$

$$\leq \frac{\left|\det\left(\overline{\mathbf{A}}_q\right)\right|}{\prod_{j=1}^{N_1}\left\|\overline{\mathbf{y}}_{s_j}^{(N)}\right\|_2 \prod_{j=1}^{N_3}\left\|\mathbf{y}_{t_j}\right\|_2}\left(\sqrt{\frac{2\pi\sigma^2}{n}}\right)^n \prod_{j=1}^{N_1}\left(\sqrt{\frac{1}{c_{j,q}}}erf\left(\sqrt{\frac{nc_{j,q}}{2\sigma^2}}\left\|\overline{\mathbf{y}}_{s_j}^{(N)}\right\|_2\right)\right) \quad (5.18)$$

$$= \left|\det\left(\overline{\mathbf{U}}_q\right)\right|\left(\sqrt{\frac{2\pi\sigma^2}{n}}\right)^n \prod_{j=1}^{N_1}\left(\sqrt{\frac{1}{c_{j,q}}}erf\left(\sqrt{\frac{nc_{j,q}}{2\sigma^2}}\left\|\overline{\mathbf{y}}_{s_j}^{(N)}\right\|_2\right)\right),$$

where

$$\overline{\mathbf{U}}_q \triangleq \left[\frac{\overline{\mathbf{y}}_{s_1}^{(N)}}{\left\|\overline{\mathbf{y}}_{s_1}^{(N)}\right\|_2}, \cdots, \frac{\overline{\mathbf{y}}_{s_{N_1}}^{(N)}}{\left\|\overline{\mathbf{y}}_{s_{N_1}}^{(N)}\right\|_2}, \frac{\mathbf{y}_{t_1}}{\left\|\mathbf{y}_{t_1}\right\|_2}, \cdots, \frac{\mathbf{y}_{t_{N_3}}}{\left\|\mathbf{y}_{t_{N_3}}\right\|_2}, \mathbf{n}_1, \cdots, \mathbf{n}_{n-N+1}\right] \in \mathbb{R}^{n \times n} \quad (5.19)$$

is obtained by normalizing the first $N-1$ columns of $\overline{\mathbf{A}}_q$ to be unit-norm. With (5.1) and (5.18), we reach

$$\Pr(\{\mathbf{y}_N \in \Phi_q \mid \mathbf{y}_1, \mathbf{y}_2, \cdots, \mathbf{y}_{N-1}\})$$

$$= \left(\sqrt{\frac{2\pi\sigma^2}{n}}\right)^{-n} \int_{\mathcal{B}_N} \rho(\mathbf{x}_N) p(\mathbf{x}_N) d\mathbf{x}_N$$

$$\stackrel{(a)}{\leq} \left(\sqrt{\frac{2\pi\sigma^2}{n}}\right)^{-n} \int_{\mathcal{B}_N} \left[\left|\det\left(\overline{\mathbf{U}}_q\right)\right|\left(\sqrt{\frac{2\pi\sigma^2}{n}}\right)^n \prod_{j=1}^{N_1}\left(\sqrt{\frac{1}{c_{j,q}}}erf\left(\sqrt{\frac{nc_{j,q}}{2\sigma^2}}\left\|\overline{\mathbf{y}}_{s_j}^{(N)}\right\|_2\right)\right)\right] p(\mathbf{x}_N) d\mathbf{x}_N \quad (5.20)$$

$$= \left|\det\left(\overline{\mathbf{U}}_q\right)\right| \prod_{j=1}^{N_1}\left(\sqrt{\frac{1}{c_{j,q}}}erf\left(\sqrt{\frac{nc_{j,q}}{2\sigma^2}}\left\|\overline{\mathbf{y}}_{s_j}^{(N)}\right\|_2\right)\right) \underbrace{\int_{\mathcal{B}_N} p(\mathbf{x}_N) d\mathbf{x}_N}_{=1}$$

where (a) holds by (5.18). Recall from (3.1) that the effect of data weighting is to scale the vector $\mathbf{y}_j$ to $(1/w_j^{(N)})\mathbf{y}_j$: this amounts to scaling $\overline{\mathbf{y}}_j^{(N)}$ to $w_j^{(N)}\overline{\mathbf{y}}_j^{(N)}$ (see (2.8)). Hence, with data weighting the upper bound in (5.20) accordingly becomes

$$\Pr(\{\mathbf{y}_N \in \Phi_q \mid \mathbf{y}_1, \mathbf{y}_2, \cdots, \mathbf{y}_{N-1}\}) \leq |\det(\overline{\mathbf{U}}_q)| \prod_{j=1}^{N_1} \left( \sqrt{\frac{1}{c_{j,q}}} erf\left( \sqrt{\frac{nc_{j,q}}{2\sigma^2}} w_{s_j}^{(N)} \left\| \overline{\mathbf{y}}_{s_j}^{(N)} \right\|_2 \right) \right)$$
$$= C_q \prod_{j=1}^{N_1} \left( erf\left(g_{s_j} w_{s_j}^{(N)}\right) \right), \quad (5.21)$$

in which

$$C_q \triangleq |\det(\overline{\mathbf{U}}_q)| \prod_{j=1}^{N_1} \sqrt{\frac{1}{c_{j,q}}} \quad \text{and} \quad g_{s_j} \triangleq \sqrt{\frac{nc_{j,q}}{2\sigma^2}} \left\| \overline{\mathbf{y}}_{s_j}^{(N)} \right\|_2. \quad (5.22)$$

The proof is thus completed.

### B. Proof of Theorem 3.2

*Proof of Part (a):*

Recall that $\{t_1, \cdots, t_{m_N}\}$ and $\{f_1, \cdots, f_{(N-1)-m_N}\}$ are the index subsets for the data points originating from, respectively, the same and different clusters as $\mathbf{y}_N$. Therefore, $\Pr(\{|\mathcal{T}| < k_t\} \cap \{|\mathcal{F}| \leq k_f\})$ can be expressed as

$$\Pr(\{|\mathcal{T}| < k_t\} \cap \{|\mathcal{F}| \leq k_f\}) = \sum_{q \in \mathcal{J}_{k_t,k_f}} \Pr(\{\mathbf{y}_N \in \Phi_q\}), \quad (5.23)$$

where

$$\mathcal{J}_{k_t,k_f} \triangleq \left\{ q \mid \left|(\mathcal{I}_{q,+} \cup \mathcal{I}_{q,-}) \cap \{t_1, \cdots, t_{m_N}\}\right| < k_t, \ \left|(\mathcal{I}_{q,+} \cup \mathcal{I}_{q,-}) \cap \{f_1, \cdots, f_{(N-1)-m_N}\}\right| \leq k_f \right\} \quad (5.24)$$

is the index subset associated with those $\Phi_q$'s yielding less than $k_t$ true discoveries and at most $k_f$ false discoveries, in which $\mathcal{I}_{q,+}$ and $\mathcal{I}_{q,-}$ are defined in (2.4) and (2.5), respectively. To proceed, we shall first find an upper bound for $\Pr(\{\mathbf{y}_N \in \Phi_q\})$, namely, the probability that $\mathbf{y}_N$ is contained in $\Phi_q$, where $1 \leq q \leq 3^{N-1}$ is arbitrary and fixed. By definition we have

$$\Pr(\{\mathbf{y}_N \in \Phi_q\})$$
$$= \int_{\mathbb{R}^n} \cdots \int_{\mathbb{R}^n} \Pr(\{\mathbf{y}_N \in \Phi_q \mid \mathbf{y}_1, \mathbf{y}_2, \cdots, \mathbf{y}_{N-1}\}) p(\mathbf{y}_1, \mathbf{y}_2, \cdots, \mathbf{y}_{N-1}) d\mathbf{y}_1 \cdots d\mathbf{y}_{N-1}$$
$$= \int_{\mathbb{R}^n \times \cdots \times \mathbb{R}^n} \int \Pr(\{\mathbf{y}_N \in \Phi_q \mid \mathbf{y}_1, \mathbf{y}_2, \cdots, \mathbf{y}_{N-1}\}) p(\mathbf{y}_1, \mathbf{y}_2, \cdots, \mathbf{y}_{N-1}) d\mathbf{y}_1 \cdots d\mathbf{y}_{N-1}$$
$$= \int_{\mathcal{K}_{M,k}} \int \Pr(\{\mathbf{y}_N \in \Phi_q \mid \mathbf{y}_1, \mathbf{y}_2, \cdots, \mathbf{y}_{N-1}\}) p(\mathbf{y}_1, \mathbf{y}_2, \cdots, \mathbf{y}_{N-1}) d\mathbf{y}_1 \cdots d\mathbf{y}_{N-1} \quad (5.25)$$
$$+ \int_{(\mathbb{R}^n \times \cdots \times \mathbb{R}^n) \setminus \mathcal{K}_{M,k}} \int \Pr(\{\mathbf{y}_N \in \Phi_q \mid \mathbf{y}_1, \mathbf{y}_2, \cdots, \mathbf{y}_{N-1}\}) p(\mathbf{y}_1, \mathbf{y}_2, \cdots, \mathbf{y}_{N-1}) d\mathbf{y}_1 \cdots d\mathbf{y}_{N-1}$$
$$\stackrel{(a)}{\leq} \underbrace{\int_{\mathcal{K}_{M,k}} \int |\det(\overline{\mathbf{U}}_q)| \prod_{j=1}^{N_1} \left( \sqrt{1/c_{j,q}} erf\left( \sqrt{nc_{j,q}/2\sigma^2} w_{s_j}^{(N)} \left\| \overline{\mathbf{y}}_{s_j}^{(N)} \right\|_2 \right) \right) p(\mathbf{y}_1, \mathbf{y}_2, \cdots, \mathbf{y}_{N-1}) d\mathbf{y}_1 \cdots d\mathbf{y}_{N-1}}_{\triangleq \delta_4}$$
$$+ \underbrace{\int_{(\mathbb{R}^n \times \cdots \times \mathbb{R}^n) \setminus \mathcal{K}_{M,k}} \int p(\mathbf{y}_1, \mathbf{y}_2, \cdots, \mathbf{y}_{N-1}) d\mathbf{y}_1 \cdots d\mathbf{y}_{N-1}}_{\triangleq \delta_5} = \delta_4 + \delta_5,$$

where (a) holds by using (5.21) and

$$\mathcal{K}_{M,k} \triangleq \mathcal{K}_M \setminus \{\mathbf{y}_1,\cdots,\mathbf{y}_{N-1} : |\det(\overline{\mathbf{U}}_q)| < 1/k\} \tag{5.26}$$

and

$$\mathcal{K}_M \triangleq \{\mathbf{x} + \mathbf{e} : \mathbf{x} \in \mathcal{B}_1, \|\mathbf{e}\|_2 \leq M\sigma\} \times \cdots \times \{\mathbf{x} + \mathbf{e} : \mathbf{x} \in \mathcal{B}_{N-1}, \|\mathbf{e}\|_2 \leq M\sigma\}, \tag{5.27}$$

with $\mathcal{B}_i$ denoting the intersection of the unit sphere in $\mathbb{R}^n$ with the subspace containing the noiseless signal point $\mathbf{x}_i$. Here, by splitting the integral in (5.25) into a sum of two terms, the integral is upper bounded by $\delta_4 + \delta_5$, whose magnitudes can be readily controlled and proved to be small. An upper bound for $\Pr(\{\mathbf{y}_N \in \Phi_q\})$ can therefore be obtained by deriving upper bounds for the two terms $\delta_4$ and $\delta_5$ defined in (5.25).

We will first find an upper bound for the term $\delta_4$. Since $|\det(\overline{\mathbf{U}}_q)|$ ($\overline{\mathbf{U}}_q$ is specified in (5.19)) is a continuous function from $\mathbb{R}^n \times \cdots \times \mathbb{R}^n$ into $\mathbb{R}$ and $\{x \in \mathbb{R} | x < 1/k\}$ is open in $\mathbb{R}$, the inverse image $\{\mathbf{y}_1,\cdots,\mathbf{y}_{N-1} : |\det(\overline{\mathbf{U}}_q)| < 1/k\}$ is open in $\mathbb{R}^n \times \cdots \times \mathbb{R}^n$. Also, as $\mathcal{K}_M$ in (5.27) is compact and the complement of $\{\mathbf{y}_1,\cdots,\mathbf{y}_{N-1} : |\det(\overline{\mathbf{U}}_q)| < 1/k\}$ is closed, it follows from (5.26) that $\mathcal{K}_{M,k}$ is again compact. Since $|\det(\overline{\mathbf{U}}_q)|$, $\sqrt{1/c_{j,q}}$ and $\sqrt{nc_{j,q}/2\sigma^2} \|\overline{\mathbf{y}}_{s_j}^{(N)}\|_2$ ($c_{j,q}$ is defined in (C.8) in Appendix C) are continuous on the compact set $\mathcal{K}_{M,k}$, by the extreme value theorem [37] there exist $\alpha_q$, $\beta_{j,q}$ and $\gamma_{j,q}$ such that

$$|\det(\overline{\mathbf{U}}_q)| \leq \alpha_q, \quad \sqrt{1/c_{j,q}} \leq \beta_{j,q} \quad \text{and} \quad \sqrt{nc_{j,q}/2\sigma^2} \|\overline{\mathbf{y}}_{s_j}^{(N)}\|_2 \leq \gamma_{j,q} \tag{5.28}$$

for all $(\mathbf{y}_1,\cdots,\mathbf{y}_{N-1}) \in \mathcal{K}_{M,k} \subset \mathbb{R}^n \times \cdots \times \mathbb{R}^n$. Therefore,

$$\begin{aligned}
\delta_4 &= \int_{\mathcal{K}_{M,k}} \cdots \int |\det(\overline{\mathbf{U}}_q)| \prod_{j=1}^{N_1} \left(\sqrt{1/c_{j,q}} \, erf\left(w_{s_j}^{(N)} \sqrt{nc_{j,q}/2\sigma^2} \|\overline{\mathbf{y}}_{s_j}^{(N)}\|_2\right)\right) p(\mathbf{y}_1,\mathbf{y}_2,\cdots,\mathbf{y}_{N-1}) d\mathbf{y}_1 \cdots d\mathbf{y}_{N-1} \\
&\stackrel{(a)}{\leq} \int_{\mathcal{K}_{M,k}} \cdots \int \alpha_q \prod_{j=1}^{N_1} \left(\beta_{j,q} \, erf\left(\gamma_{j,q} w_{s_j}^{(N)}\right)\right) p(\mathbf{y}_1,\mathbf{y}_2,\cdots,\mathbf{y}_{N-1}) d\mathbf{y}_1 \cdots d\mathbf{y}_{N-1} \\
&= \alpha_q \prod_{j=1}^{N_1} \left(\beta_{j,q} \, erf\left(\gamma_{j,q} w_{s_j}^{(N)}\right)\right) \int_{\mathcal{K}_{M,k}} \cdots \int p(\mathbf{y}_1,\mathbf{y}_2,\cdots,\mathbf{y}_{N-1}) d\mathbf{y}_1 \cdots d\mathbf{y}_{N-1} \\
&\leq \alpha_q \prod_{j=1}^{N_1} \left(\beta_{j,q} \, erf\left(\gamma_{j,q} w_{s_j}^{(N)}\right)\right) = \widehat{C}_q \prod_{j=1}^{N_1} erf\left(\gamma_{j,q} w_{s_j}^{(N)}\right),
\end{aligned} \tag{5.29}$$

where (a) holds by (5.28) and

$$\widehat{C}_q \triangleq \alpha_q \prod_{j=1}^{N_1} \beta_{j,q}. \tag{5.30}$$

Next, we go on to find an upper bound for $\delta_5$. Note that the domain of integration can be expressed as

$$\begin{aligned}
&(\mathbb{R}^n \times \cdots \times \mathbb{R}^n) \setminus \mathcal{K}_{M,k} \\
&\stackrel{(a)}{=} (\mathbb{R}^n \times \cdots \times \mathbb{R}^n) \setminus (\mathcal{K}_M \setminus \{\mathbf{y}_1,\cdots,\mathbf{y}_{N-1} : |\det(\overline{\mathbf{U}}_q)| < 1/k\}) \\
&\stackrel{(b)}{=} ((\mathbb{R}^n \times \cdots \times \mathbb{R}^n) \setminus \mathcal{K}_M) \cup \mathcal{Z}_{M,k} \\
&\stackrel{(c)}{\subset} (\mathcal{L}_{1,M} \cup \cdots \cup \mathcal{L}_{N-1,M}) \cup \mathcal{Z}_{M,k},
\end{aligned} \tag{5.31}$$

where (a) follows from the definition of $\mathcal{K}_{M,k}$ in (5.26), and (b) can be obtained by simple set algebra with

$$\mathcal{Z}_{M,k} \triangleq \mathcal{K}_M \cap \{\mathbf{y}_1,\cdots,\mathbf{y}_{N-1} : |\det(\overline{\mathbf{U}}_q)| < 1/k\}, \tag{5.32}$$

and (c) directly follows also from simple set algebra and with

$$\mathcal{L}_{i,M} \triangleq \underbrace{\mathbb{R}^n \times \cdots \times \mathbb{R}^n}_{(i-1)-\text{fold}} \times \underbrace{\{\mathbf{y}_i \in \mathbb{R}^n : \|\mathbf{y}_i - \mathbf{x}_i\|_2 > M\sigma, \ \mathbf{x}_i \in \mathcal{B}_i\}}_{\triangleq \mathcal{H}_{i,M}} \times \underbrace{\mathbb{R}^n \times \cdots \times \mathbb{R}^n}_{(N-1-i)-\text{fold}}. \tag{5.33}$$

Since $p(\mathbf{y}_1, \mathbf{y}_2, \cdots, \mathbf{y}_{N-1})$ is continuous on the compact set $\mathcal{K}_M$, again by the extreme value theorem there exists $c$ such that

$$p(\mathbf{y}_1, \mathbf{y}_2, \cdots, \mathbf{y}_{N-1}) \leq c, \text{ for all } (\mathbf{y}_1, \cdots, \mathbf{y}_{N-1}) \in \mathcal{K}_M. \tag{5.34}$$

Accordingly, we can obtain

$$\begin{aligned}
\delta_5 &= \int_{(\mathbb{R}^n \times \cdots \times \mathbb{R}^n) \setminus \mathcal{K}_{M,k}} \cdots \int p(\mathbf{y}_1, \mathbf{y}_2, \cdots, \mathbf{y}_{N-1}) d\mathbf{y}_1 \cdots d\mathbf{y}_{N-1} \\
&\stackrel{(a)}{\leq} \left(\sum_{i=1}^{N-1} \int_{\mathcal{L}_{i,M}} \cdots \int p(\mathbf{y}_1, \mathbf{y}_2, \cdots, \mathbf{y}_{N-1}) d\mathbf{y}_{N-1} \cdots d\mathbf{y}_1\right) + \int_{\mathcal{Z}_{M,k}} \cdots \int p(\mathbf{y}_1, \mathbf{y}_2, \cdots, \mathbf{y}_{N-1}) d\mathbf{y}_{N-1} \cdots d\mathbf{y}_1 \\
&\stackrel{(b)}{=} \left(\sum_{i=1}^{N-1} \int_{\mathcal{L}_{i,M}} \cdots \int p(\mathbf{y}_1) \cdots p(\mathbf{y}_{N-1}) d\mathbf{y}_{N-1} \cdots d\mathbf{y}_1\right) + \int_{\mathcal{Z}_{M,k}} \cdots \int p(\mathbf{y}_1, \mathbf{y}_2, \cdots, \mathbf{y}_{N-1}) d\mathbf{y}_{N-1} \cdots d\mathbf{y}_1 \\
&\stackrel{(c)}{=} \left(\sum_{i=1}^{N-1} \int_{\mathcal{H}_{i,M}} p(\mathbf{y}_i) d\mathbf{y}_i\right) + \int_{\mathcal{Z}_{M,k}} \cdots \int p(\mathbf{y}_1, \mathbf{y}_2, \cdots, \mathbf{y}_{N-1}) d\mathbf{y}_{N-1} \cdots d\mathbf{y}_1 \\
&\stackrel{(d)}{\leq} \left(\sum_{i=1}^{N-1} \int_{\mathcal{H}_{i,M}} p(\mathbf{y}_i) d\mathbf{y}_i\right) + \int_{\mathcal{Z}_{M,k}} \cdots \int c \, d\mathbf{y}_{N-1} \cdots d\mathbf{y}_1 \\
&= \left(\sum_{i=1}^{N-1} \int_{\mathcal{H}_{i,M}} p(\mathbf{y}_i) d\mathbf{y}_i\right) + c|\mathcal{Z}_{M,k}|,
\end{aligned} \tag{5.35}$$

where (a) follows from (5.31), (b) holds since $\mathbf{y}_i$'s are independent, (c) is true due to (5.33) and (d) holds from (5.34). By Assumptions 1 and 2 made in Section III-A, it follows that, conditioned on an $\mathbf{x}_i$, $\mathbf{y}_i$ is a Gaussian random vector; this together with (5.33) implies

$$\lim_{M \to \infty} \int_{\mathcal{H}_{i,M}} p(\mathbf{y}_i \mid \mathbf{x}_i) d\mathbf{y}_i = 0, \tag{5.36}$$

where $\mathcal{H}_{i,M}$ is defined in (5.33). Recalling that $\mathcal{B}_i$ is the intersection of the unit sphere of $\mathbb{R}^n$ with the subspace containing the signal point $\mathbf{x}_i$, we can obtain

$$\begin{aligned}
&\lim_{M \to \infty} \int_{\mathcal{H}_{i,M}} p(\mathbf{y}_i) d\mathbf{y}_i \\
&= \lim_{M \to \infty} \int_{\mathcal{H}_{i,M}} \left(\int_{\mathcal{B}_i} p(\mathbf{y}_i \mid \mathbf{x}_i) p(\mathbf{x}_i) d\mathbf{x}_i\right) d\mathbf{y}_i \\
&= \int_{\mathcal{B}_i} \left(\lim_{M \to \infty} \int_{\mathcal{H}_{i,M}} p(\mathbf{y}_i \mid \mathbf{x}_i) d\mathbf{y}_i\right) p(\mathbf{x}_i) d\mathbf{x}_i \\
&\stackrel{(a)}{=} \int_{\mathcal{B}_i} (0) p(\mathbf{x}_i) d\mathbf{x}_i = 0,
\end{aligned} \tag{5.37}$$

where (a) follows from (5.36). Also, it can be shown that (see Appendix D) $|\mathcal{Z}_{M,k}|$ converges to zero as

$k$ approaches infinity, i.e.,

$$\lim_{k \to \infty} |\mathcal{Z}_{M,k}| = 0. \tag{5.38}$$

Now, choose $\varepsilon = (3 \times 10)^{-(N-1)}$. Then by (5.37), there exists $\widehat{M} \in \mathbb{N}$ such that

$$\int_{\mathcal{H}_{i,\widehat{M}}} p(\mathbf{y}_i) d\mathbf{y}_i < \varepsilon/(2(N-1)). \tag{5.39}$$

Also, with (5.38) there exists $\hat{k} \in \mathbb{N}$ such that

$$\left|\mathcal{Z}_{\widehat{M},\hat{k}}\right| \leq \varepsilon/2c, \tag{5.40}$$

where $c$ is defined in (5.34). With (5.35), (5.39) (5.40), for the particular $\widehat{M}$ and $\hat{k}$ it follows

$$\delta_5 \leq \left(\sum_{i=1}^{N-1} \int_{\mathcal{H}_{i,\widehat{M}}} p(\mathbf{y}_i) d\mathbf{y}_i\right) + c\left|\mathcal{Z}_{\widehat{M},\hat{k}}\right| \leq \varepsilon = (3 \times 10)^{-(N-1)}. \tag{5.41}$$

By combining the upper bound of $\delta_4$ and $\delta_5$, we can obtain

$$\Pr(\{\mathbf{y}_N \in \Phi_q\}) \leq \delta_4 + \delta_5 \leq \widehat{C}_q \prod_{j=1}^{N_1} erf\left(\gamma_{j,q} w_{s_j}^{(N)}\right) + (3 \times 10)^{-(N-1)}. \tag{5.42}$$

Now we are in a place to derive an upper bound for $\sum_{q \in \mathcal{J}_{k_t,k_f}} \Pr(\{\mathbf{y}_N \in \Phi_q\})$. Recall that the data points $\{\mathbf{y}_{t_1}, \cdots, \mathbf{y}_{t_{m_N}}\}$ are in the same cluster as $\mathbf{y}_N$ and the corresponding weighting coefficients are sorted in an increasing order $w_{t_1}^{(N)} \leq w_{t_2}^{(N)} \leq \cdots \leq w_{t_{m_N}}^{(N)}$; similarly, for those data points $\{\mathbf{y}_{f_1}, \cdots, \mathbf{y}_{f_{(N-1)-m_N}}\}$ from different clusters the corresponding weighting coefficients are likewise sorted according to $w_{f_1}^{(N)} \leq w_{f_2}^{(N)} \leq \cdots \leq w_{f_{(N-1)-m_N}}^{(N)}$. With $\gamma_{j,q}$ and $\mathcal{J}_{k_t,k_f}$ given in, respectively, (5.28) and (5.24), we define

$$\gamma_j = \max_{q \in \mathcal{J}_{k_t,k_f}} \{\gamma_{j,q}\}. \tag{5.43}$$

Note that, for $q \in \mathcal{J}_{k_t,k_f}$, $\Phi_q$ are associated with at least $|\mathcal{I}_{k_t,k_f}|$ ($\mathcal{I}_{k_t,k_f}$ is specified in (3.5)) inactive constraints, thereby $|\mathcal{I}_{k_t,k_f}| \leq |\mathcal{I}_{q,0}| = N_1$. We then have

$$\prod_{j=1}^{N_1} erf\left(\gamma_{j,q} w_{s_j}^{(N)}\right) \overset{(a)}{\leq} \prod_{j=1}^{N_1} erf\left(\gamma_j w_{s_j}^{(N)}\right) \overset{(b)}{\leq} \prod_{j \in \mathcal{I}_{k_t,k_f}} erf\left(\gamma_j w_j^{(N)}\right), \tag{5.44}$$

where (a) is true since $erf(\cdot)$ is increasing and using (5.43), and (b) holds since $|\mathcal{I}_{k_t,k_f}| \leq N_1$ and $\{w_j^{(N)} \mid j \in \mathcal{I}_{k_t,k_f}\}$ consists of the $m_N - (k_t - 1)$ largest weights of the correct neighbors and the $N - 1 - m_N - k_f$ largest weights of the incorrect neighbors. Hence, by setting $C = \sum_{q \in \mathcal{J}_{k_t,k_f}} \widehat{C}_q$, we can obtain the following desired upper bound for $\Pr(\{|\mathcal{T}| < k_t\} \cap \{|\mathcal{F}| \leq k_f\})$:

$$
\begin{aligned}
\Pr(\{|\mathcal{T}| < k_t\} \cap \{|\mathcal{F}| \leq k_f\}) &\stackrel{(a)}{=} \sum_{q \in \mathcal{J}_{k_t,k_f}} \Pr(\{\mathbf{y}_N \in \Phi_q\}) \\
&\stackrel{(b)}{\leq} \sum_{q \in \mathcal{J}_{k_t,k_f}} \left[ \widehat{C}_q \prod_{j=1}^{|\mathcal{I}_{q,0}|} \mathrm{erf}\left(\gamma_{j,q} w_{s_j}^{(N)}\right) + (3 \times 10)^{-(N-1)} \right] \\
&\stackrel{(c)}{\leq} \sum_{q \in \mathcal{J}_{k_t,k_f}} \left[ \widehat{C}_q \prod_{j \in \mathcal{I}_{k_t,k_f}} \mathrm{erf}\left(\gamma_j w_j^{(N)}\right) + (3 \times 10)^{-(N-1)} \right] \\
&= C \prod_{j \in \mathcal{I}_{k_t,k_f}} \mathrm{erf}\left(\gamma_j w_j^{(N)}\right) + |\mathcal{J}_{k_t,k_f}| (3 \times 10)^{-(N-1)} \\
&\stackrel{(d)}{\leq} C \prod_{j \in \mathcal{I}_{k_t,k_f}} \mathrm{erf}\left(\gamma_j w_j^{(N)}\right) + 3^{N-1} (3 \times 10)^{-(N-1)} \\
&= C \prod_{j \in \mathcal{I}_{k_t,k_f}} \mathrm{erf}\left(\gamma_j w_j^{(N)}\right) + 10^{-(N-1)},
\end{aligned}
\tag{5.45}
$$

where (a) holds by (5.23), (b) holds by (5.42), (c) follows from (5.44) and (d) holds since there are totally $3^{N-1}$ regions $\Phi_q$.

*Proof of Part (b):*

Our proof in this part basically follows the theme of [10]. Associated with the correct neighbors $\{\mathbf{y}_{t_1} \cdots \mathbf{y}_{t_{m_N}}\}$, i.e., data points that lie in the same cluster as $\mathbf{y}_N$, we define

$$
\mathbf{Y}_t \triangleq [\mathbf{y}_{t_1} \cdots \mathbf{y}_{t_{m_N}}] \in \mathbb{R}^{n \times t_{m_N}} \quad \text{and} \quad \mathbf{W}_t \triangleq \mathrm{diag}\left\{1/w_{t_1}^{(N)}, \cdots, 1/w_{t_{m_N}}^{(N)}\right\} \in \mathbb{R}^{t_{m_N} \times t_{m_N}}. \tag{5.46}
$$

The following three lemmas are needed in the proof.

*Lemma 5.2*: Let $\mathbf{a} \in \mathbb{R}^{n_1}$ and $\mathbf{v} \in \mathbb{R}^{n_2}$ be two vectors drawn uniformly at random from, respectively, the unit spheres of $\mathbb{R}^{n_1}$ and $\mathbb{R}^{n_2}$, and $\Sigma \in \mathbb{R}^{n_1 \times n_2}$ be a given deterministic matrix. For a given $w > 0$, the inequality

$$
|(1/w)\mathbf{a}^T \Sigma \mathbf{v}| \leq 8 \log N \frac{\|\Sigma\|_F}{\sqrt{n_1 n_2}}, \tag{5.47}
$$

holds with probability at least $1 - 4N^{-4w}$.

*[Proof]:* See Appendix E. $\square$

*Lemma 5.3*: Suppose the entries of $\mathbf{a} \in \mathbb{R}^n$ are i.i.d. Gaussian random variables with zero mean and unit variance, and let $\mathbf{z} \in \mathbb{R}^n$ be a deterministic unit-norm vector. For a given $w > 0$, the inequality

$$
|(1/w)\mathbf{a}^T \mathbf{z}| \leq 2\sqrt{2 \log N} \tag{5.48}
$$

holds with probability at least $1 - \exp\left(-\frac{(1-\log 2)((1/w)n - 8\log n)^2}{2n(1/w)^2}\right)$.

*[Proof]:* See Appendix F. $\square$

***Lemma 5.4***: Let $\bar{\mathbf{c}}_N$ be the optimal solution of the *reduced problem*, i.e.,

$$\bar{\mathbf{c}}_N = \arg\min \ \frac{1}{2}\|\mathbf{y}_N - \mathbf{Y}_t\mathbf{W}_t\mathbf{c}_N\|_2^2 + \lambda\|\mathbf{c}_N\|_1, \tag{5.49}$$

Let $\mathbf{y}_{N,\|}$ and $\mathbf{Y}_{t,\|}$ be the projections of, respectively, the data point $\mathbf{y}_N$ and data matrix $\mathbf{Y}_t$ onto the subspace that attached to $\mathbf{y}_N$. Also, let $\mathbf{y}_{N,\perp}$ and $\mathbf{Y}_{t,\perp}$ be the projection of $\mathbf{y}_N$ onto the orthogonal complement of subspace that attached to $\mathbf{y}_N$. If $\rho_N \geq \rho^*$ and $\lambda > \frac{\sigma}{\sqrt{8d_N}}$, the following two inequalities

$$\|\mathbf{y}_{N,\perp} - \mathbf{Y}_{t,\perp}\mathbf{W}_t\bar{\mathbf{c}}_N\|_2 \leq h_1\sigma \tag{5.50}$$

and

$$\|\mathbf{y}_{N,\|} - \mathbf{Y}_{t,\|}\mathbf{W}_t\bar{\mathbf{c}}_N\|_2 \leq h_2\lambda\sqrt{d_N}, \tag{5.51}$$

hold with a probability at least $1 - e^{-\alpha_0(n-d_N)\left(1-(w_{f_1}^{(N)})^2/2\right)^2} - e^{-\sqrt{m_N d_N}} - N^{-2} - N^{-3}$, where $\alpha_0$, $h_1$ and $h_2$ are numerical constants, $\rho_N$ and $d_N$ are the sample density and the dimension of the subspace attached to $\mathbf{y}_N$.

*[Proof]:* See Appendix G. □

Recall that, for the data points $\mathbf{y}_{f_1},\cdots,\mathbf{y}_{f_{(N-1)-m_N}}$ from different clusters, the corresponding weighting coefficients are sorted according to $w_{f_1}^{(N)} \leq w_{f_2}^{(N)} \leq \cdots \leq w_{f_{(N-1)-m_N}}^{(N)}$. We then define

$$\mathbf{Y}_f \triangleq [\mathbf{y}_{f_{k_f+1}},\cdots,\mathbf{y}_{f_{(N-1)-m_N}}] \in \mathbb{R}^{n\times[(N-1)-m_N-k_f]}, \tag{5.52}$$

which consists of $\mathbf{y}_{f_j}$'s corresponding to $(N-1)-m_N-k_f$ largest weighting coefficients, and

$$\mathbf{W}_f \triangleq diag\{1/w_{f_{k_f+1}}^{(N)},\cdots,1/w_{f_{(N-1)-m_N}}^{(N)}\} \in \mathbb{R}^{[(N-1)-m_N-k_f]\times[(N-1)-m_N-k_f]}. \tag{5.53}$$

Suppose there are at least $(N-1)-m_N-k_f$ false neighbors left "unrecovered", say, those with indexes $\{f_{j_1},f_{j_2},\cdots,f_{j_{N-1-m_N-k_f}}\}$ where $\{j_1,j_2,\cdots,j_{N-1-m_N-k_f}\} \subset \{1,\cdots,N-1-m_N\}$, so that the corresponding indexed constraints in the dual of weighted LASSO are inactive. As such, there must be at most $k_f$ false discoveries. For the particular case that $j_1 = k_f+1$, $j_2 = k_f+2,\cdots,j_{N-1-m_N-k_f} = N-1-m_N$, Lemma 8.6 in [10] asserts the corresponding indexed $N-1-m_N-k_f$ constraints are inactive if

$$\|\mathbf{W}_f\mathbf{Y}_f^T(\mathbf{y}_N - \mathbf{Y}_t\mathbf{W}_t\bar{\mathbf{c}}_N)\|_\infty < \lambda. \tag{5.54}$$

Therefore, a lower bound for $\Pr(\{|\mathcal{F}| \leq k_f\})$ can be obtained as

$$\Pr(\{|\mathcal{F}| \leq k_f\}) \geq \Pr(\{\|\mathbf{W}_f\mathbf{Y}_f^T(\mathbf{y}_N - \mathbf{Y}_t\mathbf{W}_t\bar{\mathbf{c}}_N)\|_\infty < \lambda\}). \tag{5.55}$$

To ease analysis, we use the data model (3.2) to write $\mathbf{Y}_f = \mathbf{X}_f + \mathbf{E}_f$, where $\mathbf{X}_f$ and $\mathbf{E}_f$ are, respectively, the noiseless data matrix and noise matrix, so that

$$\begin{aligned}
\left\|\mathbf{W}_f \mathbf{Y}_f^T (\mathbf{y}_N - \mathbf{Y}_t \mathbf{W}_t \bar{\mathbf{c}}_N)\right\|_\infty &\leq \left\|\mathbf{W}_f \mathbf{X}_f^T (\mathbf{y}_{N,\|} - \mathbf{Y}_{t,\|} \mathbf{W}_t \bar{\mathbf{c}}_N)\right\|_\infty + \left\|\mathbf{W}_f \mathbf{X}_f^T (\mathbf{y}_{N,\perp} - \mathbf{Y}_{t,\perp} \mathbf{W}_t \bar{\mathbf{c}}_N)\right\|_\infty \\
&+ \left\|\mathbf{W}_f \mathbf{E}_f^T (\mathbf{y}_{N,\|} - \mathbf{Y}_{t,\|} \mathbf{W}_t \bar{\mathbf{c}}_N)\right\|_\infty + \left\|\mathbf{W}_f \mathbf{E}_f^T (\mathbf{y}_{N,\perp} - \mathbf{Y}_{t,\perp} \mathbf{W}_t \bar{\mathbf{c}}_N)\right\|_\infty.
\end{aligned} \quad (5.56)$$

In what follows we leverage Lemmas 5.2, 5.3, and 5.4 to bound each infinite-norm term on the RHS of (5.56), and then prove (5.54) holds with a high probability; this in turn proves (3.6).

Assume that (5.50) and (5.51) in Lemma 5.4 hold and that, for $\mathbf{y}_N = \mathbf{x}_N + \mathbf{e}_N$, $\mathbf{x}_N$ belongs to the $i$th subspace $\mathcal{S}_i$; also, let $\mathbf{U}_i$ be a matrix whose columns form an orthonormal basis for $\mathcal{S}_i$. To find an upper bound for $\left\|\mathbf{W}_f \mathbf{X}_f^T (\mathbf{y}_{N,\|} - \mathbf{Y}_{t,\|} \mathbf{W}_t \bar{\mathbf{c}}_N)\right\|_\infty$, assume that $\mathbf{x}_j$ belongs to the $k$th subspace $\mathcal{S}_k$ whose orthonormal basis are columns of $\mathbf{U}_k$, $i \neq k$. We can directly deduce from Lemma 5.2 that the following inequality

$$\begin{aligned}
\left|(1/w_j^{(N)}) \mathbf{x}_j^T (\mathbf{y}_{N,\|} - \mathbf{Y}_{t,\|} \mathbf{W}_t \bar{\mathbf{c}}_N)\right| &\leq 8 \log N \frac{\|\mathbf{U}_i^T \mathbf{U}_k\|_F}{\sqrt{d_i d_k}} \|(\mathbf{y}_{N,\|} - \mathbf{Y}_{t,\|} \mathbf{W}_t \bar{\mathbf{c}}_N)\|_2 \\
&\stackrel{(a)}{\leq} 8 \log N \frac{aff(\mathcal{S}_i, \mathcal{S}_k)}{\sqrt{d_i}} \|(\mathbf{y}_{N,\|} - \mathbf{Y}_{t,\|} \mathbf{W}_t \bar{\mathbf{c}}_N)\|_2,
\end{aligned} \quad (5.57)$$

where (a) is obtained by definition of affinity $aff(\mathcal{S}_i, \mathcal{S}_k) = \|\mathbf{U}_i^T \mathbf{U}_k\|_F / \sqrt{\min(d_i, d_k)}$, holds with a probability at least $1 - 4N^{-4w_j^{(N)}}$. Since $\mathbf{y}_j$'s are independent, it follows

$$\left\|\mathbf{W}_f \mathbf{X}_f^T (\mathbf{y}_{N,\|} - \mathbf{Y}_{t,\|} \mathbf{W}_t \bar{\mathbf{c}}_N)\right\|_\infty \leq \frac{8 \log N}{\sqrt{d_i}} \max_{k \neq i} (aff(\mathcal{S}_i, \mathcal{S}_k)) \|(\mathbf{y}_{N,\|} - \mathbf{Y}_{t,\|} \mathbf{W}_t \bar{\mathbf{c}}_N)\|_2 \quad (5.58)$$

holds with probability at least $1 - 4N^{-4w_{f_{k_f}+1}^{(N)}+1}$. Based on (5.58) and (5.51), we obtain

$$\left\|\mathbf{W}_f \mathbf{X}_f^T (\mathbf{y}_{N,\|} - \mathbf{Y}_{t,\|} \mathbf{W}_t \bar{\mathbf{c}}_N)\right\|_\infty \leq \lambda \overline{C}_1 \log N aff(\mathcal{S}_i, \mathcal{S}_k) \triangleq \lambda I_1, \quad (5.59)$$

in which $\overline{C}_1$ is a numerical constant. By following similar arguments, the following inequality

$$\begin{aligned}
\left\|\mathbf{W}_f \mathbf{X}_f^T (\mathbf{y}_{N,\perp} - \mathbf{Y}_{t,\perp} \mathbf{W}_t \bar{\mathbf{c}}_N)\right\|_\infty &\stackrel{(a)}{\leq} \frac{8 \log N}{\sqrt{n - d_i}} \max_{k \neq i} (aff(\mathcal{S}_i, \mathcal{S}_k)) \|\mathbf{y}_{N,\perp} - \mathbf{Y}_{t,\perp} \mathbf{W}_t \bar{\mathbf{c}}_N\|_2 \\
&\stackrel{(b)}{\leq} \overline{C}_2 \log N \frac{\sigma}{\sqrt{n - d_i}} \triangleq I_2,
\end{aligned} \quad (5.60)$$

where $\overline{C}_2$ is a numerical constant, (a) follows from Lemma 5.2 and (b) is due to (5.50), holds with probability at least $1 - 4N^{-4w_{f_{k_f}+1}^{(N)}+1}$. Similarly, we have

$$\begin{aligned}
\left\|\mathbf{W}_f \mathbf{E}_f^T (\mathbf{y}_{N,\|} - \mathbf{Y}_{t,\|} \mathbf{W}_t \bar{\mathbf{c}}_N)\right\|_\infty &\stackrel{(a)}{\leq} 2\sigma \sqrt{\frac{2 \log N}{n}} \|\mathbf{y}_{N,\|} - \mathbf{Y}_{t,\|} \mathbf{W}_t \bar{\mathbf{c}}_N\|_2 \\
&\stackrel{(b)}{\leq} \lambda \overline{C}_3 \sigma \sqrt{\frac{d_i \log N}{n}} \triangleq \lambda I_3,
\end{aligned} \quad (5.61)$$

where $\overline{C}_3$ is a numerical constant, (a) holds by Lemma 5.3 and (b) follows from (5.51), holds with probability at least $1 - Ne^{-\alpha n}$ for some constant $\alpha$. Finally, it can be shown using similar arguments that

$$\left\|\mathbf{W}_f \mathbf{E}_f^T (\mathbf{y}_{N,\perp} - \mathbf{Y}_{t,\perp} \mathbf{W}_t \bar{\mathbf{c}}_N)\right\|_\infty \leq \sigma^2 \overline{C}_4 \sqrt{\frac{\log N}{n}} \triangleq I_4 \tag{5.62}$$

where $\overline{C}_4$ is a numerical constant, holds with probability at least $1 - Ne^{-\alpha n}$. Combining (5.59)~(5.62), we conclude that the following inequality

$$\left\|\mathbf{W}_f \mathbf{Y}_f^T (\mathbf{y}_N - \mathbf{Y}_t \mathbf{W}_t \bar{\mathbf{c}}_N)\right\|_\infty \leq \lambda (I_1 + I_3) + (I_2 + I_4) \tag{5.63}$$

holds with a probability exceeding $1 - e^{-\alpha_0 (n - d_N)\left(1 - (w_{t_1}^{(N)})^2/2\right)^2} - e^{-\sqrt{m_N d_N}} - N^{-2} - N^{-3} - 2Ne^{-\alpha_1 n} - 8N^{-4w_{f k_f + 1}^{(N)} + 1}$. It suffices to show $\lambda (I_1 + I_3) + (I_2 + I_4) < \lambda$. By setting $K = (1 - 1/\sqrt{3})/\max(\overline{C}_1, \overline{C}_3)$ in Assumption 3 made in Section III-A, it can be verified that $(I_1 + I_3) < (1 - 1/\sqrt{3})$. Hence, it remains to check if $\lambda > \sqrt{3}(I_2 + I_4)$ holds. This inequality is typically true since the ambient dimension $n$ is very large. Hence we can obtain

$$\begin{aligned}
\Pr(\{|\mathcal{F}| \leq k_f\}) &\geq \Pr\left(\left\{\left\|\mathbf{W}_f \mathbf{Y}_f^T (\mathbf{y}_N - \mathbf{Y}_t \mathbf{W}_t \bar{\mathbf{c}}_N)\right\|_\infty < \lambda\right\}\right) \\
&\geq 1 - e^{-\alpha_0 (n - d_N)\left(1 - (w_{t_1}^{(N)})^2/2\right)^2} - e^{-\sqrt{m_N d_N}} - N^{-2} - N^{-3} - 2Ne^{-\alpha_1 n} - 8N^{-4w_{f k_f + 1}^{(N)} + 1},
\end{aligned} \tag{5.64}$$

which proves (3.6).

*Proof of Part (c):*

Since

$$\begin{aligned}
\Pr(\{|\mathcal{T}| \geq k_t\} \cap \{|\mathcal{F}| \leq k_f\}) &= \Pr(\{|\mathcal{F}| \leq k_f\} \setminus \{\{|\mathcal{T}| \geq k_t\}^c \cap \{|\mathcal{F}| \leq k_f\}\}) \\
&= \Pr(\{|\mathcal{F}| \leq k_f\} \setminus \{\{|\mathcal{T}| < k_t\} \cap \{|\mathcal{F}| \leq k_f\}\}) \\
&= \Pr(\{|\mathcal{F}| \leq k_f\}) - \Pr(\{|\mathcal{T}| < k_t\} \cap \{|\mathcal{F}| \leq k_f\}),
\end{aligned} \tag{5.65}$$

The lower bound (3.7) follows directly from (3.4), (3.6) and (5.65). The proof of Theorem 3.2 is completed.

## VI. CONCLUSION

Design and analysis of sparsity-assisted neighbor identification schemes is a fundamental issue in the study of SSC. In this paper, we propose a new solution based on two-step reweighted $\ell_1$-minimization. Our approach generalizes the two-step algorithm proposed in [10] by employing data weighting in the second step to better exploit the side neighbor information conveyed by the computed coarse sparse representation vector in the first step. The major contribution of this paper is the development of a general analytic framework for provable neighbor recovery rate analysis. Our formulation is built on the dual problem of weighted LASSO and exploits the underlying geometry of "projection onto polyhedron". We establish an interesting connection between the identities of the recovered neighbors, determined based on the optimal sparse solution to the primal weighted LASSO, and the indexes of the active constraints of the dual problem. In our setting, identification of correct neighbors amounts to judging which constraints of the dual problem are "correctly activated". The proposed formulation enjoys the following two distinctive features. Firstly, it allows us to consider general neighbor recovery events, say, there exist at least $k_t > 0$ correct neighbors and at most $k_f \geq 0$ incorrect neighbors, without any restrictions on $k_t$ and $k_f$. Secondly, it allows us to obtain provable neighbor recovery rates specified by various analytic probability lower/upper bounds. Our analytic bounds confirm that the proposed approach can recover many correct neighbors and only few incorrect neighbors with a high probability; moreover, thanks to data weighting, it can improve the neighbor recovery rates, and therefore the global data clustering accuracy, as compared to [10] whenever the prior neighbor information acquired in the first step is accurate enough. Simulation results using both synthetic data and real human face data evidence our analytic study and the effectiveness of the proposed scheme. Above all, our paper is the first in the SSC literature that proposes two-step reweighted $\ell_1$-minimization for neighbor identification and, more importantly, offers a solid and rigorous analytic framework for developing provable recovery rate results. Meanwhile, our work in the context of SSC also complements the study of reweighted $\ell_1$-minimization in CS-based sparse signal reconstruction. We conclude this paper with some possible future works as stated in the following:

I. *Weighting Coefficient Design:* The optimal design of the weighting coefficients (the parameter $\varepsilon$ in (1.4)), particularly from an analytic perspective, is for sure an important future work. It would be interesting to formulate the prior neighbor information using a probabilistic-based model similar to existing works in the CS literature [38-39].

II. *Joint Design of Regularization Factor and Weighting Coefficients:* A more challenging task is to consider joint design/optimization of the regularization factor $\lambda$ and the weighting coefficients in the considered two-step $\ell_1$-minimization setup.

III. *Impacts of Data Ordering:* Our current algorithm conducts neighbor identification on a sample-by-sample basis, say, starting from $\mathbf{y}_1$ until $\mathbf{y}_N$, without considering any specific data ordering rule. Suppose the algorithm starts with $\mathbf{y}_1$, and the computed optimal sparse representation vector $\mathbf{c}_1^* = [c_{1,2}^* \cdots c_{1,N}^*]^T$ in the second step yields $5 = \arg\max_{2 \leq j \leq N} |c_{1,j}^*|$, meaning that $\mathbf{y}_5$ is most certain to be a neighbor of $\mathbf{y}_1$. It is reasonable to expect that, next to $\mathbf{y}_1$, the algorithm shall move on to identify the neighbors for $\mathbf{y}_5$ (rather than for $\mathbf{y}_2$); conceptually, knowledge about neighbors of $\mathbf{y}_1$ and $\mathbf{y}_5$ can be updated and fused to successively improve the quality of prior neighbor information. To what extent a data ordering rule of this kind, once implemented, can further impact the neighbor recovery performance, as well as the design and selection of the corresponding weighting coefficients, is an interesting topic (a.k.a., joint data ordering and weighting) deserving further investigation.

# APPENDIX

## A. Proof of Lemma 2.1

*Proof of Part (a):*

It is equivalent to show that $\left|(1/w_j^{(N)})\mathbf{y}_j^T\mathbf{z}^*(\mathbf{y}_N)\right| < \lambda$ only if $c_{N,j}^* = 0$. Recall the weighted LASSO can be equivalently formulated as

$$\text{Minimize } \lambda\|\mathbf{c}_N\|_1 + \frac{1}{2}\|\mathbf{r}\|_2^2 \text{ subject to } \mathbf{r} = \mathbf{Y}_{-N}\mathbf{W}_N^{-1}\mathbf{c}_N - \mathbf{y}_N. \tag{A.1}$$

The associated Lagrangian then reads

$$L(\mathbf{c}_N, \mathbf{r}, \mathbf{z}) = \lambda\|\mathbf{c}_N\|_1 + \frac{1}{2}\|\mathbf{r}\|_2^2 + \mathbf{z}^T(\mathbf{Y}_{-N}\mathbf{W}_N^{-1}\mathbf{c}_N - \mathbf{y}_N - \mathbf{r}), \tag{A.2}$$

where $\mathbf{z} \in \mathbb{R}^n$ is the Lagrange multiplier. Denote by $\partial f(x)$ the subdifferential of $f$ at $x$. Then the subdifferential of $L$ at $\mathbf{c}_N = \mathbf{c}_N^*$ obeys[12]

$$\begin{aligned}\mathbf{0} &\overset{(a)}{\in} \partial(\lambda\|\mathbf{c}_N^*\|_1) + \partial\left(\left(\mathbf{W}_N^{-1}\mathbf{Y}_{-N}^T\mathbf{z}^*(\mathbf{y}_N)\right)^T\mathbf{c}_N^*\right) \\ &\overset{(b)}{=} \partial(\lambda\|\mathbf{c}_N^*\|_1) + \{\mathbf{W}_N^{-1}\mathbf{Y}_{-N}^T\mathbf{z}^*(\mathbf{y}_N)\},\end{aligned} \tag{A.3}$$

where (a) is true because the subdifferential at the optimal solution contains the zero vector [40, Theorem 28.3], and (b) holds since the subdifferential reduces to the differential whenever the function is differentiable [40, Theorem 25.1]. The condition (A.3) guarantees the existence of an $\mathbf{s} = [s_1 \cdots s_{N-1}]^T \in \partial(\|\mathbf{c}_N^*\|_1)$, with

$$s_j = \begin{cases} +1, & \text{if } c_{N,j}^* > 0 \\ -1, & \text{if } c_{N,j}^* < 0 \\ [-1,1], & \text{if } c_{N,j}^* = 0, \end{cases} \tag{A.4}$$

such that $\lambda\mathbf{s} + \mathbf{W}_N^{-1}\mathbf{Y}_{-N}^T\mathbf{z}^*(\mathbf{y}_N) = \mathbf{0}$, or equivalently,

$$\lambda s_j + (1/w_j^{(N)})\mathbf{y}_j^T\mathbf{z}^*(\mathbf{y}_N) = 0 \text{ for all } 1 \leq j \leq N-1. \tag{A.5}$$

If the $j$th constraint is inactive, meaning that $(1/w_j^{(N)})|\mathbf{y}_j^T\mathbf{z}^*(\mathbf{y}_N)| < \lambda$, (A.5) implies $|s_j| < 1$, which occurs only when $c_{N,j}^* = 0$ thanks to (A.4). The proof of part (a) is thus completed. □

*Proof of Part (b):*

---

12. The addition of two sets $\mathcal{A}$ and $\mathcal{B}$ of the Euclidean space is defined as $\mathcal{A} + \mathcal{B} \triangleq \{a + b \mid a \in \mathcal{A}, b \in \mathcal{B}\}$. If $\mathcal{A}$ and $\mathcal{B}$ are subspaces, $\mathcal{A} \oplus \mathcal{B}$ denotes the sum of $\mathcal{A}$ and $\mathcal{B}$.

Based on (A.2), the solution to the dual problem of (A.1) can be expressed as

$$\hat{\mathbf{z}}(\mathbf{y}_N) = \arg\max_{\mathbf{z}} \left[ \inf_{\mathbf{c}_N, \mathbf{r}} L(\mathbf{c}_N, \mathbf{r}, \mathbf{z}) \right] = \arg\max_{\mathbf{z}} \left[ \inf_{\mathbf{c}_N} \left( \lambda \|\mathbf{c}_N\|_1 + \mathbf{z}^T \mathbf{Y}_{-N} \mathbf{W}_N^{-1} \mathbf{c}_N \right) + \inf_{\mathbf{r}} \left( \frac{1}{2} \|\mathbf{r}\|_2^2 - \mathbf{z}^T \mathbf{r} - \mathbf{z}^T \mathbf{y}_N \right) \right]. \quad (A.6)$$

By some manipulations, we have

$$\inf_{\mathbf{r}} \left( \frac{1}{2} \|\mathbf{r}\|_2^2 - \mathbf{z}^T \mathbf{r} \right) = -\frac{1}{2} \mathbf{z}^T \mathbf{z} \quad (A.7)$$

and

$$\begin{aligned}
&\inf_{\mathbf{c}_N} \left( \lambda \|\mathbf{c}_N\|_1 + \mathbf{z}^T \mathbf{Y}_{-N} \mathbf{W}_N^{-1} \mathbf{c}_N \right) \\
&= -\sup_{\mathbf{c}_N} \left( -\lambda \|\mathbf{c}_N\|_1 - \mathbf{z}^T \mathbf{Y}_{-N} \mathbf{W}_N^{-1} \mathbf{c}_N \right) \\
&= -\lambda \sup_{\mathbf{c}_N} \left( -\left((1/\lambda) \mathbf{W}_N^{-1} \mathbf{Y}_{-N}^T \mathbf{z}\right)^T \mathbf{c}_N - \|\mathbf{c}_N\|_1 \right) \\
&\stackrel{(a)}{=} \begin{cases} 0, & \|\mathbf{W}_N^{-1} \mathbf{Y}_{-N}^T \mathbf{z}\|_\infty \leq \lambda \\ -\infty, & \text{otherwise} \end{cases},
\end{aligned} \quad (A.8)$$

where (a) holds by the definition of conjugate function [41, p.93] and the dual norm of $\|\cdot\|_1$ is $\|\cdot\|_\infty$. Thus, (A.6) then reads

$$\begin{aligned}
\hat{\mathbf{z}}(\mathbf{y}_N) &= \arg\max_{\mathbf{z}} \ -\frac{1}{2}\mathbf{z}^T\mathbf{z} - \mathbf{z}^T \mathbf{y}_N \quad \text{s.t.} \quad \|\mathbf{W}_N^{-1} \mathbf{Y}_{-N}^T \mathbf{z}\|_\infty \leq \lambda \\
&= \arg\min_{\mathbf{z}} \ \|\mathbf{z} + \mathbf{y}_N\|_2^2 \quad \text{s.t.} \quad \|\mathbf{W}_N^{-1} \mathbf{Y}_{-N}^T \mathbf{z}\|_\infty \leq \lambda.
\end{aligned} \quad (A.9)$$

With (2.1) and (A.9), it can be directly seen that $\mathbf{z}^*(\mathbf{y}_N) = -\hat{\mathbf{z}}(\mathbf{y}_N)$. Besides, with the optimal $\hat{\mathbf{z}}(\mathbf{y}_N)$, the corresponding optimal $\mathbf{r}^*$ is immediately obtained as (cf. (A.6))

$$\mathbf{r}^* = \arg\inf_{\mathbf{r}} \left( \frac{1}{2}\|\mathbf{r}\|_2^2 - \hat{\mathbf{z}}(\mathbf{y}_N)^T \mathbf{r} - \hat{\mathbf{z}}(\mathbf{y}_N)^T \mathbf{y}_N \right) = \hat{\mathbf{z}}(\mathbf{y}_N) = -\mathbf{z}^*(\mathbf{y}_N). \quad (A.10)$$

Since the optimal solution $\mathbf{c}_N^*$ to (A.1) satisfies $\mathbf{r}^* = \mathbf{Y}_{-N} \mathbf{W}_N^{-1} \mathbf{c}_N^* - \mathbf{y}_N$, using (A.10) we can obtain

$$\begin{aligned}
\mathbf{y}_N &= -\mathbf{r}^* + \mathbf{Y}_{-N} \mathbf{W}_N^{-1} \mathbf{c}_N^* \\
&= \mathbf{z}^*(\mathbf{y}_N) + \mathbf{Y}_{-N} \mathbf{W}_N^{-1} \mathbf{c}_N^*.
\end{aligned} \quad (A.11)$$

Suppose that there are totally $k$ active constraints indexed by $\{j_1, \cdots, j_k\} \subset \{1, \cdots, N-1\}$. Without loss of generality, we assume that

$$(1/w_{j_l}^{(N)}) \mathbf{y}_{j_l}^T \mathbf{z}^*(\mathbf{y}_N) = \lambda, \ 1 \leq l \leq k. \quad (A.12)$$

According to part (a), we have $c_{N,j}^* = 0$, for all $j \in \{1, \cdots, N-1\} \setminus \{j_1, \cdots, j_k\}$. Hence, it can be deduced from (A.11) that

$$\mathbf{y}_N = \mathbf{z}^*(\mathbf{y}_N) + \sum_{l=1}^{k} (1/w_{j_l}^{(N)}) c_{N,j_l}^* \mathbf{y}_{j_l} \quad (A.13)$$

By (A.12), $\mathbf{z}^*(\mathbf{y}_N)$ is contained in the intersection of $k$ hyperplanes, i.e.,

$$\mathbf{z}^*(\mathbf{y}_N) \in \bigcap_{l=1}^{k} \{\mathbf{z} \mid (1/w_{j_l}^{(N)})\mathbf{y}_{j_l}^T \mathbf{z} = \lambda\}, \tag{A.14}$$

which is a translation of the subspace $\bigcap_{l=1}^{k}\{\mathbf{z} \mid (1/w_{j_l}^{(N)})\mathbf{y}_{j_l}^T\mathbf{z} = 0\}$ of dimension $n-k$. In case $c_{N,j_m}^* = 0$ for some $j_m \in \{j_1, \cdots, j_k\}$, it then follows from (A.13) and (A.14) that

$$\mathbf{y}_N \in \underbrace{\bigcap_{l=1}^{k}\{\mathbf{z} \mid (1/w_{j_l}^{(N)})\mathbf{y}_{j_l}^T\mathbf{z} = \lambda\} + Span\{\mathbf{y}_{j_l}\}_{l \neq m}}_{\triangleq \mathcal{H}}. \tag{A.15}$$

Note that the set $\mathcal{H}$ in (A.15) is a translation of the subspace $\bigcap_{l=1}^{k}\{\mathbf{z} \mid (1/w_{j_l}^{(N)})\mathbf{y}_{j_l}^T\mathbf{z} = 0\} \oplus Span\{\mathbf{y}_{j_l}\}_{l \neq m}$, which is of dimension $n-1$. Since a subspace (of $\mathbb{R}^n$) with dimension less than $n$ is measure zero [35] and the measure of a set is invariant under translation [35], $\mathcal{H}$ is of measure zero. The proof is completed. $\square$

*B. Proof of Lemma 2.2*

The following two lemmas are needed to prove Lemma 2.2; to ease reading, their proofs are relegated to the end of this appendix.

*Lemma B.1:* For the identity weighting matrix, i.e., $\mathbf{W}_N = \mathbf{I}$, the feasible set of (2.1) can be expressed as

$$\{\mathbf{z} : \|\mathbf{Y}_{-N}^T \mathbf{z}\|_\infty \leq \lambda\} = \left\{\sum_{j=1}^{N-1} d_j \overline{\mathbf{y}}_j^{(N)} + \sum_{k=1}^{n-N+1} h_k \mathbf{n}_k, \ |d_j| \leq 1, \ h_k \in \mathbb{R}\right\}. \tag{B.1}$$

$\square$

*Lemma B.2:* Assume that $\mathbf{y}_N \in \Phi_q$. Then we have

$$\mathbf{z}^*(\mathbf{y}_N) = \sum_{j \in \mathcal{I}_{q,0}} a_j \overline{\mathbf{y}}_j^{(N)} + \sum_{j \in \mathcal{I}_{q,+}} \overline{\mathbf{y}}_j^{(N)} - \sum_{i \in \mathcal{I}_{q,-}} \overline{\mathbf{y}}_j^{(N)} + \sum_{l=1}^{n-N+1} h_k \mathbf{n}_k, \tag{B.2}$$

where $|a_j| < 1$ and $h_k \in \mathbb{R}$, and

$$\mathbf{y}_N = \mathbf{z}^*(\mathbf{y}_N) + \sum_{j \in \mathcal{I}_{q,+}} b_j \mathbf{y}_j - \sum_{j \in \mathcal{I}_{q,-}} b_j \mathbf{y}_j, \tag{B.3}$$

where $b_j \geq 0$. $\square$

Based on Lemmas B.1 and B.2, we start to prove Lemma 2.2. Pick $\mathbf{y}_N \in \Phi_q$, and we first claim that $\mathbf{y}_N$ belongs to the set specified on the RHS of (2.9). Using (B.2) and (B.3) we can obtain

$$\mathbf{y}_N = \sum_{j \in \mathcal{I}_{q,0}} a_j \overline{\mathbf{y}}_j^{(N)} + \sum_{j \in \mathcal{I}_{q,+}} (\overline{\mathbf{y}}_j^{(N)} + b_j \mathbf{y}_j) - \sum_{j \in \mathcal{I}_{q,-}} (\overline{\mathbf{y}}_j^{(N)} + b_j \mathbf{y}_j) + \sum_{k=1}^{n-N-1} h_k \mathbf{n}_k, \tag{B.4}$$

where $|a_j| < 1$, $b_j \geq 0$, and $h_k \in \mathbb{R}$. Hence, $\mathbf{y}_N$ belongs to the set on the RHS of (2.9). Now, we assume that $\mathbf{y}_N$ admits the following form

$$\mathbf{y}_N = \sum_{j \in \mathcal{I}_{q,0}} a_j \overline{\mathbf{y}}_j^{(N)} + \sum_{j \in \mathcal{I}_{q,+}} (\overline{\mathbf{y}}_j^{(N)} + b_j \mathbf{y}_j) - \sum_{j \in \mathcal{I}_{q,-}} (\overline{\mathbf{y}}_j^{(N)} + b_j \mathbf{y}_j) + \sum_{k=1}^{n-N-1} h_k \mathbf{n}_k, \tag{B.5}$$

where $|a_j| < 1$, $b_j \geq 0$ and $h_k \in \mathbb{R}$, and our purpose is to show $\mathbf{y}_N \in \Phi_q$. For this we denote by $f \triangleq \|\mathbf{z} - \mathbf{y}_N\|_2^2$, the objective function of (2.1), and

$$\mathbf{dom} f \triangleq \{\mathbf{z} : \|\mathbf{Y}_{-N}^T \mathbf{z}\|_\infty \leq \lambda\}. \tag{B.6}$$

Choose

$$\mathbf{x} = \sum_{j \in \mathcal{I}_{q,0}} a_j \overline{\mathbf{y}}_j^{(N)} + \sum_{j \in \mathcal{I}_{q,+}} \overline{\mathbf{y}}_j^{(N)} - \sum_{j \in \mathcal{I}_{q,-}} \overline{\mathbf{y}}_j^{(N)} + \sum_{k=1}^{n-N-1} h_k \mathbf{n}_k. \tag{B.7}$$

Since the objective function $f$ is differentiable in $\mathbf{z}$, $\mathbf{x}$ is the optimal solution to (2.1) if and only if $\mathbf{x} \in \mathbf{dom} f$ and $\nabla_{\mathbf{z}} f(\mathbf{x})^T (\mathbf{z} - \mathbf{x}) \geq 0$ for all $\mathbf{z} \in \mathbf{dom} f$ [41]. With (B.1), it is obvious that $\mathbf{x} \in \mathbf{dom} f$. Using (B.1), (B.5), and (B.7), we obtain

$$\nabla_{\mathbf{z}} f(\mathbf{x})^T (\mathbf{z} - \mathbf{x})$$
$$= 2(\mathbf{x} - \mathbf{y}_N)^T (\mathbf{z} - \mathbf{x})$$
$$= 2 \left( \underbrace{-\sum_{j \in \mathcal{I}_{q,+}} b_j \mathbf{y}_j + \sum_{j \in \mathcal{I}_{q,-}} b_j \mathbf{y}_j}_{\text{by (B.5) and (B.7)}} \right)^T \left( \underbrace{\sum_{j=1}^{N-1} d_j \overline{\mathbf{y}}_j^{(N)}}_{\text{(B.1)}} - \underbrace{\sum_{j \in \mathcal{I}_{q,0}} a_j \overline{\mathbf{y}}_j^{(N)} - \sum_{j \in \mathcal{I}_{q,+}} \overline{\mathbf{y}}_j^{(N)} + \sum_{i \in \mathcal{I}_{q,-}} \overline{\mathbf{y}}_j^{(N)} + \sum_{k=1}^{n-N+1} h_k \mathbf{n}_k}_{\text{by (B.7)}} \right)$$
$$= 2 \left( -\sum_{j \in \mathcal{I}_{q,+}} b_j \mathbf{y}_j + \sum_{j \in \mathcal{I}_{q,-}} b_j \mathbf{y}_j \right)^T \left( -\sum_{j \in \mathcal{I}_{q,+}} (1 - d_j) \overline{\mathbf{y}}_j^{(N)} + \sum_{j \in \mathcal{I}_{q,-}} (1 + d_j) \overline{\mathbf{y}}_j^{(N)} - \sum_{j \in \mathcal{I}_{q,0}} (c_j - d_j) \overline{\mathbf{y}}_j^{(N)} + \sum_{k=1}^{n-N+1} h_k \mathbf{n}_k \right)$$
$$= 2 \left( \sum_{j \in \mathcal{I}_{q,+}} b_j (1 - d_j) \mathbf{y}_j^T \overline{\mathbf{y}}_j^{(N)} + \sum_{j \in \mathcal{I}_{q,-}} b_j (1 + d_j) \mathbf{y}_j^T \overline{\mathbf{y}}_j^{(N)} \right) \overset{(a)}{\geq} 0,$$

$$\tag{B.8}$$

where (a) holds since $b_j \geq 0$, $|d_j| \leq 1$ and using (2.8). Therefore, $\mathbf{x}$ is the optimal solution to (2.1), i.e., $\mathbf{x} = \mathbf{z}^*(\mathbf{y}_N)$. Based on (2.8) and (B.7) it is straightforward to see

$$\mathbf{y}_j^T \mathbf{z}^*(\mathbf{y}_N) = \lambda, \ \forall j \in \mathcal{I}_{q,+} \tag{B.9}$$

$$|\mathbf{y}_j^T \mathbf{z}^*(\mathbf{y}_N)| < \lambda, \ \forall j \in \mathcal{I}_{q,0} \tag{B.10}$$

$$\mathbf{y}_j^T \mathbf{z}^*(\mathbf{y}_N) = -\lambda, \ \forall j \in \mathcal{I}_{q,-}. \tag{B.11}$$

Hence, $\mathbf{y}_N \in \Phi_q$. The proof of Lemma 2.2 is thus completed. $\square$

*[Proof of Lemma B.1]:* Suppose that $\mathbf{z} = \sum_{j=1}^{N-1} d_j \overline{\mathbf{y}}_j^{(N)} + \sum_{k=1}^{n-N+1} h_k \mathbf{n}_k$ and there exists $d_j$ such that $|d_j| > 1$, then it follows from (2.8) that $|\mathbf{y}_j^T \mathbf{z}| > \lambda$, therefore $\mathbf{z} \notin \{\mathbf{z} : \|\mathbf{Y}_{-N}^T \mathbf{z}\|_\infty \leq \lambda\}$. On the other hand, assume $\mathbf{z} = \sum_{j=1}^{N-1} d_j \overline{\mathbf{y}}_j^{(N)} + \sum_{k=1}^{n-N+1} h_k \mathbf{n}_k$ with $|d_j| \leq 1$ and $h_k \in \mathbb{R}$. It is straightforward to see $|\mathbf{y}_j^T \mathbf{z}| \leq \lambda$ for all $1 \leq j \leq N - 1$, thereby $\mathbf{z} \in \{\mathbf{z} : \|\mathbf{Y}_{-N}^T \mathbf{z}\|_\infty \leq \lambda\}$. $\square$

*[Proof of Lemma B.2]:* Suppose $\mathbf{y}_N \in \Phi_q$. Note that

$$\mathbb{R}^n = Span\{\mathbf{y}_1, \cdots, \mathbf{y}_{N-1}, \mathbf{n}_1, \cdots, \mathbf{n}_{n-N+1}\}$$
$$\overset{(a)}{=} Span\{\overline{\mathbf{y}}_1^{(N)}, \cdots, \overline{\mathbf{y}}_{N-1}^{(N)}, \mathbf{n}_1, \cdots, \mathbf{n}_{n-N+1}\} \tag{B.12}$$

where (a) holds due to (2.8). To prove (B.2), using (B.12) we write

$$\mathbf{z}^*(\mathbf{y}_N) = \sum_{j \in \mathcal{I}_{q,0}} a_j \overline{\mathbf{y}}_j^{(N)} + \sum_{j \in \mathcal{I}_{q,+}} b_j \overline{\mathbf{y}}_j^{(N)} - \sum_{j \in \mathcal{I}_{q,-}} b_j \overline{\mathbf{y}}_j^{(N)} + \sum_{k=1}^{n-N+1} h_k \mathbf{n}_k. \tag{B.13}$$

If there exists $j \in \mathcal{I}_{q,0}$ such that $|a_j| \geq 1$, it can be checked that $|\mathbf{y}_j^T \mathbf{z}^*(\mathbf{y}_N)| \geq \lambda$, which implies $\mathbf{y}_N \notin \Phi_q$. Therefore, $|a_j| < 1$ for all $j \in \mathcal{I}_{q,0}$. Similarly, If there exists $j \in \mathcal{I}_{q,+} \cup \mathcal{I}_{q,-}$ such that $b_j \neq 1$, it can be readily deduced using (B.9)~(B.11) that $\mathbf{y}_N \notin \Phi_q$. This then proves (B.2).

To prove (B.3), we use (B.12) to write

$$\mathbf{y}_N = \mathbf{z}^*(\mathbf{y}_N) + \sum_{j \in \mathcal{I}_{q,0}} a_j \mathbf{y}_j + \sum_{j \in \mathcal{I}_{q,+}} b_j \mathbf{y}_j - \sum_{j \in \mathcal{I}_{q,-}} b_j \mathbf{y}_j + \sum_{k=1}^{n-N+1} h_k \mathbf{n}_k. \tag{B.14}$$

It then suffices to show $a_j = 0$, $b_j \geq 0$ $h_k = 0$ in (B.14). Towards this end, we recall that $\mathbf{z}^*(\mathbf{y}_N)$ is the optimal solution if and only if [41]

$$\mathbf{z}^*(\mathbf{y}_N) \in \mathbf{dom} f \text{ and } \nabla_{\mathbf{z}} f(\mathbf{z}^*(\mathbf{y}_N))^T (\mathbf{z} - \mathbf{z}^*(\mathbf{y}_N)) \geq 0 \text{ for all } \mathbf{z} \in \mathbf{dom} f. \tag{B.15}$$

We will first show, if there exists some $a_j \neq 0$, the condition (B.15) is violated; hence we must have $a_j = 0$ for all $j \in \mathcal{I}_{q,0}$. For this observe that

$$\nabla_{\mathbf{z}} f(\mathbf{z}^*(\mathbf{y}_N))^T (\mathbf{z} - \mathbf{z}^*(\mathbf{y}_N))$$
$$= 2(\mathbf{z}^*(\mathbf{y}_N) - \mathbf{y}_N)^T (\mathbf{z} - \mathbf{z}^*(\mathbf{y}_N))$$
$$= 2 \underbrace{\left( -\sum_{j \in \mathcal{I}_{q,0}} a_j \mathbf{y}_j - \sum_{j \in \mathcal{I}_{q,+}} b_j \mathbf{y}_j + \sum_{j \in \mathcal{I}_{q,-}} b_j \mathbf{y}_j - \sum_{k=1}^{n-N+1} h_k \mathbf{n}_k \right)^T}_{\text{by (B.14)}}$$
$$\underbrace{\left( \sum_{j=1}^{N-1} d_j \overline{\mathbf{y}}_j^{(N)} - \sum_{j \in \mathcal{I}_{q,0}} c_j \overline{\mathbf{y}}_j^{(N)} - \sum_{j \in \mathcal{I}_{q,+}} \overline{\mathbf{y}}_j^{(N)} + \sum_{i \in \mathcal{I}_{q,-}} \overline{\mathbf{y}}_j^{(N)} - \sum_{k=1}^{n-N+1} \hat{h}_k \mathbf{n}_k \right)}_{\text{by (B.1) and (B.2)}}$$
$$= 2 \left( -\sum_{j \in \mathcal{I}_{q,0}} a_j \mathbf{y}_j - \sum_{j \in \mathcal{I}_{q,+}} b_j \mathbf{y}_j + \sum_{j \in \mathcal{I}_{q,-}} b_j \mathbf{y}_j - \sum_{k=1}^{n-N+1} h_k \mathbf{n}_k \right)^T$$
$$\left( -\sum_{j \in \mathcal{I}_{q,0}} (c_j - d_j) \overline{\mathbf{y}}_j^{(N)} - \sum_{j \in \mathcal{I}_{q,+}} (1 - d_j) \overline{\mathbf{y}}_j^{(N)} + \sum_{j \in \mathcal{I}_{q,-}} (1 + d_j) \overline{\mathbf{y}}_j^{(N)} - \sum_{k=1}^{n-N+1} \hat{h}_k \mathbf{n}_k \right)$$
$$= 2 \left( \sum_{j \in \mathcal{I}_{q,0}} a_j (c_j - d_j) \mathbf{y}_j^T \overline{\mathbf{y}}_j^{(N)} + \sum_{j \in \mathcal{I}_{q,+}} b_j (1 - d_j) \mathbf{y}_j^T \overline{\mathbf{y}}_j^{(N)} + \sum_{j \in \mathcal{I}_{q,-}} b_j (1 + d_j) \mathbf{y}_j^T \overline{\mathbf{y}}_j^{(N)} + \sum_{k=1}^{n-N+1} h_k \hat{h}_k \right),$$
(B.16)

where $|d_j| \leq 1$ and $\hat{h}_k \in \mathbb{R}$ are arbitrary, and $|c_j| < 1$ is fixed. Assume there exists $j_1 \in \mathcal{I}_{q,0}$ with $a_{j_1} \neq 0$. We choose $d_{j_1} \neq c_{j_1}$ with which $a_{j_1}(c_{j_1} - d_{j_1}) < 0$, $d_j = c_j$ for $j \in \mathcal{I}_{q,0} \setminus \{j_1\}$, $d_j = 1$ for $j \in \mathcal{I}_{q,+}$, $d_j = -1$ for $j \in \mathcal{I}_{q,-}$ and $\hat{h}_k = 0$ for $1 \leq k \leq n-N+1$. Then, with the particular $\tilde{\mathbf{z}} = d_{j_1} \overline{\mathbf{y}}_{j_1}^{(N)} + \sum_{j \in \mathcal{I}_{q,0} \setminus \{j_1\}} c_j \overline{\mathbf{y}}_j^{(N)} + \sum_{j \in \mathcal{I}_{q,+}} \overline{\mathbf{y}}_j^{(N)} - \sum_{j \in \mathcal{I}_{q,-}} \overline{\mathbf{y}}_j^{(N)} \in \mathbf{dom} f$, it can be verified using (B.16) that $\nabla_{\mathbf{z}} f(\mathbf{z}^*(\mathbf{y}_N))^T (\tilde{\mathbf{z}} - \mathbf{z}^*(\mathbf{y}_N)) < 0$, contradicting with (B.15). Similarly, if there exists $j_2 \in \mathcal{I}_{q,+} \cup \mathcal{I}_{q,-}$, say $j_2 \in \mathcal{I}_{q,+}$ such that $b_{j_2} < 0$, choosing $\overline{\mathbf{z}} = \sum_{j \in \mathcal{I}_{q,0}} c_j \overline{\mathbf{y}}_j^{(N)} + \sum_{j \in \mathcal{I}_{q,+}\setminus\{j_2\}} \overline{\mathbf{y}}_j^{(N)} + \frac{1}{2}\overline{\mathbf{y}}_{j_2}^{(N)} - \sum_{j \in \mathcal{I}_{q,-}} \overline{\mathbf{y}}_j^{(N)} \in \mathbf{dom} f$ yields $\nabla_{\mathbf{z}} f(\mathbf{z}^*(\mathbf{y}_N))^T (\overline{\mathbf{z}} - \mathbf{z}^*(\mathbf{y}_N)) < 0$, contradicting with (B.15). Hence, we must have $b_j \geq 0$. Again, by using the same procedures, we can show that $h_k = 0$, for $1 \leq k \leq n - N + 1$. The assertion (B.3) thus follows.

To prove Corollary 2.3, using Lemma 2.2 we have

$$\Phi_q + \mathbf{w} = \left\{ \sum_{j \in \mathcal{I}_{q,0}} a_j \overline{\mathbf{y}}_j^{(N)} + \sum_{j \in \mathcal{I}_{q,+}} \left(\overline{\mathbf{y}}_j^{(N)} + b_j \mathbf{y}_j\right) - \sum_{j \in \mathcal{I}_{q,-}} \left(\overline{\mathbf{y}}_j^{(N)} + b_j \mathbf{y}_j\right) + \sum_{k=1}^{n-N-1} h_k \mathbf{n}_k + \mathbf{w},\ |a_j| < 1,\ b_j \geq 0,\ h_k \in \mathbb{R} \right\}. \tag{B.17}$$

Note that

$$\begin{aligned}
\mathbb{R}^n &= Span\{\mathbf{y}_1, \cdots, \mathbf{y}_{N-1}, \mathbf{n}_1, \cdots, \mathbf{n}_{n-N+1}\} \\
&= Span\{\mathbf{y}_j\}_{j \in \mathcal{I}_{q,+}} \oplus Span\{\mathbf{y}_j\}_{j \in \mathcal{I}_{q,-}} \oplus Span\{\mathbf{y}_j\}_{j \in \mathcal{I}_{q,0}} \oplus \{\mathbf{n}_1, \cdots, \mathbf{n}_{n-N+1}\} \\
&\stackrel{(a)}{=} Span\{\mathbf{y}_j\}_{j \in \mathcal{I}_{q,+}} \oplus Span\{\mathbf{y}_j\}_{j \in \mathcal{I}_{q,-}} \oplus Span\{\overline{\mathbf{y}}_j^{(N)}\}_{j \in \mathcal{I}_{q,0}} \oplus \{\mathbf{n}_1, \cdots, \mathbf{n}_{n-N+1}\},
\end{aligned} \tag{B.18}$$

where (a) holds due to (2.8). With (B.18) we can write

$$\sum_{j \in \mathcal{I}_{q,+}} \overline{\mathbf{y}}_j^{(N)} - \sum_{j \in \mathcal{I}_{q,-}} \overline{\mathbf{y}}_j^{(N)} + \mathbf{w} = \sum_{j \in \mathcal{I}_{q,0}} \alpha_j \overline{\mathbf{y}}_j^{(N)} + \sum_{j \in \mathcal{I}_{q,+}} \beta_j \mathbf{y}_j - \sum_{j \in \mathcal{I}_{q,-}} \beta_j \mathbf{y}_j + \sum_{k=1}^{n-N+1} \gamma_k \mathbf{n}_k, \tag{B.19}$$

which together with (B.17) implies

$$\begin{aligned}
\Phi_q + \mathbf{w} &= \left\{ \sum_{j \in \mathcal{I}_{q,0}} (a_j + \alpha_j) \overline{\mathbf{y}}_j^{(N)} + \sum_{j \in \mathcal{I}_{q,+}} (b_j + \beta_j) \mathbf{y}_j - \sum_{j \in \mathcal{I}_{q,-}} (b_j + \beta_j) \mathbf{y}_j + \sum_{k=1}^{n-N+1} (h_k + \gamma_k) \mathbf{n}_k,\ |a_j| < 1,\ b_j \geq 0,\ h_k \in \mathbb{R} \right\} \\
&= \left\{ \sum_{j \in \mathcal{I}_{q,0}} \hat{a}_j \overline{\mathbf{y}}_j^{(N)} + \sum_{j \in \mathcal{I}_{q,+}} \hat{b}_j \mathbf{y}_j - \sum_{j \in \mathcal{I}_{q,-}} \hat{b}_j \mathbf{y}_j + \sum_{k=1}^{n-N+1} \hat{h}_k \mathbf{n}_k,\ -1 + \alpha_j < \hat{a}_j < 1 + \alpha_j,\ \hat{b}_j \geq \beta_j,\ \hat{h}_k \in \mathbb{R} \right\}
\end{aligned} \tag{B.20}$$

Therefore, the proof of Corollary 2.3 is completed. □

*C. Derivations of (5.16) and (5.17)*

The following lemma (whose proof is placed at the end of this appendix) is needed for deriving (5.16) and (5.17).

**Lemma C.1:** For $a > 0$, the following two inequalities hold.

(1). $\int_{-1+\gamma}^{1+\gamma} \exp\left(-(ax^2 + bx)\right) dx \leq \sqrt{\pi/a} \exp\left(b^2/4a\right) erf(\sqrt{a})$. (C.1)

(2) $\int_t^\infty \exp\left(-(ax^2 + bx)\right) dx \leq \sqrt{\pi/a}$. (C.2)

□

*Part I: Upper Bound for $\delta_2$ in (5.16):*

The proof is done by repeatedly using part (1) of Lemma C.1 together with induction to deduce the claimed result in (5.16). By definition of $\delta_2$ in (5.13), we first conduct integration with respect to (w.r.t.) the variable $a_{s_1}$ to obtain

$$\delta_2 = \int_{-1+\alpha_{s_1}}^{1+\alpha_{s_1}} \cdots \int_{-1+\alpha_{sN_1}}^{1+\alpha_{sN_1}} \exp\left(\frac{-n}{2\sigma^2}\left\|\sum_{j=1}^{N_1} a_{s_j}\overline{\mathbf{y}}_{s_j}^{(N)}\right\|_2^2\right) da_{sN_1} \cdots da_{s_1}$$

$$= \int_{-1+\alpha_{s_1}}^{1+\alpha_{s_1}} \cdots \int_{-1+\alpha_{sN_1}}^{1+\alpha_{sN_1}} \exp\left(\frac{-n}{2\sigma^2}\left(\sum_{j=1}^{N_1}\left\|\overline{\mathbf{y}}_{s_j}^{(N)}\right\|_2^2 a_{s_j}^2 + 2\sum_{j=1}^{N_1}\sum_{i=j+1}^{N_1}\left(\left\langle\overline{\mathbf{y}}_{s_j}^{(N)},\overline{\mathbf{y}}_{s_i}^{(N)}\right\rangle a_{s_j}a_{s_i}\right)\right)\right) da_{sN_1} \cdots da_{s_1}$$

$$\overset{(a)}{\leq} \sqrt{\frac{2\pi\sigma^2}{n\left\|\overline{\mathbf{y}}_{s_1}^{(N)}\right\|_2^2}} erf\left(\sqrt{\frac{n\left\|\overline{\mathbf{y}}_{s_1}^{(N)}\right\|_2^2}{2\sigma^2}}\right) \times \int_{-1+\alpha_{s_2}}^{1+\alpha_{s_2}} \cdots \int_{-1+\alpha_{sN_1}}^{1+\alpha_{sN_1}}$$

$$\exp\left(\frac{n\left(\sum_{i=2}^{N_1}\left(\left\langle\overline{\mathbf{y}}_{s_1}^{(N)},\overline{\mathbf{y}}_{s_i}^{(N)}\right\rangle a_{s_i}\right)\right)^2}{\left\|\overline{\mathbf{y}}_{s_1}^{(N)}\right\|_2^2 2\sigma^2}\right)\exp\left(\frac{-n}{2\sigma^2}\left(\sum_{j=2}^{N_1}\left\|\overline{\mathbf{y}}_{s_j}^{(N)}\right\|_2^2 a_{s_j}^2 + 2\sum_{j=2}^{N_1}\sum_{i=j+1}^{N_1}\left\langle\overline{\mathbf{y}}_{s_j}^{(N)},\overline{\mathbf{y}}_{s_i}^{(N)}\right\rangle a_{s_j}a_{s_i}\right)\right) da_{sN_1} \cdots da_{s_2}$$

$$\overset{(b)}{=} \sqrt{\frac{2\pi\sigma^2}{n\left\|\overline{\mathbf{y}}_{s_1}^{(N)}\right\|_2^2}} erf\left(\sqrt{\frac{n\left\|\overline{\mathbf{y}}_{s_1}^{(N)}\right\|_2^2}{2\sigma^2}}\right) \times \int_{-1+\alpha_{s_2}}^{1+\alpha_{s_2}} \cdots \int_{-1+\alpha_{sN_1}}^{1+\alpha_{sN_1}}$$

$$\exp\left(\frac{-n}{2\sigma^2}\left(\sum_{j=2}^{N_1}\left(\left\|\overline{\mathbf{y}}_{s_j}^{(N)}\right\|_2^2 - \left\langle\frac{\overline{\mathbf{y}}_{s_1}^{(N)}}{\left\|\overline{\mathbf{y}}_{s_1}^{(N)}\right\|_2},\overline{\mathbf{y}}_{s_j}^{(N)}\right\rangle^2\right)a_{s_j}^2 + 2\sum_{j=2}^{N_1}\sum_{i=j+1}^{N_1}\left(\left\langle\overline{\mathbf{y}}_{s_j}^{(N)},\overline{\mathbf{y}}_{s_i}^{(N)}\right\rangle - \left\langle\frac{\overline{\mathbf{y}}_{s_1}^{(N)}}{\left\|\overline{\mathbf{y}}_{s_1}^{(N)}\right\|_2},\overline{\mathbf{y}}_{s_j}^{(N)}\right\rangle\left\langle\frac{\overline{\mathbf{y}}_{s_1}^{(N)}}{\left\|\overline{\mathbf{y}}_{s_1}^{(N)}\right\|_2},\overline{\mathbf{y}}_{s_i}^{(N)}\right\rangle\right)a_{s_j}a_{s_i}\right)\right) da_{sN_1} \cdots da_{s_2}$$

(C.3)

where (a) holds by using (C.1) w.r.t. the variable $a_{s_1}$, and (b) follows after some straightforward manipulations. Compared to (5.13), we can see from (C.3) that the coefficients of the rest variables $a_{s_2}^2, \cdots, a_{s_{N_1}}^2$ in the exponent admit the form $\frac{-n\left(\left\|\overline{\mathbf{y}}_{s_j}^{(N)}\right\|_2^2 - \left\langle\overline{\mathbf{y}}_{s_1}^{(N)}/\left\|\overline{\mathbf{y}}_{s_1}^{(N)}\right\|_2,\overline{\mathbf{y}}_{s_j}^{(N)}\right\rangle^2\right)}{2\sigma^2}$, which can be shown to be negative. Actually, it can be proved by induction that, when applying (C.1) to the $j$th variable $a_{s_j}$, the coefficient of the next variable $a_{s_{j+1}}^2$ in the exponent is always negative (detailed given at the end of Part I). This ensures we can successively apply (C.1) to derive the desired upper bound. For an integer $p > 1$, and repeatedly applying (C.1), starting from the variable $a_{s_1}$ to $a_{s_p}$, it can be shown by induction that (details given later)

$$\delta_2 \leq \left(\prod_{j=1}^p g_j\right)\int_{-1+\alpha_{s(p+1)}}^{1+\alpha_{s(p+1)}} \cdots \int_{-1+\alpha_{sN1}}^{1+\alpha_{sN1}} e^{\frac{-n}{2\sigma^2}\left(\sum_{j=p+1}^{N_1}\left(\delta_{j,j,p+1}a_{s_j}^2\right)+2\sum_{j=p+1}^{N_1}\sum_{i=j+1}^{N_1}\left(\delta_{i,j,p+1}a_{s_i}a_{s_j}\right)\right)} da_{sN_1}\cdots da_{s(p+1)}, \text{(C.4)}$$

where

$$g_j \triangleq \sqrt{\frac{2\pi\sigma^2}{n\delta_{j,j,j}}}erf\left(\sqrt{\frac{n\delta_{j,j,j}}{2\sigma^2}}\right), \tag{C.5}$$

and

$$\delta_{i,j,k} \triangleq \left\langle\overline{\mathbf{y}}_{s_i}^{(N)},\overline{\mathbf{y}}_{s_j}^{(N)}\right\rangle - \sum_{q=1}^{k-1}\frac{\delta_{i,q,q}\delta_{j,q,q}}{\delta_{q,q,q}}, \text{ for } i,j \geq k. \tag{C.6}$$

Based on (C.4), and applying (C.1) till $p = N_1$ we can obtain

$$\delta_2 \leq \prod_{j=1}^{N_1} g_j = \prod_{j=1}^{N_1} \sqrt{\frac{2\pi\sigma^2}{n\delta_{j,j,j}}} erf\left(\sqrt{\frac{n\delta_{j,j,j}}{2\sigma^2}}\right)$$
$$= \prod_{j=1}^{N_1} \left( \sqrt{\frac{2\pi\sigma^2}{nc_{j,q}\left\|\overline{\mathbf{y}}_{s_j}^{(N)}\right\|_2^2}} erf\left(\sqrt{\frac{nc_{j,q}}{2\sigma^2}}\left\|\overline{\mathbf{y}}_{s_j}^{(N)}\right\|_2\right)\right), \quad \text{(C.7)}$$

in which we have defined

$$c_{j,q} \triangleq \delta_{j,j,j} / \left\|\overline{\mathbf{y}}_{s_j}^{(N)}\right\|_2^2. \quad \text{(C.8)}$$

The proof of (5.16) is thus completed. □

Now we will show by induction that (C.4) is true, based on which we can deduce again using induction that the coefficients of $a_{s_j}^2$ are always negative when successively applying (C.1) to obtain the bound. To prove (C.4), the case with $p = 1$ is true, as can be directly seen from (C.3). Assume the claim holds for $p = k - 1$. Then we have

$$\delta_2 \leq \left(\prod_{j=1}^{k-1} g_j\right) \int_{-1+\alpha_{s_k}}^{1+\alpha_{s_k}} \cdots \int_{-1+\alpha_{sN_1}}^{1+\alpha_{sN_1}} \exp\left(\frac{-n}{2\sigma^2}\left(\sum_{j=k}^{N_1}\left(\delta_{j,j,k}a_{s_j}^2\right) + 2\sum_{j=k}^{N_1}\sum_{i=j+1}^{N_1}\left(\delta_{i,j,k}a_{s_i}a_{s_j}\right)\right)\right) da_{sN_1}\cdots da_{s_k}$$
$$\overset{(a)}{\leq} \left(\prod_{j=1}^{k-1} g_j\right) \sqrt{\frac{2\pi\sigma^2}{n\delta_{k,k,k}}} erf\left(\sqrt{\frac{n\delta_{k,k,k}}{2\sigma^2}}\right)$$
$$\times \int_{-1+\alpha_{s_{k+1}}}^{1+\alpha_{s_{k+1}}} \cdots \int_{-1+\alpha_{sN_1}}^{1+\alpha_{sN_1}} \exp\left(\frac{n\left(\sum_{i=k+1}^{N_1}\delta_{i,k,k}a_{s_i}\right)^2}{\delta_{k,k,k}2\sigma^2}\right)\exp\left(\frac{-n}{2\sigma^2}\left(\sum_{j=k+1}^{N_1}\left(\delta_{j,j,k}a_{s_j}^2\right) + 2\sum_{j=k+1}^{N_1}\sum_{i=j+1}^{N_1}\left(\delta_{i,j,k}a_{s_i}a_{s_j}\right)\right)\right) da_{sN_1}\cdots da_{s_{k+1}}, \quad \text{(C.9)}$$

where (a) holds by applying (C.1) w.r.t. the variable $a_{s_k}$. Since

$$\exp\left(\frac{n\left(\sum_{j=k+1}^{N_1}(\delta_{j,k,k}a_{s_j})\right)^2}{\delta_{k,k,k}2\sigma^2}\right) = \exp\left(\frac{n}{\delta_{k,k,k}2\sigma^2}\left(\sum_{j=k+1}^{N_1}(\delta_{j,k,k})^2 a_{s_j}^2 + 2\sum_{j=k+1}^{N_1}\sum_{i=j+1}^{N_1}(\delta_{i,k,k}\delta_{j,k,k})a_{s_i}a_{s_j}\right)\right), \quad \text{(C.10)}$$

it follows from (C.9) and (C.10) that

$$\delta_2 \leq \left[\left(\prod_{j=1}^{k-1} g_j\right) \sqrt{\frac{2\pi\sigma^2}{n\delta_{k,k,k}}} erf\left(\sqrt{\frac{n\delta_{k,k,k}}{2\sigma^2}}\right)\right]$$
$$\times \int_{-1+\alpha_{s_{k+1}}}^{1+\alpha_{s_{k+1}}} \cdots \int_{-1+\alpha_{sN_1}}^{1+\alpha_{sN_1}} \exp\left(\frac{-n}{2\sigma^2}\left(\sum_{j=k+1}^{N_1}\left(\delta_{j,j,k} - \frac{(\delta_{j,k,k})^2}{\delta_{k,k,k}}\right)a_{s_j}^2 + 2\sum_{j=k+1}^{N_1}\sum_{i=j+1}^{N_1}\left(\delta_{i,j,k} - \frac{\delta_{i,k,k}\delta_{j,k,k}}{\delta_{k,k,k}}\right)a_{s_i}a_{s_j}\right)\right) da_{sN_1}\cdots da_{s_{k+1}}$$
$$= \prod_{j=1}^{k}(g_j) \int_{-1+\alpha_{s_{k+1}}}^{1+\alpha_{s_{k+1}}} \cdots \int_{-1+\alpha_{sN_1}}^{1+\alpha_{sN_1}} \exp\left(\frac{-n}{2\sigma^2}\left(\sum_{j=k+1}^{N_1}\left(\delta_{j,j,k+1}a_{s_j}^2\right) + 2\sum_{j=k+1}^{N_1}\sum_{i=j+1}^{N_1}\left(\delta_{i,j,k+1}a_{s_i}a_{s_j}\right)\right)\right) da_{sN_3}\cdots da_{s_{k+1}}.$$
$$\quad \text{(C.11)}$$

This implies that (C.4) holds for the case with $p = k$. Hence the proof of (C.4) using induction is completed. □

Finally, we will prove $(-n\delta_{j,j,j})/2\sigma^2 < 0$, that is, the coefficient of the quadratic term of $a_{s_j}^2$ when applying inequality (C.1) to the $j$th variable $a_{s_j}$ is negative; as mentioned before, this condition is required for ensuring that we can repeatedly use (C.1) to derive the upper bound for $\delta_2$. It is equivalent to show $\delta_{j,j,j} > 0$. Below we will show $\delta_{j,j,j} \geq 0$ by induction, and the equality never holds. By definition (C.6), the statement is obviously true for $j = 1$ since $\delta_{1,1,1} = \|\overline{\mathbf{y}}_{s_1}^{(N)}\|_2^2 \geq 0$. Assume that the statement holds for the case with $j = p$, meaning that

$$\delta_{p,p,p} = \|\overline{\mathbf{y}}_{s_p}^{(N)}\|_2^2 \left(1 - \sum_{q=1}^{p-1} \frac{\delta_{p,q,q}^2}{\delta_{q,q,q} \|\overline{\mathbf{y}}_{s_p}^{(N)}\|_2^2}\right) \geq 0. \tag{C.12}$$

Based on (C.4), repeatedly applying (C.1) starting from the variable $a_{s_1}$ to $a_{s_{p-1}}$ we can obtain

$$\delta_2 \leq \left(\prod_{j=1}^{p-1} g_j\right) \int_{-1+\alpha_{s_p}}^{1+\alpha_{s_p}} \cdots \int_{-1+\alpha_{s_{N_1}}}^{1+\alpha_{s_{N_1}}} \exp\left(\frac{-n}{2\sigma^2}\left(\sum_{j=p}^{N_1}\left(\delta_{j,j,p} a_{s_j}^2\right) + 2\sum_{j=p}^{N_1}\sum_{i=j+1}^{N_1}\left(\delta_{i,j,p} a_{s_i} a_{s_j}\right)\right)\right) da_{s_{N_1}} \cdots da_{s_p}. \tag{C.13}$$

Hence, the condition (C.12) is equivalent to the assertion that the coefficient $(-n\delta_{p,p,p})/2\sigma^2$ of the quadratic term $a_{s_p}^2$ in the exponent on the RHS of (C.13) is never positive. For $j = p+1$ our purpose is to prove $\delta_{p+1,p+1,p+1} \geq 0$. By definition (C.6), it then suffices to show $\left(1 - \sum_{q=1}^{p-1} \frac{\delta_{p,q,q}^2}{\delta_{q,q,q} \|\overline{\mathbf{y}}_{s_p}^{(N)}\|_2^2}\right)\left(1 - \sum_{q=1}^{p-1} \frac{\delta_{p+1,q,q}^2}{\delta_{q,q,q} \|\overline{\mathbf{y}}_{s_{p+1}}^{(N)}\|_2^2}\right) - \left(\frac{\delta_{p+1,p,p}}{\|\overline{\mathbf{y}}_{s_p}^{(N)}\|_2 \|\overline{\mathbf{y}}_{s_{p+1}}^{(N)}\|_2}\right)^2 \geq 0$. Using vector notation, this is equivalent to establishing the following condition:

$$f(\mathbf{a}, \mathbf{b}, c) \triangleq \left(1 - \|\mathbf{a}\|_2^2\right)\left(1 - \|\mathbf{b}\|_2^2\right) - \left(c - \mathbf{a}^T \mathbf{b}\right)^2 \geq 0, \tag{C.14}$$

where

$$\mathbf{a} \triangleq \left[\frac{\delta_{p,1,1}}{\sqrt{\delta_{1,1,1}} \|\overline{\mathbf{y}}_{s_p}^{(N)}\|_2} \cdots \frac{\delta_{p,p-1,p-1}}{\sqrt{\delta_{p-1,p-1,p-1}} \|\overline{\mathbf{y}}_{s_p}^{(N)}\|_2}\right]^T, \quad \mathbf{b} \triangleq \left[\frac{\delta_{p+1,1,1}}{\sqrt{\delta_{1,1,1}} \|\overline{\mathbf{y}}_{s_{p+1}}^{(N)}\|_2} \cdots \frac{\delta_{p+1,p-1,p-1}}{\sqrt{\delta_{p-1,p-1,p-1}} \|\overline{\mathbf{y}}_{s_{p+1}}^{(N)}\|_2}\right]^T, \quad c \triangleq \left\langle \frac{\overline{\mathbf{y}}_{s_p}^{(N)}}{\|\overline{\mathbf{y}}_{s_p}^{(N)}\|_2}, \frac{\overline{\mathbf{y}}_{s_{p+1}}^{(N)}}{\|\overline{\mathbf{y}}_{s_{p+1}}^{(N)}\|_2}\right\rangle. \tag{C.15}$$

By definition (C.15), the three variables $(\mathbf{a}, \mathbf{b}, c)$ of the function $f(\cdot)$ in (C.14) are subject to certain constraints. Specifically, it is straightforward to rewrite (C.12) as $\|\overline{\mathbf{y}}_{s_p}^{(N)}\|_2^2 \left(1 - \|\mathbf{a}\|_2^2\right) \geq 0$, which implies $\left(1 - \|\mathbf{a}\|_2^2\right) \geq 0$. Also, note that if we conduct a $p$-fold integration of (5.13) using inequality (C.1), except obeying the order $a_{s_1}, a_{s_2}, \cdots, a_{s_{p-1}}, a_{s_{p+1}}$ (i.e., the last integration variable is $a_{s_{p+1}}$ rather than $a_{s_p}$), the induction assumption that the statement holds for $j = p$ implies the coefficient of variable $a_{s_{p+1}}^2$ in the exponent on the RHS of (C.13) is never positive, meaning that

$$(-n\delta_{p+1,p+1,p})/2\sigma^2 = \frac{-n\|\overline{\mathbf{y}}_{s_{p+1}}^{(N)}\|_2^2}{2\sigma^2}\left(1 - \sum_{q=1}^{p-1} \frac{\delta_{p+1,q,q}^2}{\delta_{q,q,q} \|\overline{\mathbf{y}}_{s_{p+1}}^{(N)}\|_2^2}\right) = \frac{-n\|\overline{\mathbf{y}}_{s_{p+1}}^{(N)}\|_2^2}{2\sigma^2}\left(1 - \|\mathbf{b}\|_2^2\right) \leq 0, \tag{C.16}$$

which in turn implies $\left(1 - \|\mathbf{b}\|_2^2\right) \geq 0$. Obviously, $|c| \leq 1$. To prove (C.14), the basic idea is to show the minimum of $f(\mathbf{a}, \mathbf{b}, c)$, subject to $\left(1 - \|\mathbf{a}\|_2^2\right) \geq 0$, $\left(1 - \|\mathbf{b}\|_2^2\right) \geq 0$, and $|c| \leq 1$, is never negative. Since $f(\mathbf{a}, \mathbf{b}, c)$ is concave in $\mathbf{a}$, $\mathbf{b}$ and $c$, the minimum of $f(\mathbf{a}, \mathbf{b}, c)$ is achieved by some points on the boundary of the feasible set; hence we have

$$\left(1-\|\mathbf{a}^*\|_2^2\right)=0,\ \left(1-\|\mathbf{b}^*\|_2^2\right)=0 \text{ and } |c^*|=1, \tag{C.17}$$

where $[\mathbf{a}^*,\mathbf{b}^*,c^*]$ is the optimal solution. Note that for $i,j\geq k$, $\delta_{i,j,k}$ defined in (C.6) can be expressed as

$$\delta_{i,j,k}=\left\|\overline{\mathbf{y}}_{s_i}^{(N)}\right\|_2\left\|\overline{\mathbf{y}}_{s_j}^{(N)}\right\|_2 d_{i,j,k}, \tag{C.18}$$

where

$$d_{i,j,k}\triangleq\left\langle\frac{\overline{\mathbf{y}}_{s_i}^{(N)}}{\left\|\overline{\mathbf{y}}_{s_i}^{(N)}\right\|_2},\frac{\overline{\mathbf{y}}_{s_j}^{(N)}}{\left\|\overline{\mathbf{y}}_{s_j}^{(N)}\right\|_2}\right\rangle-\sum_{q=1}^{k-1}\frac{d_{i,q,q}d_{j,q,q}}{d_{q,q,q}} \tag{C.19}$$

(this can be easily prove by induction and the definition of $\delta_{i,j,k}$ in (C.6)). With (C.18), $\mathbf{a}$ and $\mathbf{b}$ defined in (C.15) can be expressed as

$$\mathbf{a}=\left[\frac{d_{p,1,1}}{\sqrt{d_{1,1,1}}}\cdots\frac{d_{p,p-1,p-1}}{\sqrt{d_{p-1,p-1,p-1}}}\right]^T \text{ and } \mathbf{b}=\left[\frac{d_{p+1,1,1}}{\sqrt{d_{1,1,1}}}\cdots\frac{d_{p+1,p-1,p-1}}{\sqrt{d_{p-1,p-1,p-1}}}\right]^T. \tag{C.20}$$

Moreover, it can be shown using (C.19) (details given later) that, if $\overline{\mathbf{y}}_{s_i}^{(N)}/\|\overline{\mathbf{y}}_{s_i}^{(N)}\|_2=\overline{\mathbf{y}}_{s_m}^{(N)}/\|\overline{\mathbf{y}}_{s_m}^{(N)}\|_2$, then

$$d_{i,j,k}=d_{m,j,k}; \tag{C.21}$$

similarly, if $\overline{\mathbf{y}}_{s_i}^{(N)}/\|\overline{\mathbf{y}}_{s_i}^{(N)}\|_2=-\overline{\mathbf{y}}_{s_n}^{(N)}/\|\overline{\mathbf{y}}_{s_n}^{(N)}\|_2$, then

$$d_{i,j,k}=-d_{n,j,k}. \tag{C.22}$$

In the case that $c^*=1$, we have $\overline{\mathbf{y}}_{s_p}^{(N)}/\|\overline{\mathbf{y}}_{s_p}^{(N)}\|_2=\overline{\mathbf{y}}_{s_{p+1}}^{(N)}/\|\overline{\mathbf{y}}_{s_{p+1}}^{(N)}\|_2$, which together with (C.21) implies $d_{p,k,k}=d_{p+1,k,k}$ for all $1\leq k\leq p-1$. From (C.20), we then obtain $\mathbf{a}^*=\mathbf{b}^*$, which together with (C.17) implies $(c^*-(\mathbf{a}^*)^T\mathbf{b}^*)=1-1=0$. Similarly, for $c^*=-1$, we have $\overline{\mathbf{y}}_{s_p}^{(N)}/\|\overline{\mathbf{y}}_{s_p}^{(N)}\|_2=-\overline{\mathbf{y}}_{s_{p+1}}^{(N)}/\|\overline{\mathbf{y}}_{s_{p+1}}^{(N)}\|_2$. By (C.20) and (C.22), we can obtain $\mathbf{a}^*=-\mathbf{b}^*$, which together with (C.17) implies $(c^*-(\mathbf{a}^*)^T\mathbf{b}^*)=0$. We have shown the minimum of $f(\mathbf{a},\mathbf{b},c)$ is zero, thereby $f(\mathbf{a},\mathbf{b},c)\geq 0$. Hence, the assertion holds for $j=p+1$, which then completes the induction proof.

Finally, we note that $|c|=1$ only when $\overline{\mathbf{y}}_{s_p}^{(N)}$ and $\overline{\mathbf{y}}_{s_{p+1}}^{(N)}$ are co-linear (see (C.15)), which never happens in the SSC context. Hence, $f(\mathbf{a},\mathbf{b},c)>0$, yielding $(-n\delta_{j,j,j})/2\sigma^2<0$. $\square$

*[Proof of (C.21)]:* We will prove (C.21) by induction. Assume that $\overline{\mathbf{y}}_{s_i}^{(N)}/\|\overline{\mathbf{y}}_{s_i}^{(N)}\|_2=\overline{\mathbf{y}}_{s_m}^{(N)}/\|\overline{\mathbf{y}}_{s_m}^{(N)}\|_2$. By definition of $d_{i,j,k}$ in (C.19), the statement is true for the case with $k=1$ since $d_{i,j,1}=\left\langle\frac{\overline{\mathbf{y}}_{s_i}^{(N)}}{\|\overline{\mathbf{y}}_{s_i}^{(N)}\|_2},\frac{\overline{\mathbf{y}}_{s_j}^{(N)}}{\|\overline{\mathbf{y}}_{s_j}^{(N)}\|_2}\right\rangle=\left\langle\frac{\overline{\mathbf{y}}_{s_m}^{(N)}}{\|\overline{\mathbf{y}}_{s_m}^{(N)}\|_2},\frac{\overline{\mathbf{y}}_{s_j}^{(N)}}{\|\overline{\mathbf{y}}_{s_j}^{(N)}\|_2}\right\rangle=d_{m,j,1}$. Assume that the statement is true for $k=p$, that is,

$$d_{i,j,p}=d_{m,j,p}, \text{ for all } j\geq p. \tag{C.23}$$

For $k = p+1$ our purpose is to prove $d_{i,j,p+1} = d_{m,j,p+1}$. By definition of $d_{i,j,k}$, we can obtain

$$d_{i,j,p+1} = \left\langle \frac{\overline{\mathbf{y}}_{s_i}^{(N)}}{\|\overline{\mathbf{y}}_{s_i}^{(N)}\|_2}, \frac{\overline{\mathbf{y}}_{s_j}^{(N)}}{\|\overline{\mathbf{y}}_{s_j}^{(N)}\|_2} \right\rangle - \sum_{q=1}^{p} \frac{d_{i,q,q} d_{j,q,q}}{d_{q,q,q}} = d_{i,j,p} - \frac{d_{i,p,p} d_{j,p,p}}{d_{p,p,p}}$$

$$\overset{(a)}{=} d_{m,j,p} - \frac{d_{m,p,p} d_{j,p,p}}{d_{p,p,p}} = \left\langle \frac{\overline{\mathbf{y}}_{s_m}^{(N)}}{\|\overline{\mathbf{y}}_{s_m}^{(N)}\|_2}, \frac{\overline{\mathbf{y}}_{s_j}^{(N)}}{\|\overline{\mathbf{y}}_{s_j}^{(N)}\|_2} \right\rangle - \sum_{q=1}^{p} \frac{d_{i,q,q} d_{j,q,q}}{d_{q,q,q}} = d_{m,j,p+1},$$

(C.24)

where (a) holds by (C.23). The proof of (C.21) is thus completed. □

*Part II: Upper Bound for $\delta_3$ in (5.17):*

Since

$$\frac{-n}{2\sigma^2}\left(\left\|\sum_{j=1}^{N_2} b_{t_j} \mathbf{y}_{t_j} - \sum_{j=N_2+1}^{N_3} b_{t_j} \mathbf{y}_{t_j}\right\|_2^2\right) = \frac{-n}{2\sigma^2}\left(\sum_{j=1}^{N_3} \|\mathbf{y}_{t_j}\|_2^2 b_{t_j}^2 + 2\sum_{j=1}^{N_3}\left\langle b_{t_j}\mathbf{y}_{t_j}, \left(\sum_{i=j+1}^{N_2}(b_{t_i}\mathbf{y}_{t_i}) - \sum_{i=N_2+1}^{N_3}(b_{t_i}\mathbf{y}_{t_i})\right)\right\rangle\right),$$ (C.25)

the coefficient of the quadratic term $b_{t_j}^2$ in the exponent of $\delta_3$ (defined in (5.14)) is $-(n\|\mathbf{y}_{t_j}\|^2/2\sigma^2)$, which is negative. Hence, repeatedly using part (2) of Lemma C.1 (one for each variable $b_{t_j}$) leads to

$$\delta_3 \leq \prod_{j=1}^{N_3} \sqrt{\frac{2\pi\sigma^2}{n\|\mathbf{y}_{t_j}\|_2^2}}.$$ (C.26)

The proof of (5.17) is thus completed. □

We end this appendix by proving Lemma C.1.

*[Proof of Lemma C.1]:* To prove part (1), we have

$$\int_{-1+\gamma}^{1+\gamma} \exp\left(-\left(ax^2 + bx\right)\right) dx$$

$$= \int_{-1+\gamma}^{1+\gamma} \exp\left(-a\left(x + \frac{b}{2a}\right)^2 + \frac{b^2}{4a}\right) dx$$

$$\overset{(a)}{=} \int_{\sqrt{a}(-1+\gamma+b/2a)}^{\sqrt{a}(1+\gamma+b/2a)} \exp\left(-v^2 + \frac{b^2}{4a}\right) \frac{1}{\sqrt{a}} dv$$

$$\leq 2\exp\left(\frac{b^2}{4a}\right) \frac{1}{\sqrt{a}} \int_0^{\sqrt{a}} \exp\left(-v^2\right) dv = \exp\left(\frac{b^2}{4a}\right) \sqrt{\frac{\pi}{a}} \, erf\left(\sqrt{a}\right),$$

(C.27)

where (a) follows from the change of variable. (C.1) then follows from (C.27). To prove (C.2), we have

$$\int_t^\infty \exp\left(-(ax^2+bx)\right) dx$$

$$= \int_0^\infty \exp\left(-(ax^2+bx)\right) dx + \int_t^0 \exp\left(-(ax^2+bx)\right) dx$$

$$< \int_0^\infty \exp\left(-(ax^2+bx)\right) dx + \int_{-\infty}^0 \exp\left(-(ax^2+bx)\right) dx$$

$$\overset{(a)}{=} \sqrt{\frac{\pi}{4a}} \left\{\exp\left(\frac{b^2}{4a}\right)\left[1 - erf\left(\frac{b}{2\sqrt{a}}\right)\right] + \exp\left(\frac{b^2}{4a}\right)\left[erf\left(\frac{b}{2\sqrt{a}}\right) + 1\right]\right\}$$

$$= \sqrt{\frac{\pi}{4a}} \left\{\exp\left(\frac{b^2}{4a}\right)\left[erfc\left(\frac{b}{2\sqrt{a}}\right)\right] + \exp\left(\frac{b^2}{4a}\right)\left[erfc\left(\frac{-b}{2\sqrt{a}}\right)\right]\right\}$$

$$\overset{(b)}{\leq} \sqrt{\frac{\pi}{4a}} \{1 + 1\} = \sqrt{\frac{\pi}{a}}$$

(C.28)

where (a) follows from the change of variable and (b) holds since the complementary error function $erfc(x) \leq \exp(-x^2)$. □

## D. Derivations of (5.38)

Since $\mathcal{K}_M$ defined in (5.27) is closed and $\{\mathbf{y}_1,\cdots,\mathbf{y}_{N-1}: |\det(\overline{\mathbf{U}}_q)| < 1/k\}$ is open, both are measureable. Hence $\mathcal{Z}_{M,k}$ defined in (5.32), being an intersection of two measurable sets, is measurable. In addition, the sequence of measurable sets $\{\mathcal{Z}_{M,k}\}_{k=1}^{\infty}$ is decreasing such that $\mathcal{Z}_{M,k} \searrow \mathcal{Z}$, where

$$\mathcal{Z} \triangleq \mathcal{K}_M \cap \{\mathbf{y}_1,\cdots,\mathbf{y}_{N-1}: |\det(\overline{\mathbf{U}}_q)| = 0\}. \tag{D.1}$$

Note that

$$\{\mathbf{y}_1,\cdots,\mathbf{y}_{N-1}: |\det(\overline{\mathbf{U}}_q)| = 0\} = \bigcup_{j=1}^{N-1}\left(\bigcup_{k=1}^{N-2}\left\{\mathbf{y}_1,\cdots,\mathbf{y}_{N-1}: \mathbf{y}_j = \sum_{i=1}^{k} a_i \mathbf{y}_{s_i},\ s_i \neq j,\ a_i \in \mathbb{R}\right\}\right) \triangleq \bigcup_{j=1}^{N-1}\bigcup_{k=1}^{N-2} \mathbf{A}_{j,k}, \tag{D.2}$$

which is a finite union of subsets $\mathbf{A}_{j,k} \subset \underbrace{\mathbb{R}^n \times \cdots \times \mathbb{R}^n}_{(N-1)-\text{fold}}$. Notably, each $\mathbf{A}_{j,k}$ can also be decomposed into a finite union of subsets, each contained in a subspace with dimension less than $n(N-1)$. For example,

$$\begin{aligned}\mathbf{A}_{1,1} &= \bigcup_{m=2}^{N-1} \{\mathbf{y}_1,\cdots,\mathbf{y}_{N-1}: \mathbf{y}_1 = a\mathbf{y}_m,\ a \in \mathbb{R}\} \\ &= \bigcup_{m=2}^{N-1}\left(\{\mathbf{y}_1: \mathbf{y}_1 = a\mathbf{y}_m,\text{ where } a \in \mathbb{R}\} \times \underbrace{\mathbb{R}^n \times \cdots \times \mathbb{R}^n}_{N-2}\right) \triangleq \bigcup_{m=2}^{N-1} \mathbf{A}_{1,1}^{(m)}\end{aligned} \tag{D.3}$$

is a union of $N-2$ subsets $\mathbf{A}_{1,1}^{(m)}$, each contained in a subspace with dimension $n(N-2)+1 < n(N-1)$, and is thus of measure zero [35, Chap. 2]. By following similar arguments, it can be verified that each $\mathbf{A}_{j,k}$ (as a finite union of subsets, each contained in a subspace with dimension $n(N-2)+k < n(N-1)$) is of measure zero. This then implies $\{\mathbf{y}_1,\cdots,\mathbf{y}_{N-1}: \det(\overline{\mathbf{U}}_q) = 0\}$ defined in (D.2) is measure zero, thereby $|\mathcal{Z}| = 0$. Moreover, since $|\mathcal{Z}_{M,k}| \leq |\mathcal{K}_M| < \infty$, the it follows immediately from [35, Thm. 3.26] that

$$\lim_{k \to \infty} |\mathcal{Z}_{M,k}| = |\mathcal{Z}| = 0. \tag{D.4}$$

The proof of (5.38) is thus completed. □

## E. Proof of Lemma 5.2

Under the setting of Lemma 5.2 and using [11, Lemma 7.5], it follows

$$\Pr\left\{|\mathbf{a}^T \Sigma \mathbf{v}| \leq 8(\log N)\frac{\|\Sigma\|_F}{\sqrt{n_1 n_2}}\right\} \geq 1 - 4N^{-4}. \tag{E.1}$$

Hence, for $c > 0$ we can get

$$\Pr\left\{|c\mathbf{a}^T \sum \mathbf{v}| \leq 8 \log N \frac{\|\Sigma\|_F}{\sqrt{n_1 n_2}}\right\}$$

$$= \Pr\left\{|\mathbf{a}^T \sum \mathbf{v}| \leq 8 \frac{\log N}{c} \frac{\|\Sigma\|_F}{\sqrt{n_1 n_2}}\right\} \quad\quad\quad\quad (\text{E.2})$$

$$= \Pr\left\{|\mathbf{a}^T \sum \mathbf{v}| \leq 8 \log(N^{1/c}) \frac{\|\Sigma\|_F}{\sqrt{n_1 n_2}}\right\} \geq 1 - 4N^{-4/c}$$

The proof of Lemma 5.2 is thus completed. $\square$

## F. Proof of Lemma 5.3:

From [10, eq. (A.1)], a Chi-square random variable $\mathcal{X}_n^2$ with $n$ degrees of freedom obeys

$$\Pr\{\mathcal{X}_n^2 < (1+\kappa)n\} \geq 1 - \exp\left(\frac{-(1-\log 2)}{2} n\kappa^2\right). \quad\quad\quad (\text{F.1})$$

By Cauchy-Schwarz inequality, we have $|\mathbf{a}^T \mathbf{z}| \leq \|\mathbf{a}\|_2 \|\mathbf{z}\|_2 = \|\mathbf{a}\|_2$. Next, by setting $\kappa = \frac{8 \log n}{nc} - 1$ for $c > 0$ we can obtain

$$\begin{aligned}
\Pr\left\{c|\mathbf{a}^T \mathbf{z}| \leq 2\sqrt{2 \log N}\right\} &\geq \Pr\left\{c\|\mathbf{a}\|_2 \leq 2\sqrt{2 \log N}\right\} \\
&= \Pr\left\{\|\mathbf{a}\|_2 \leq \frac{2\sqrt{2\log N}}{c}\right\} \\
&= \Pr\left\{\|\mathbf{a}\|_2^2 \leq (1+\kappa)n\right\} \\
&\overset{(a)}{\geq} 1 - \exp\left(\frac{-(1-\log 2)}{2}n\kappa^2\right) = 1 - \exp\left(-\frac{(1-\log 2)(cn - 8\log n)^2}{2nc^2}\right)
\end{aligned} \quad\quad (\text{F.2})$$

where (a) follows from (F.1). The proof of Lemma 5.3 is completed. $\square$

## G. Proof of Lemma 5.4:

The proof basically follows the structure of the proof of Lemma 8.5 in [10]. Let $\hat{\mathbf{c}}_N$ be the optimal solution of the *projected problem*, namely,

$$\hat{\mathbf{c}}_N = \arg\min \frac{1}{2}\left\|(\mathbf{y}_{N,\|} - \mathbf{Y}_{t,\|}\mathbf{W}_t \mathbf{c}_N)\right\|_2^2 + \lambda \|\mathbf{c}_N\|_1. \quad\quad\quad (\text{G.1})$$

The following lemma (whose proof is placed at the end of this appendix) is needed.

***Lemma G.1:*** The inequality

$$\left\|(\mathbf{y}_{N,\perp} - \mathbf{Y}_{t,\perp}\mathbf{W}_t \hat{\mathbf{c}}_N)\right\|_2^2 \leq \sigma^2 \left(1 - \frac{d_N}{n}\right)(1 + \|\hat{\mathbf{c}}_N\|_2^2) \quad\quad\quad (\text{G.2})$$

holds with a probability at least $1 - \exp\left(\frac{-(1-\log 2)}{2}(n-d_N)\left(1 - (w_{t_1}^{(N)})^2/2\right)^2\right)$. In addition, there exists a constant $\eta_1$ such that

$$\|\hat{\mathbf{c}}_N\|_2 \leq \eta_1 \quad\quad\quad (\text{G.3})$$

with a probability at least $1 - N^{-3}$. $\square$

Recall that $\bar{\mathbf{c}}_N$ defined in (5.49) is the optimal solution to the reduced problem, therefore

$$\frac{1}{2}\|(\mathbf{y}_N - \mathbf{Y}_t\mathbf{W}_t\bar{\mathbf{c}}_N)\|_2^2 + \lambda\|\bar{\mathbf{c}}_N\|_1 \leq \frac{1}{2}\|(\mathbf{y}_N - \mathbf{Y}_t\mathbf{W}_t\hat{\mathbf{c}}_N)\|_2^2 + \lambda\|\hat{\mathbf{c}}_N\|_1. \tag{G.4}$$

Since $\hat{\mathbf{c}}_N$ is the optimal solution to the projected problem, we have

$$\frac{1}{2}\|(\mathbf{y}_{N,\|} - \mathbf{Y}_{t,\|}\mathbf{W}_t\hat{\mathbf{c}}_N)\|_2^2 + \lambda\|\hat{\mathbf{c}}_N\|_1 \leq \frac{1}{2}\|(\mathbf{y}_{N,\|} - \mathbf{Y}_{t,\|}\mathbf{W}_t\bar{\mathbf{c}}_N)\|_2^2 + \lambda\|\bar{\mathbf{c}}_N\|_1. \tag{G.5}$$

Using the Pythagorean theorem we obtain

$$\|(\mathbf{y}_N - \mathbf{Y}_t\mathbf{W}_t\hat{\mathbf{c}}_N)\|_2^2 = \|(\mathbf{y}_{N,\|} - \mathbf{Y}_{t,\|}\mathbf{W}_t\hat{\mathbf{c}}_N)\|_2^2 + \|(\mathbf{y}_{N,\perp} - \mathbf{Y}_{t,\perp}\mathbf{W}_t\hat{\mathbf{c}}_N)\|_2^2 \tag{G.6}$$

and

$$\|(\mathbf{y}_N - \mathbf{Y}_t\mathbf{W}_t\bar{\mathbf{c}}_N)\|_2^2 = \|(\mathbf{y}_{N,\|} - \mathbf{Y}_{t,\|}\mathbf{W}_t\bar{\mathbf{c}}_N)\|_2^2 + \|(\mathbf{y}_{N,\perp} - \mathbf{Y}_{t,\perp}\mathbf{W}_t\bar{\mathbf{c}}_N)\|_2^2. \tag{G.7}$$

With (G.4)~(G.7), direct manipulations show

$$\|(\mathbf{y}_{N,\perp} - \mathbf{Y}_{t,\perp}\mathbf{W}_t\bar{\mathbf{c}}_N)\|_2^2 \leq \|(\mathbf{y}_{N,\perp} - \mathbf{Y}_{t,\perp}\mathbf{W}_t\hat{\mathbf{c}}_N)\|_2^2. \tag{G.8}$$

Based on (G.2), (G.3), and (G.8), it can be deduced that the following inequality

$$\|(\mathbf{y}_{N,\perp} - \mathbf{Y}_{t,\perp}\mathbf{W}_t\bar{\mathbf{c}}_N)\|_2^2 \leq \sigma^2\left(1 - \frac{d_N}{n}\right)(1 + \eta_1^2) \tag{G.9}$$

holds with a probability at least $1 - \exp\left(\frac{-(1-\log 2)}{2}(n - d_N)\left(1 - (w_{t_1}^{(N)})^2/2\right)^2\right) - N^{-3}$. Therefore, the proof of (5.50) is completed. Under the assumption that (G.9) holds, the inequality in (5.51) holds with a probability at least $1 - e^{-\sqrt{m_N d_N}} - N^{-2}$ (the arguments are similar to the proof of [10, Lemma 8.5] and the details are thus omitted). Hence, (5.51) holds with a probability at least $1 - e^{\frac{-(1-\log 2)}{2}(n-d_N)\left(1-(w_{t_1}^{(N)})^2/2\right)^2} - e^{-\sqrt{m_N d_N}} - N^{-2} - N^{-3}$. The proof of Lemma 5.4 is thus completed. Finally, we prove Lemma G.1.

*[Proof of Lemma G.1]:*

$$\Pr\left\{\|\mathbf{y}_{N,\perp} - \mathbf{Y}_{t,\perp}\mathbf{W}_t\hat{\mathbf{c}}_N\|_2^2 \leq \frac{\sigma^2}{2}\left(1 - \frac{d_N}{n}\right)(1 + \|\hat{\mathbf{c}}_N\|_2^2)\right\}$$

$$\overset{(a)}{\geq} \Pr\left\{\|(1/w_{t_1}^{(N)})\mathbf{y}_{N,\perp} - (1/w_{t_1}^{(N)})\mathbf{Y}_{t,\perp}\hat{\mathbf{c}}_N\|_2^2 \leq \frac{\sigma^2}{2}\left(1 - \frac{d_N}{n}\right)(1 + \|\hat{\mathbf{c}}_N\|_2^2)\right\}$$

$$= \Pr\left\{\left(1/w_{t_1}^{(N)}\right)^2 \|y_{N,\perp} - \mathbf{Y}_{t,\perp}\hat{c}_N\|_2^2 \leq \frac{\sigma^2}{2}\left(1 - \frac{d_N}{n}\right)(1 + \|\hat{\mathbf{c}}_N\|_2^2)\right\} \tag{G.10}$$

$$\overset{(b)}{=} \Pr\left\{\left(1/w_{t_1}^{(N)}\right)^2 \frac{\sigma^2}{n}(1 + \|\hat{\mathbf{c}}_N\|_2^2)\mathcal{X}_{n-d_N}^2 \leq \frac{\sigma^2}{2}\left(1 - \frac{d_N}{n}\right)(1 + \|\hat{\mathbf{c}}_N\|_2^2)\right\}$$

$$= \Pr\left\{\mathcal{X}_{n-d_N}^2 \leq (n - d_N)(w_{t_1}^{(N)})^2/2\right\} \overset{(c)}{\geq} 1 - \exp\left(\frac{-(1-\log 2)}{2}(n - d_N)\left(1 - (w_{t_1}^{(N)})^2/2\right)^2\right)$$

where (a) holds since $1/w_{t_1}^{(N)} > 1$, (b) follows from Lemma 8.4 in [10] and (c) holds from (F.1). Inequality (G.2) is thus proved.

To establish (G.3) we note that, for fixed $\mathbf{Y}_{t,\parallel}$ and $\mathbf{W}_t$, $\hat{\mathbf{c}}_N$ in (G.2) is a continuous function of $\mathbf{y}_{N,\parallel}$. Since the norm $\|\cdot\|_2$ is also continuous, $\|\hat{\mathbf{c}}_N\|_2$ is a continuous function of $\mathbf{y}_{N,\parallel}$. Therefore, if the domain of $\|\hat{\mathbf{c}}_N\|_2$ is restricted to $\{\mathbf{x}: \|\mathbf{x}\|_2 \in [3/4, 5/4]\}$, which is a compact set, by the extreme value theorem there exists a constant $\eta_1$ such that $\|\hat{\mathbf{c}}_N\|_2 \leq \eta_1$. By Lemma A.4 in [10], the restriction holds with a probability at least $1 - N^{-3}$. The proof of (G.3) is completed. □